\documentclass[a4paper,fleqn,usenatbib]{mnras}

\usepackage{newtxtext,newtxmath}
\usepackage{enumitem}

\usepackage[T1]{fontenc}
\usepackage{ae,aecompl}
\usepackage{multirow}

\usepackage{graphicx}	
\usepackage{amsmath}	
\usepackage{amssymb}	

\usepackage{color}
\usepackage{BibDef}
\def\lsim{\mathrel{\rlap{\lower 3pt \hbox{$\sim$}} \raise 2.0pt \hbox{$<$}}}
\def\gsim{\mathrel{\rlap{\lower 3pt \hbox{$\sim$}} \raise 2.0pt \hbox{$>$}}}
\def\msun{\rm {M_\odot}}

\def\mach{\mathcal{M}}

\def\grav{\rm G}
\def\gizmo{\textsc{gizmo}}

\title[High-redshift QSOs] 
{High-redshift quasars and their host galaxies I: kinematical and dynamical properties and their tracers}
\author[A. Lupi et al.]{Alessandro Lupi,$^{1}$\thanks{E-mail:
alessandro.lupi@sns.it} Marta Volonteri,$^2$ Roberto Decarli,$^3$ Stefano Bovino,$^4$\newauthor  Joseph Silk,$^{2,5,6,7}$ and Jacqueline Bergeron$^2$\\
$^1$Scuola Normale Superiore, Piazza dei Cavalieri 7, I-56126 Pisa, Italy\\
$^2$Sorbonne Universit\`{e}s, UPMC Univ Paris 6 et CNRS, UMR 7095, Institut d'Astrophysique de Paris, 98 bis bd Arago, F-75014 Paris, France\\
$^3$INAF -- Osservatorio Astronomico di Bologna, via Gobetti 93/3, I-40129, Bologna, Italy\\
$^4$Departamento de Astronomía, Faculdad Ciencias Físicas y Matemáticas, Universidad de Concepción, Av. Esteban Iturra s/n Barrio Universitario, \\Casilla 160, Concepción, Chile\\
$^5$AIM-Paris-Saclay, CEA/DSM/IRFU, CNRS, Univ Paris 7, F-91191 Gif-sur-Yvette, France\\
$^6$Department of Physics and Astronomy, The Johns Hopkins University, Baltimore, MD 21218, USA\\\
$^7$BIPAC, University of Oxford,1 Keble Road, Oxford OX1 3RH, UK
}

\begin{document}

\date{Draft \today}

\pagerange{\pageref{firstpage}--\pageref{lastpage}} \pubyear{2018}

\maketitle

\label{firstpage}

\begin{abstract}

Observations of high-redshift quasars provide information on the massive black holes (MBHs) powering them and the galaxies hosting them. Current observations of $z \gtrsim 6$ hosts, at sub-mm wavelengths, trace the properties of cold gas, and these are used to compare with the correlations between MBHs and galaxies characterising the $z=0$ population.  The relations at $z=0$, however, rely on stellar-based tracers of the galaxy properties. 
We perform a very-high resolution cosmological zoom-in simulation of a $z=7$ quasar including state-of-the-art non-equilibrium chemistry, MBH formation, growth and feedback, to assess the evolution of the galaxy host and the central MBH, and compare the results with recent ALMA observations of high-redshift quasars. 
We measure both the stellar-based quantities used to establish the $z=0$ correlations, as well as the gas-based quantities available in $z \gtrsim 6$ observations, adopting the same assumptions and techniques used in observational studies.   The high-redshift studies argued that MBHs at high redshift deviate from the local MBH-galaxy correlations.  In our analysis of the single galaxy we evolve, we find that the high-redshift population sits on the same correlations as the local one, when using the same tracers used at $z=0$. When using the gas-based tracers, however, MBHs appear to be over-massive.  The discrepancy between local and high-redshift MBHs seems caused by the different tracers employed, and necessary assumptions, and not by an intrinsic difference. Better calibration of the tracers, higher resolution data and availability of facilities that can probe the stellar population will be crucial to assess precisely and accurately high-redshift quasar hosts.

\end{abstract}

\section{Introduction}

Massive black holes (MBHs) are ubiquitously observed in the centre of massive galaxies at all redshifts. By spanning a mass range that goes from about $10^5\,\msun$ up to $10^{9-10}\,\msun$, they also represent the most massive compact objects in the Universe.  Observationally, they are usually detected during accretion events, when a fraction of the energy of the accreting gas is released as radiation, making them shine as active galactic nuclei (AGN), or produces powerful collimated jets. According to  Soltan's argument \citep{soltan82}, these MBHs must have gained most of their mass via gas accretion, and they should have formed as smaller objects (called `seeds') in the early Universe.

Recent observations of quasars at high redshift have shown that MBHs with masses around 10$^9\,\msun$ were already present when the Universe was less than 1 Gyr old \citep{fan06,mortlock11,banados18}, posing tight constraints on the mechanisms proposed to explain their formation and early growth. A crucial requirement to improve our understanding of how these MBHs formed and managed to grow so quickly is to properly resolve the environment around these objects and the properties of their galaxy hosts. Unfortunately, while in the nearby Universe the contribution of the galaxy and the central MBH to the total emission can be easily separated, at high redshift we are mainly limited to the most luminous sources, which, because of the intrinsic angular size of the galaxy and the limited resolution, are dominated by the central source emission, preventing us from properly resolving the galaxy host. 

The advent of current facilities like ALMA has been crucial for our understanding of these objects. By probing far-infrared lines that are uncontaminated by the quasar light and unaffected by dust extinction, in particular the [CII] line at 158~$\mu$m, i.e., the main coolant of the cold (30\,K $< T <$ 3000\,K) interstellar medium (ISM) and the CO line emission, ALMA is giving important information about the kinematics and dynamics of molecular gas within the quasar hosts. In the near future, the James Webb Space Telescope (JWST) will extend these capabilities to the near-infrared band, providing us with a unique set of tools that will finally help us to shed light on the peculiar conditions in which these MBHs form and evolve.

Compared to the local population, MBHs at high redshift seem to be overmassive by about an order of magnitude at a given galaxy mass \citep{walter04,venemans17,decarli18}, and they appear to deviate from the low-redshift $M$-$\sigma$ relation \citep{shields06,coppin08a,wang10}. The trend of increasing ratio between MBH and galaxy mass with redshift appears also in lower redshift samples \citep[e.g.,][]{merloni10,decarli10}. However, intrinsic limitations in observational techniques and selection criteria can affect the estimates of the different properties or their statistical significance in particular  at high redshift, where only the most massive/luminous objects in relatively common galaxies can be observed \citep[e.g.][]{lauer07,vestergaard08,volonteri11a}.

Recently, many authors have investigated this peculiar class of objects with semi-analytic models \citep[e.g.][]{valiante11,valiante14,pezzulli16} and numerical simulations \citep{costa14,richardson16,dimatteo17,smidt18,barai18}, employing different techniques and addressing different questions. For instance, \citet{dimatteo17} focus on the conditions for the efficient growth of these MBHs, while \citet{costa14} and \citet{smidt18} assess the role of the AGN feedback and X-ray AGN radiation in suppressing star formation (SF) in the MBH vicinity and in creating an HII region around the galaxy. \citet{richardson16} and \citet{barai18} investigate the impact of AGN feedback in a $z=5$ proto-cluster of galaxies and in a $z=6$  halo of $10^{12}\,\msun$, respectively.

In this study, we investigate the evolution of a quasar host at high-redshift through very-high resolution numerical simulations of a massive galaxy and its central MBH ($M_{\rm halo} = 1.5\times 10^{12}\,\msun$ at $z=7$), employing a detailed sub-grid modelling that also includes non-equilibrium chemistry of the primordial species, to  follow directly  the kinematics and dynamics of molecular hydrogen in the ISM.

This is the first of a series of paper addressing properties of high-redshift quasar hosts and their MBHs. In this paper, we present and discuss the main evolution of the target galaxy and its central MBH, focussing on the stellar and gas tracers (total gas and [CII] emission). In paper II (Lupi et al. in prep.) we will focus on the molecular hydrogen, directly traced in the simulation, and in Paper III (Lupi et al. in prep.) we will discuss the evolution of the entire MBH population forming during the simulation, and possible outflows from the MBHs.


\section{Simulation setup}
\label{sec:setup}
We study the evolution of a high-redshift quasar host by means of a zoom-in cosmological simulation performed with the hydrodynamic code \gizmo{} \citep{hopkins15}.

\subsection{The hydrodynamic code GIZMO}
\textsc{gizmo} \citep{hopkins15} is a particle-based code that descends from \textsc{Gadget3} and \textsc{Gadget2} \citep{springel05}. It implements a new method to solve the hydrodynamic equations that exhibits at the same time the intrinsic adaptivity and almost perfect conservation of angular momentum typical of smoothed particle hydrodynamics codes, and the excellent shock-capturing properties of grid-based codes.

The code is based on a partition scheme to sample the simulated volume via a set of tracer `particles' corresponding to unstructured cells. Compared to moving-mesh codes \citep[e.g. \textsc{arepo};][]{springel10}, in \textsc{gizmo} the effective volume of each cell is computed via a kernel smoothing procedure. Hydrodynamics is then treated like in mesh-based codes, with the Riemann problem solved with a Godunov-like method. Here, we employ the mesh-less finite-mass method, i.e. the mass preserving one available in the code, with a cubic-spline kernel, for which we set the desired number of neighbours to 32.
Gravity is solved via a Barnes-Hut tree, as in \textsc{Gadget3} and \textsc{Gadget2}. We keep the gravitational softening of dark matter and stars fixed at 40 and 10~pc~h$^{-1}$. For gas we employ the fully adaptive gravitational softening, where the softening length is maintained equal to the kernel size of the particle in an intrinsically adaptive way. This avoids the main issues arising when two different resolutions are used \citep{hopkins15}. The maximum spatial resolution for gas, determined by the minimum allowed softening, is 2.5~pc~h$^{-1}$, corresponding to an inter-particle spacing of $\sim 5$~pc. However,  this value is reached only in the very high density clumps just before a SF event.

We now summarise the sub-grid prescriptions employed in the simulation.
\begin{itemize}
\item The chemical network and cooling/heating processes, modelled via \textsc{krome} \citep{grassi14}, are the same as in \citet{lupi18a}, and include non-equilibrium chemistry for 9 primordial species (H,H$^+$,He,He$^+$,He$^{++}$,H$^-$,H$_2$, H$_2^+$, and e$^-$) with H$_2$ formation via H$^-$ associative detachment and on dust \citep[see][for details]{bovino16}, a metagalactic ultraviolet (UV) background \citet{haardt12}, and look-up metal cooling tables tabulated by \citet{shen13} and obtained with Cloudy \citep{ferland13}. The metal cooling tables take into account the first 30 elements in the periodic table, and assume photo-ionisation equilibrium of an optically thin gas with the extragalactic UV background as a function of $z$ \citep{shen10}. Because of the limited resolution, we also include a clumping factor $C_\rho = \exp(\sigma_s^2) = 1+b^2\mach^2$ in the H$_2$ formation rate on dust, where $\mach$ is the Mach number.

\item SF is implemented using a stochastic prescription that converts gas particles into stellar particles, where the star formation rate (SFR) density is defined as
\begin{equation}
\dot{\rho}_{\rm SF} = \varepsilon \frac{\rho_{\rm g}}{t_{\rm ff}},
\end{equation}
with $\varepsilon$ the SF efficiency parameter, $t_{\rm ff}=\sqrt{3{\rm \pi}/(32 G \rho_{\rm g})}$ the free-fall time, $\rho_{\rm g}$ the local gas density, and $G$ the gravitational constant.
$\varepsilon$ is computed according to the theoretical studies of turbulent magnetised clouds by \citet{padoan11}, based on the assumption of a Log-Normal gas density distribution with average density $\rho_{\rm g}$ and the width of the underlaying gaussian distribution $\sigma_s = \ln(1+b^2\mach^2)$. The $b$ parameter, representing the ratio between solenoidal and compressive modes, is set to 0.4, that corresponds to a statistical mixture of the two modes \citep{federrath12}. $\varepsilon$ can be expressed as \citep{federrath12}
\begin{equation}
\varepsilon=\frac{\varepsilon_\star}{2\phi_{\rm t}}\exp\left({\frac{3}{8}\sigma_s^2}\right)\left[1+{\rm erf}\left({\frac{\sigma_s^2-s_{\rm crit}}{\sqrt{2\sigma_s^2}}}\right)\right],
\label{eq:sfeff}
\end{equation}
where $\varepsilon_\star=0.5$ is the normalisation calibrated against observations \citep{heiderman10}, $1/\phi_{\rm t}=0.49$ is a fudge factor that takes into account the free-fall time-scale uncertainty, and $s_{\rm crit}(\alpha_{\rm vir},\mach)$ is the critical logarithmic density for SF and depends on the virial parameter $\alpha_{\rm vir}$, that determines how bound the cloud is, and $\mach$ \citep{federrath12}. 
As in \citet{lupi19}, we define
\begin{equation}
\alpha_{\rm vir} = \frac{5[\|\nabla\otimes\mathbf{v}\|^2+(c_{\rm s}/L)^2]}{\pi\grav \rho_{\rm gas}},
\end{equation} where $L\approx 0.5 h$ is the particle grid-equivalent size, with $h$ the particle smoothing length, $\sigma_{\rm eff}\equiv L\sqrt{\|\nabla\otimes\mathbf{v}\|^2+(c_{\rm s}/L)^2}$ the total support against gravitational collapse\footnote{We employ the velocity gradient before the slope-limiting procedure is applied, as in \citet{lupi19}.}, and $c_{\rm s}$ the sound speed.
SF is allowed for gas matching two criteria\footnote{As discussed in \citet{lupi18a}, the density threshold does not play any role and is kept only for numerical reasons, to avoid wasting time computing the SF rate in low-density, unbound regions.}: i) $\rho_{\rm g}> m\rm _H\, cm^{-3}$ and ii) $\mach>2$.
 
\item Stellar radiation is implemented  as in model {\it b} of \citet{lupi18a}, by collecting all the stellar sources in the gravity tree.

\item Supernova feedback is based on the mechanical feedback prescription in \citet{lupi19}, a variation of the publicly available implementation by \citet{hopkins18}, that is able to reproduce the terminal momentum of state-of-the-art high-resolution simulations \citep{kimostriker15,martizzi15} independent of resolution. However, a consensus on the terminal momentum from clustered SNe has not been reached yet, with different models varying from almost no variation with respect to the single SN case \citep{kimostriker17,gentry18} up to one order of magnitude higher values \citep{keller14,gentry17}. Here, to make SN feedback more effective, we assume a terminal momentum twice that in \citet{lupi19}.
Stellar particles represent an entire stellar population following a Chabrier initial mass function \citep[IMF;][]{chabrier03}. Here we consider both type II and type Ia SNe. SN explosions are modelled as discrete events, and not as a continuous source, as in \citet{lupi19}. We assume that stars in the range $8-40\,\msun$ explode as type II SNe, and release $10^{51}$~erg of energy and an IMF-averaged ejecta mass $M_{\rm ej}=15.1452-M_{\rm NS}=13.7452\,\msun$, with $M_{\rm NS}=1.4\,\msun$ the mass of the remnant neutron star, an oxygen mass $M_{\rm oxy}=1.2403\,\msun$ and an iron mass $M_{\rm iron}=0.10422\,\msun$. Assuming the solar ratio for the Oxygen and Iron groups, the total metal mass injected by every SN is $M_{\rm Z} = 2.09M_{\rm oxy}+1.06M_{\rm iron}=2.7028\,\msun$.  The type Ia SN rates are instead based on the delay-time distribution by 
\citet{maoz12}, and cover the stellar age interval 0.1-10~Gyr. Every type Ia SN is assumed to inject $10^{51}$~erg, $M_{\rm ej}=1.4\,\msun$ and $M_{\rm Z}=0.9604\,\msun$ ($M_{\rm oxy}=0.14\,\msun$ and $M_{\rm iron}=0.63\,\msun$).
Every time a SN explodes, we distribute mass, metals, momentum, and thermal energy among all the gas neighbours within the star particle kernel  (defined as the sphere enclosing 64 neighbours) and those whose kernel size encompasses the star particle, as in \citet{hopkins18}. Every particle receives a fraction of the total SN mass, metals and energy, according to the solid angle covered by the particle around the star, which is further symmetrised via a tensor renormalization to ensure the explosion is spherical in the stellar reference frame. The proper amount of thermal vs kinetic energy to be given to each neighbour is then determined according to the gas properties, so that the feedback is mostly thermal if the cooling mass is resolved and mostly kinetic if not (see \citet{lupi19} for details).
\item For low-mass stars, we only distribute the mass lost in slow winds and the initial momentum to the neighbours, without accounting for possible shocks with the interstellar medium. 
\end{itemize}

The total mass recycling of our stellar feedback prescription is about 42 per cent of the initial stellar mass over a Hubble time \citep[see][and references therein]{kim14}. 
We also include an additional time-step limiter for the stellar particles in the simulation, to ensure single SN events are resolved. In fact, the time-step is limited to $\sim 5\times10^4$~yr, i.e. 1/100th of a $40\,\msun$ star lifetime, for particles younger than 100 Myr (dominated by Type II SN events). This limit is instead increased to 1/10th of the stellar population age when the stellar particle is older than 100 Myr.

\subsection{Black hole physics}
Here, we describe the prescriptions we implemented in \gizmo{} to model BH seeding, accretion, feedback, and the merger of BH binaries.

\subsubsection{BH seeding}
\label{sec:seeding}
To follow BH evolution in a cosmological context, we need BHs to consistently form during the simulation, similarly to stars. The focus of this study is on the properties of the host galaxies, not on the properties of seed BHs,  therefore we devise a scheme that ensures numerical stability and results in BH masses consistent with the available observations by $z=7$. 

We seed BHs in galaxies with a stellar mass $M_{\rm star} >10^8\, \msun$, unless another BH is already present. Unlike for SF, a strictly local process, our prescription for BH seeding requires global information about entire galaxies. In order to collect this information on-the-fly in the simulation, we employ the Friends-of-Friends (FoF) algorithm available in \gizmo{}. 
We first identify DM groups assuming a linking length $d_{\rm FoF} = 0.05$,\footnote{The linking length is defined in units of the average inter-particle distance of the high-resolution DM particles in the box $\Delta x_{\rm DM} = [M_{\rm DM}/(N_{\rm DM}\rho_{z,\rm DM})]^{1/3}$, where $M_{\rm DM}$ is the total mass in high-resolution DM particles, $N_{\rm DM}$ is their number, and $\rho_{z,\rm DM}$ is the DM cosmological density at redshift $z$. Although the commonly assumed linking length for DM halo identification is $d_{\rm FoF} = 0.2$, we found that a smaller value is a better choice to avoid contamination from sub-haloes and to limit the selection to the stellar-populated region.} and then attach gas, stellar and BH particles to the group of the closest DM particle. Finally, we flag only the groups that match our criteria for seed BH formation. In all the flagged groups, a new seed is finally spawn at the position of the star surrounded by the highest gas density (typically close to the galaxy centre), with an initial mass $M_{\bullet,\rm init} = 10^6\,\msun$. The gravitational softening of BH particles is set to 5~pc.

Compared to other studies, our initial BH mass here is ten to hundred times higher. This choice is motivated by the requirement of avoiding spurious scattering of BHs out of the galaxy centre. This is a common issue in numerical simulations, when the BH mass is comparable or smaller than that of the other particles of the simulation, and can be solved by means of ad-hoc sub-grid prescriptions like i) an instantaneous re-centering \citep{springel05a,sijacki07,booth09,schaye15,barai18}, ii) a correction for unresolved dynamical friction \citep{dubois14a,tremmel15} or iii) an initially larger dynamical mass decoupled from the BH mass used for accretion \citep[e.g.][]{anglesalcazar17FIRE,biernacki17}. Unfortunately, all these prescriptions can introduce additional undesired effects in the simulation, like superluminal motions when the BH is far away (i), dynamical acceleration (ii) when the real dynamical friction is properly solved (\citealt{beckmann18}, see \citealt{tremmel15} and \citealt{pfister19}  for how to avoid this effect), or influence the galaxy dynamics (iii).
In our case, instead, we took a simpler approach and we opted for an already large initial BH mass, $\approx 12$ times larger than the DM particle mass and $\approx 60$ times larger than the gas/star particle mass. \citet{tremmel15} also advocate that small gravitational softening is needed, along with a sufficiently high ratio of MBH to particle mass, to properly resolve dynamics; in our simulation the extremely high spatial resolution, hence the small gravitational softening, is enough to correctly account for the dynamics throughout the evolution, except at most for the very early times immediately after BH formation.

\subsubsection{BH accretion}
BH accretion is implemented via the commonly adopted Bondi-Hoyle-Lyttleton formula \citep[][BHL hereafter]{hoyle39,bondi44,bondi52}, where the accretion rate for an homogeneous medium in relative motion around the BH $\dot{M}_{\rm BHL}$ is defined as
\begin{equation}
\dot{M}_{\rm BHL}= \frac{4\pi\grav^2 M_{\bullet}^2\rho_{\rm gas}}{(v_{\rm rel}^2+c_{\rm s}^2)^{3/2}},
\end{equation}
where $M_\bullet$ is the BH mass, $\rho_{\rm gas}$ is the gas density around the BH and $v_{\rm rel}$ is the gas-BH relative velocity. Because of the limited resolution, many previous studies, increased the BH accretion rate adopting a `boost' factor $\alpha$, either constant \citep[$\alpha=100$;][]{dimatteo05} or scaling with density \citep[$\alpha \propto \rho_{\rm gas}^2$;][]{booth09}. Here, thanks to the very-high resolution achieved, we can simply use the `real' BHL accretion rate. Following \citet{bellovary10}, \citet{choi12}, and \citet{tremmel17}, instead of the commonly employed kernel-weighted properties around the BH, we kernel-average the individual accretion rate for the particles within the BH kernel, obtaining: 
\begin{equation}
\dot{M}_{\rm accr}= \langle\frac{4\pi\grav^2 M_{\bullet}^2\rho_{\rm gas}}{(v_{\rm rel}^2+c_{\rm s}^2)^{3/2}}\rangle.
\end{equation}
For the BH, we consider a cubic spline with 32 neighbours. However, in order to prevent the BH accreting low density material very far away from it, the BH smoothing length is not allowed to exceed a maximum physical radius $h_{\rm max} = 150\rm\, pc\, h^{-1}$. When $h_{\rm max}$ is reached, only the particles within it are taken into account for the accretion. 

In addition, we cap the accretion rate at the Eddington limit, i.e. the maximum accretion rate for which radiation luminosity does not overcome the gravitational pull, defined as
\begin{equation}
\dot{M}_{\rm Edd}= \frac{4\pi \grav m_{\rm H}}{\eta_{\rm acc}c\sigma_{\rm T}}M_{\bullet},
\end{equation}
where $\eta_{\rm acc}$ is the accretion radiative efficiency, that we set to 0.1 \citep{soltan82}, $c$ is the speed of light, and $\sigma_{\rm T} = 6.65\times 10^{-25}\rm\, cm^2$ is the Thomson scattering cross section. 

\subsubsection{BH feedback}
Every time a BH accretes material, a fraction $\eta_{\rm acc}=0.1$ of the energy is converted into radiation, with only the remaining mass--energy actively contributing to the BH mass growth. Moreover, only part of it actually couples with the surrounding gas, heating or expelling it. Unfortunately, we do not have a unique physically-motivated value for this feedback efficiency, and different studies, depending on resolution, on the number of neighbouring particles/cells considered, and on the numerical technique employed, made different choices, with values ranging from a few $10^{-3}$ \citep{ostriker10} up to 0.05 \citep{dimatteo05} or 0.15 \citep{dubois12}. Here, we assume a moderately weak feedback, where only a fraction $\eta_{\rm fbk}=0.005$ of the energy produced is dumped on to the gas as thermal energy in a kernel-weighted fashion, among the 32 neighbours defining the kernel size around the BH. 
In Appendix \ref{app:validation}, we validate our BH model on an isolate Milky Way-galaxy simulation, to highlight how the choice of the accretion prescription and the feedback efficiency affect the BH growth with respect to the galaxy SFR.
In addition, in Appendix \ref{app:efficiency}, we also test the effect of different feedback efficiencies in the cosmological run by rerunning our simulation for $\sim 100$~Myr from $z=8$ to $z=7.2$, when the BH is already massive.

\subsubsection{BH mergers}
When two galaxies merge, the BHs hosted in their centres spiral towards the centre under the effect of dynamical friction on DM, gas and stars, until they bind in a BH binary. Then, three-body scattering or dynamical friction on gas kick in and harden the binary, down to the regime where gravitational wave emission leads to the coalescence of the two BHs. Cosmological simulations are unable to follow the entire binary formation and coalescence process, because of the huge dynamic range involved, and have to rely on sub-grid prescriptions to model BH mergers. 

The most common algorithm to treat BH binary mergers checks whether the BH separation approaches or hits the resolution limit, i.e. they are in the same cell (or in close enough cells) in mesh-based codes or they share the kernel in particle-based codes, and merges them into a new single more massive BH \citep[e.g.][]{dubois12}. Sometimes, a relative velocity/binding criterion is also employed, to prevent BHs to merge when they approach each other at very high speeds \citep[e.g.][]{sijacki09,schaye15}. 
In our case, we employ both a distance and binding criteria, merging the BHs only when their relative speed is lower than the escape velocity from the binary system. When these criteria are matched, the mass of the least massive BH is added to that of the primary BH, which is moved to the centre of mass of the pre-existing binary, and  linear momentum conservation is enforced.

\begin{figure}
\includegraphics[width=\columnwidth]{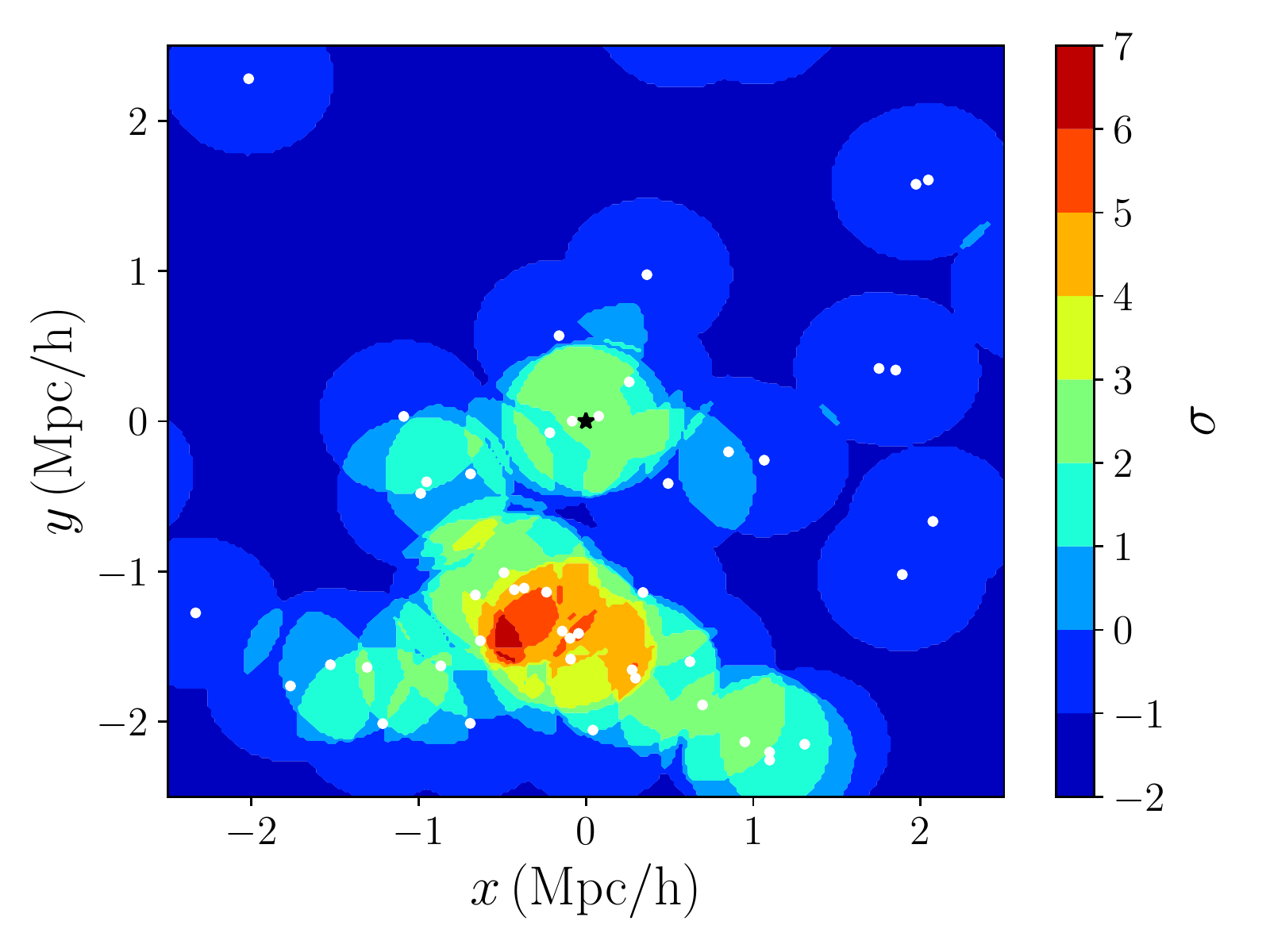}
\caption{Overdensity significance map for the candidate realisation in a box of 5~physical Mpc around the target halo, identified by a black star. The white dots correspond to the other haloes/galaxies in the box.}
\label{fig:selection}
\end{figure}

\subsection{Initial conditions}
According to recent results, high-redshift quasars are hosted in haloes with typical masses in the range $3\times 10^{11} - \times10^{13}$ \citep{dimatteo17,tenneti18}, they do not live in the most overdense regions of the Universe \citep{mazzucchelli17,uchiyama18}, and they are powered by BHs with masses above a few $10^8\,\msun$ up to $10^{10}\,\msun$.

We follow the evolution of a typical quasar host halo, with a virial mass $M_{\rm vir}\approx 3\times 10^{12}\,\msun$ at redshift $z=6$, adopting the \citet{planck16} cosmological parameters, with $\Omega_{\rm m}=0.3089$, $\Omega_\Lambda = 0.6911$, $\Omega_{\rm b}=0.0489$, $\sigma_8 = 0.8159$, $n_{\rm s} = 0.9667$, and $H_0 = 67.74\,\rm km\, s^{-1} Mpc^{-1}$, and we assume negligible contribution from both radiation and curvature.

Because of the rarity of such massive haloes, i.e. $\sim$1 in a 100~Mpc~$h^{-1}$ comoving box, where $h=H_0/(100\rm\, km\, s^{-1} Mpc^{-1}$), a very large box is necessary to guarantee the presence of at least one of these objects \citep[see, e.g.,][]{barai18}. However, this can be avoided by creating a constrained realisation in a smaller box \citep{bertschinger87,hoffman91}. Here, we consider a cosmological volume of 75~Mpc~$h^{-1}$ comoving, where we impose the collapse of a halo with $\sim 10^{12}\,\msun$ at $z=6$. The initial conditions are generated at $z=100$ by \textsc{music} \citep{hahn13}, with a coarse resolution grid of 256$^3$, that gives a DM mass resolution of $\approx 3\times 10^9\,\msun$, enough to resolve the target halo with about 1000 particles.

However, since we also want to enforce the overdensity criterion, a single realisation is not enough. Due to the constraint imposed to the random realization, the target halo is always the most massive halo in the simulation, and forms close to the centre of the box We have run several coarse-level DM-only simulations of the same box, with different random seeds, and we have identified all the haloes resolved by \textsc{amiga halo finder} \citep[][\textsc{ahf} hereafter]{knollmann09} with at least 50 particles. Then, we have used the stellar-to-halo mass relation by \citet{behroozi13} on the selected haloes to derive the rest-frame UV magnitude of the corresponding galaxies.
Since at $z\sim 5-6$, the UV emission is shifted to the local $i$-band, we followed \cite{uchiyama18} and filtered the selected haloes with $i<25$. Finally, we have computed the galaxy overdensity significance $\sigma=(N-\bar{N})/\sigma_{\rm N}$, with $N$ the number of galaxies in the aperture, and $\bar{N}$ and $\sigma_{\rm N}$ the average and the standard deviation of $N$ in all the considered region, in apertures of $0.75$~ proper Mpc ($\sim 1.8$~arcmin) around the target haloes of all realisations, and compared them with the results by \citet{uchiyama18}. {The chosen realisation corresponds to the one with $\sigma \sim 2-3$, reported in Fig.~\ref{fig:selection}, with the target halo shown as a black star.} 

\subsubsection{Refinement strategy for the high-resolution region}
To refine the zoom-in region up to the desired resolution, we follow the approach reported in \citet{fiacconi17}:
\begin{enumerate}
\item  From the results of the coarse-level DM-only simulation, we identify the halo using \textsc{ahf}. We flag all the DM particles enclosed in a sphere centred on the target halo of radius 2.5$r_{\rm vir}$, with $r_{\rm vir}$ the halo virial radius, and we trace them back in time to the initial conditions. 
\item We add one refinement level around the Lagrangian box surrounding the previously flagged DM particles, and we rerun the DM-only simulation.
\item We repeat step (i), but this time by adding two refinement levels in a convex hull region surrounding the flagged DM particles, and re-run the DM-only simulation.
\end{enumerate}

This step-by-step approach guarantees that the contamination in the high-resolution region is well below 1 per cent, and is completely absent within $r_{\rm vir}$. In the zoom-in region, the mass resolution achieved is $1.53\times 10^4\,\msun$ for gas and stellar particles, and $8.19\times 10^4\,\msun$ for DM, with a total number of particles of 67.5M in gas and 90.2M in DM.

\section{Results}
\label{sec:results}
We now present the main results of our simulation down to $z=7$. For all the following analyses, we identify the halo using \textsc{ahf}, and only select the particles within a sphere of radius $0.2 r_{\rm vir}$, to avoid contamination from satellite galaxies.

\begin{figure}
\includegraphics[width=\columnwidth]{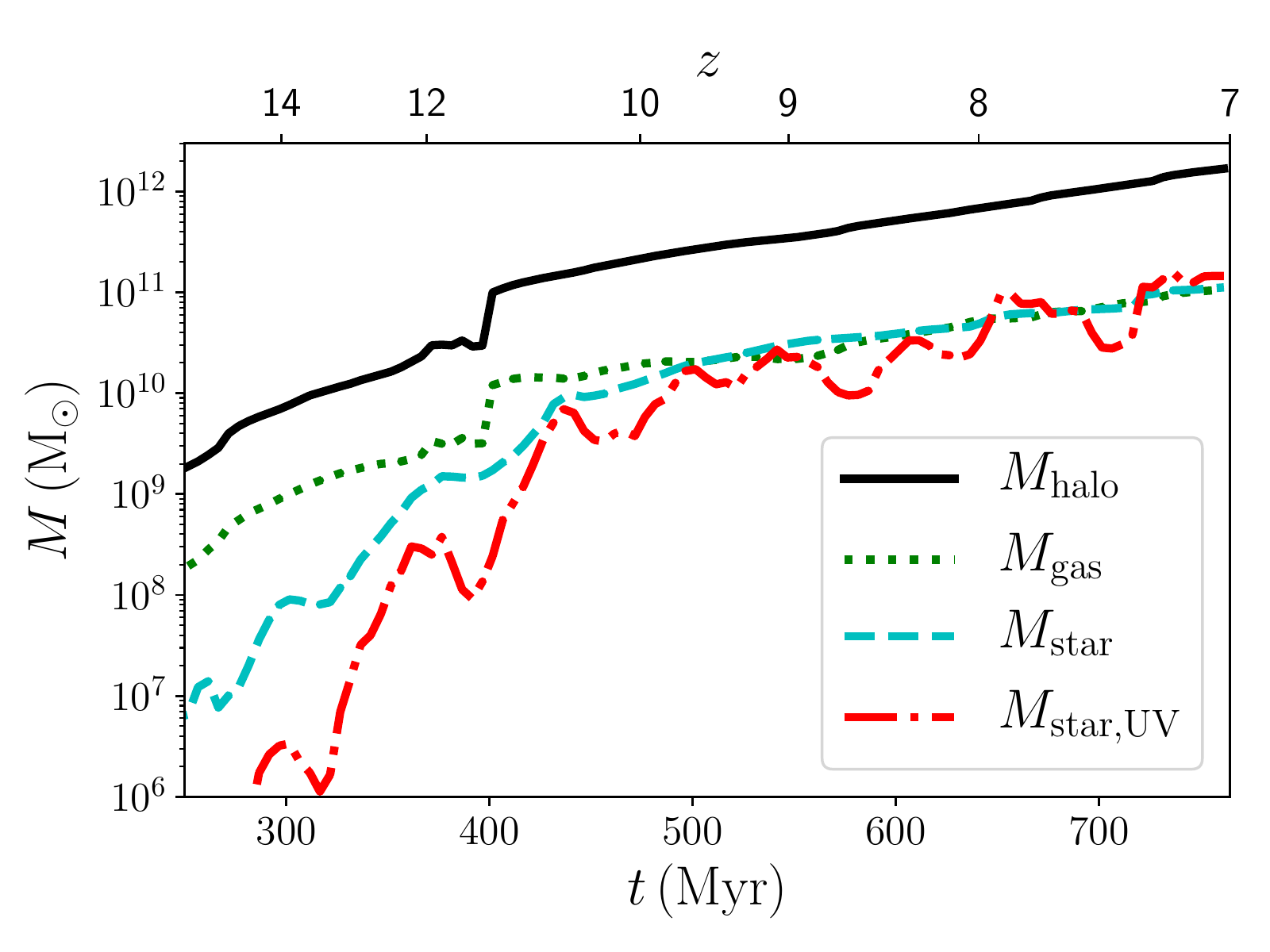}
\caption{Mass growth of the target galaxy. We show the halo mass as a black solid line, the stellar mass as a cyan dashed line and the gas mass as a green dotted line. In addition, we show the stellar mass estimate based on the rest-frame UV luminosity, as derived by \citet{song16}, as a red dot-dashed line. After a simultaneous merger of three haloes around $z\sim 11$, a massive halo dominating the local potential well emerges, marking the beginning of a phase of steady growth driven by smooth accretion or minor mergers. Except for the initial phase, when SN feedback strongly suppresses SF, gas and stars evolve in a similar way, maintaining a roughly constant mass ratio. Compared to the actual stellar mass, the UV-based estimate shows significant fluctuations, depending on the average age of the stellar population, but is generally consistent with the actual estimate. Only in the initial evolutionary stages, above $z=10$, they diverge significantly, because of the bursty SF history.}
\label{fig:mass_z}
\end{figure}

\subsection{Galaxy host and MBH evolution across cosmic time}
Here, we discuss the evolution of the target galaxy and of its central MBH during the first 800 Myr of the Universe. We defer the discussion about the other MBHs existing in the simulation to Paper III (in preparation).

\subsubsection{Halo and galaxy growth}

In Fig.~\ref{fig:mass_z} we show the halo mass $M_{\rm halo}$ (black solid), the stellar mass $M_{\rm star}$ computed using the actual mass of the stellar particles in the simulation (cyan dashed) and gas mass $M_{\rm gas}$ (green dotted) as a function of time. We also show as a comparison the stellar mass estimate based on the UV luminosity, as derived in \citet{song16}, as a red dot-dashed line \citep[see][for a discussion]{trebitsch18}. Despite the short time available, the typically high densities and the stronger turbulent motions in the interstellar medium result in a very rapid build up of the quasar host galaxy mass, up to $M_{\rm star} \sim 10^{11}\,\msun$ at $z=7$. The stellar mass at $z=7$ corresponds to a conversion efficiency of about $f_\star=M_{\rm star}/M_{\rm halo} \approx 0.06$, broadly consistent with the extrapolation of the stellar-to-halo mass relation from \citet{moster18}.

\begin{figure}
\includegraphics[width=\columnwidth]{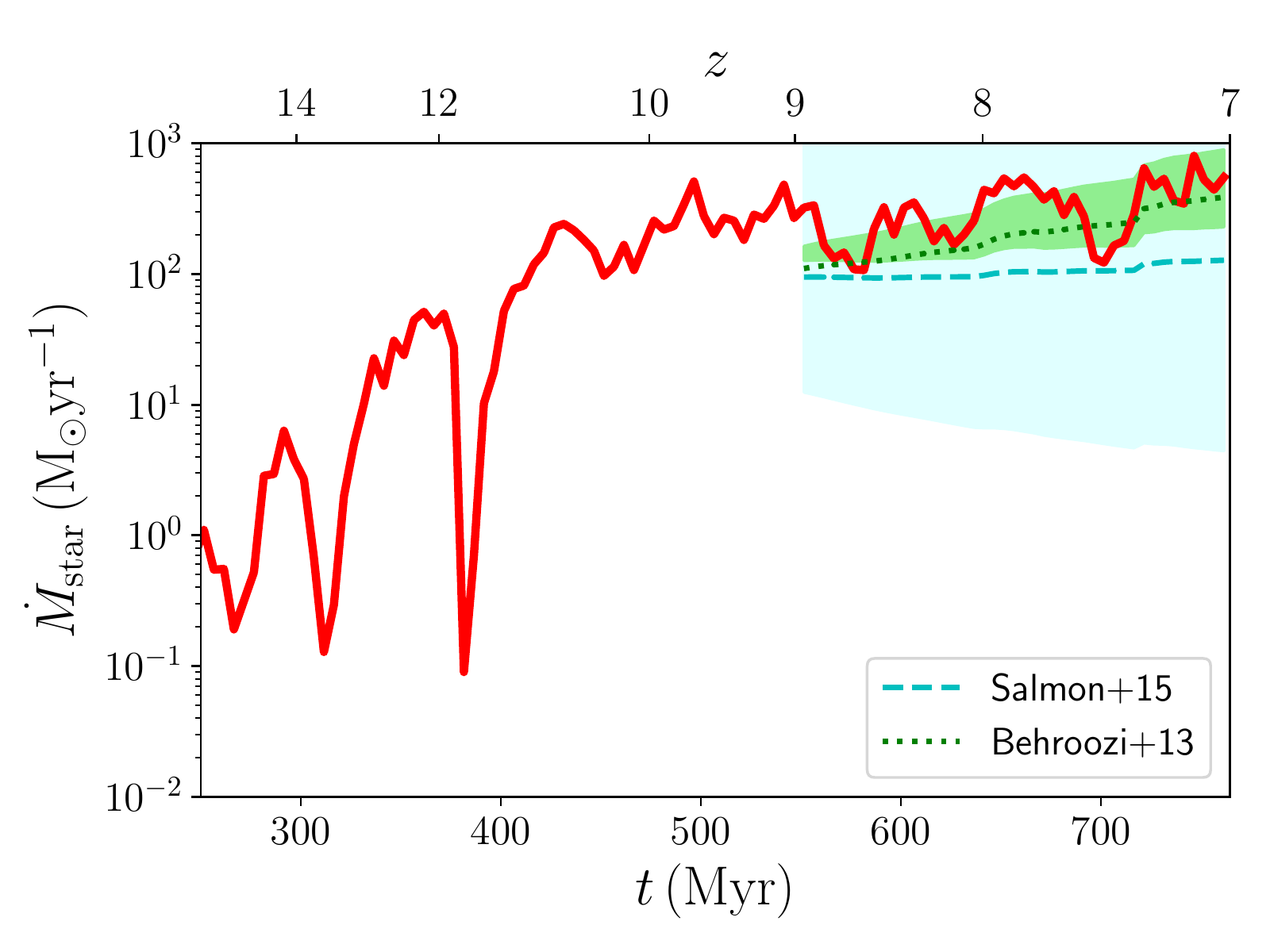}
\caption{SFR of the target galaxy as a function of time. Below $z=9$, we also show the best fit and the scatter of the expected SFR from the models by \citet{behroozi13} (empirical, extrapolated up to the stellar mass we have in the simulations) and \citet{salmon15} (observations) as a comparison. Except for the initial phase, when SN feedback strongly suppresses SF, gas and stars evolve in a similar way, maintaining a roughly constant mass ratio. Such high mass ratios are consistent with typical observations of high redshift galaxies \citep[e.g.,][]{tacconi10}.}
\label{fig:sfr}
\end{figure}

In the first 300~Myr ($z\gsim 11$), the main progenitor of the quasar host is not easily discernible from other haloes in the same region that show a comparable mass ($M_{\rm halo}\sim 2-3\times 10^{10}\,\msun$). Then, around $z\sim 11$, a significantly massive halo emerges from the simultaneous coalescence of three progenitors, reaching  $M_{\rm halo}\approx 10^{11}\,\msun$. 
 After this major event, a phase of steady growth begins, driven by smooth accretion or minor mergers, i.e., with mass ratio between the two interacting haloes less than 1:10.  

During the first 400~Myr, most of the baryons remain in the form of gas, because of the low efficiency of SF and the powerful effect of SN feedback that heats the gas up and sweeps it away. The burstiness of the star formation history in these stages is also reflected in the discrepancy between the actual mass and the UV-based diagnostic (which is a better tracer of continuous SF histories). This trend changes as soon as the galaxy reaches $M_{\rm star} \sim 10^{10}\,\msun$ \citep{dubois15,habouzit17}, when SNe become less effective and gas accumulates in the halo, settling in a disc-like structure. At this stage, self-regulation between gas cooling, SF, and SN feedback is able to stabilise the gas consumption and replenishing and yield a similar evolution, so that the gas-star ratio remains roughly constant. Similarly, below $z=10$, the UV-based stellar mass approaches the actual stellar mass, with only stronger fluctuations due to the evolution of the average age of the stellar population.

The quick build-up of a large stellar mass observed for these galaxies is the result of the massive inflows on to the halo and the very high SFRs shown in Fig.~\ref{fig:sfr}. During the first 200~Myr, multiple dips can be observed, corresponding to the ejection of gas by SN feedback. At later times, when the galaxy has become massive enough, the SFR becomes more stable with values ranging from 100 up to 800~$\rm \msun yr^{-1}$. This is in good agreement with observations of high redshift quasars by \citet[][D18 hereafter]{decarli18}, with SFRs in the range 100-2000~$\rm \msun yr^{-1}$.

At $z\sim 7$, the galaxy exhibits a SFR that is consistent with the main sequence according to the empirical model by \citet{behroozi13} (extrapolated up to the stellar mass we have), see the green shaded area and the dotted green line in Fig.~\ref{fig:sfr}. On the other hand, these SFRs are well above the best-fit of observed main sequence galaxies \citep{salmon15}, suggesting that these galaxies are more likely starburst rather than normal galaxies as shown by the shaded cyan area and the dotted cyan line in Fig.~\ref{fig:sfr}.

\begin{figure}
\includegraphics[width=\columnwidth]{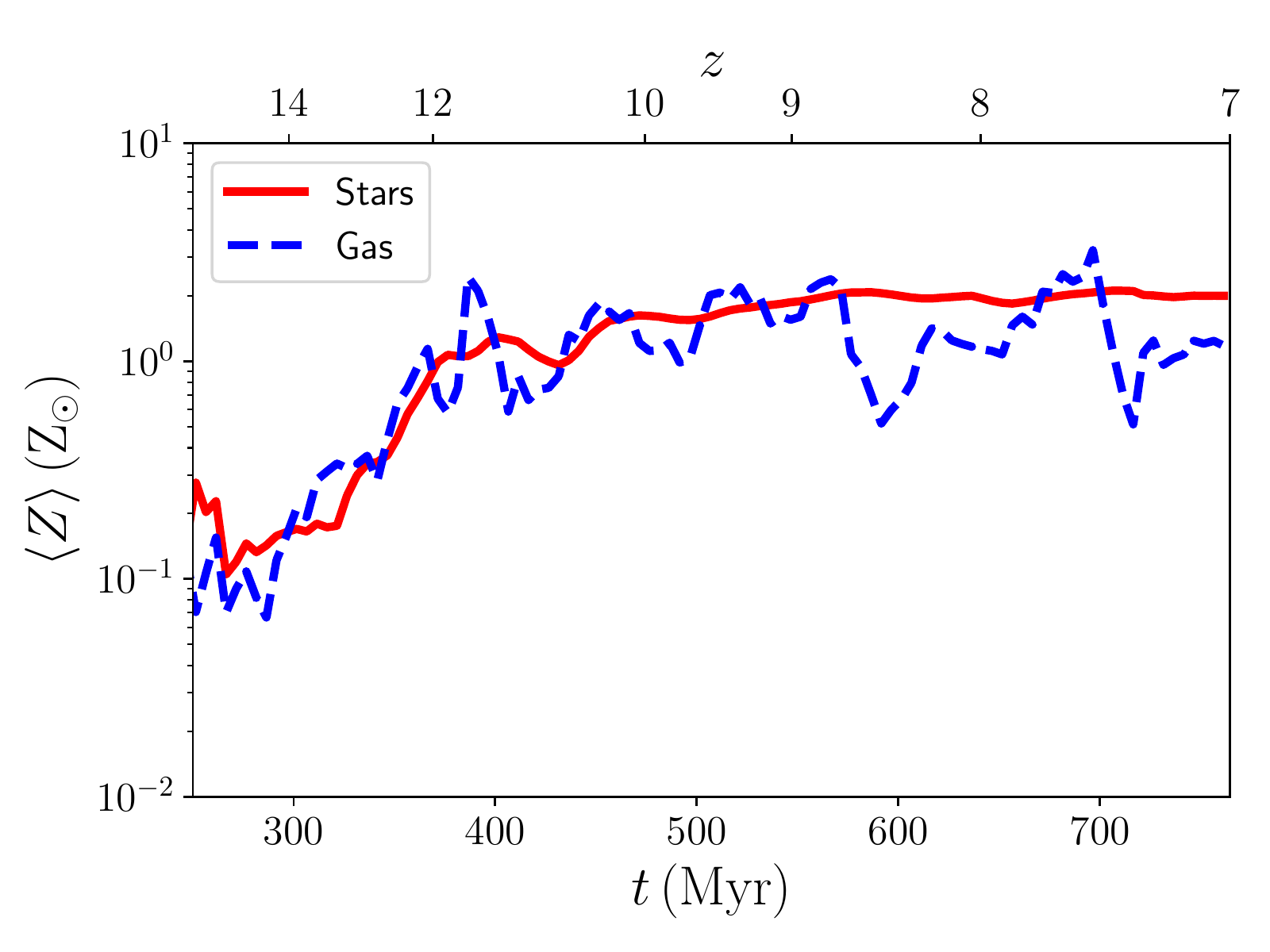}
\caption{Average metallicity of the target galaxy as a function of time in solar units. The  gas metallicity is shown as a blue dashed line, whereas the stellar metallicity as a red solid line. Because of the high SFR and the corresponding rate of SN events enriching the ISM, the metallicity steeply rises to about solar values in less than 200~Myr. At later times, the combined effect of  gas inflows and mergers result in the saturation of the metallicity  between one and two times the solar value, for both stars and gas. While the stellar metallicity exhibits a smooth evolution, the gas metallicity shows strong oscillations corresponding to large inflows of lower-metallicity gas.}
\label{fig:zave}
\end{figure}

Given the large number of SNe in the galaxy, the metallicity of the gas (and of the newly formed stars) rapidly increases  to about solar values, as shown in Fig.~\ref{fig:zave}. The stellar metallicity (red solid line) evolves smoothly with time, saturating around $Z_{\rm star}\approx 2\rm\, Z_\odot$, with small oscillations resulting from mergers with smaller lower-metallicity galaxies. The gas metallicity (blue dashed line) exhibits much stronger oscillations, mainly associated with massive inflow events of lower-metallicity gas from the filaments. Such large metallicities are consistent with the idea that quasar hosts at high redshift are already dust-rich, and produce a strong far-infrared emission of about $\sim 10^{12-13}\rm\, L_\odot$ \citep{venemans17,decarli18}.

\begin{figure*}
\Large Gas\hspace{6.6cm}Stars\\
\begin{center}
\includegraphics[width=0.38\textwidth,trim={5.7cm 4.9cm 4.5cm 5.3cm},clip]{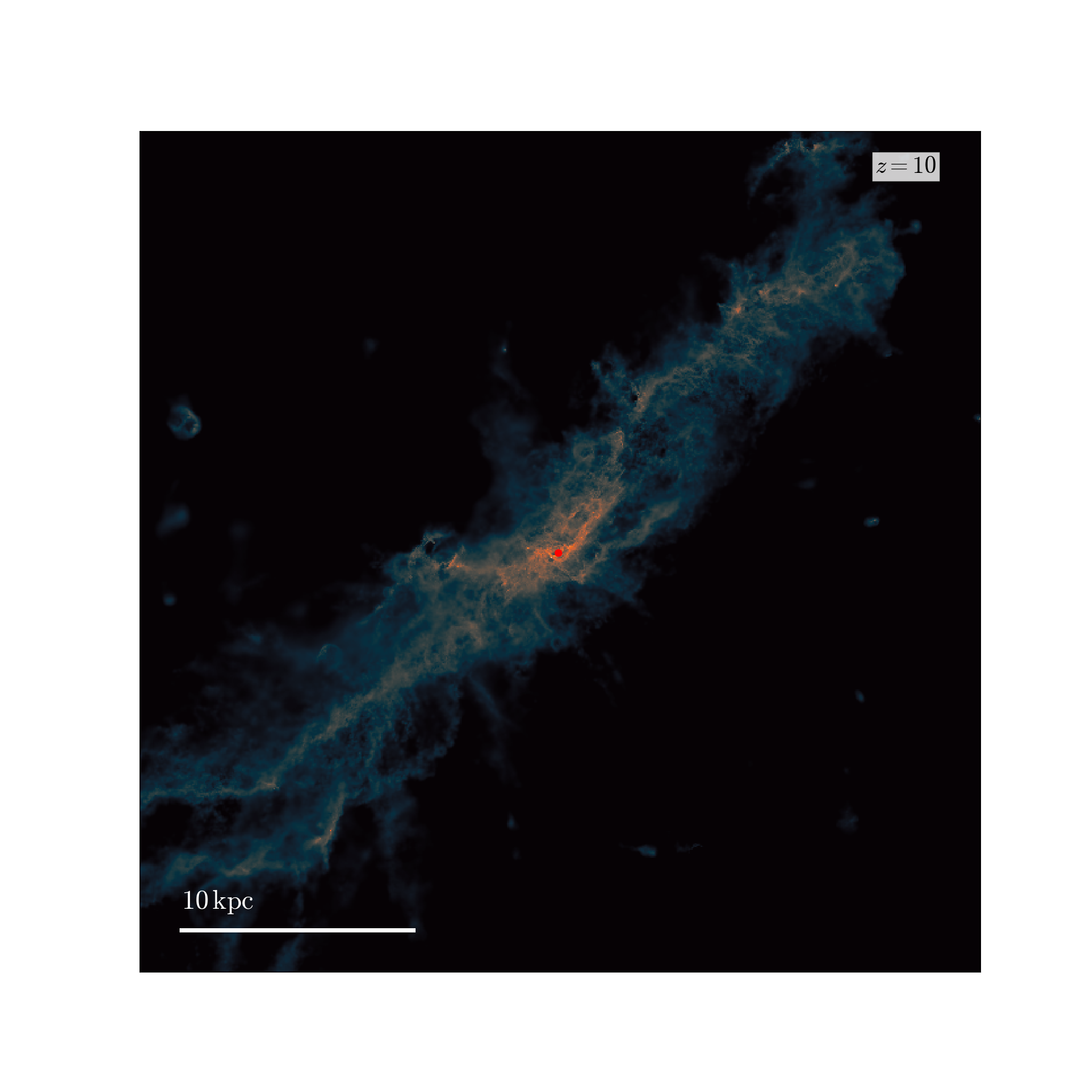}
\includegraphics[width=0.38\textwidth,trim={5.7cm 4.9cm 4.5cm 5.3cm},clip]{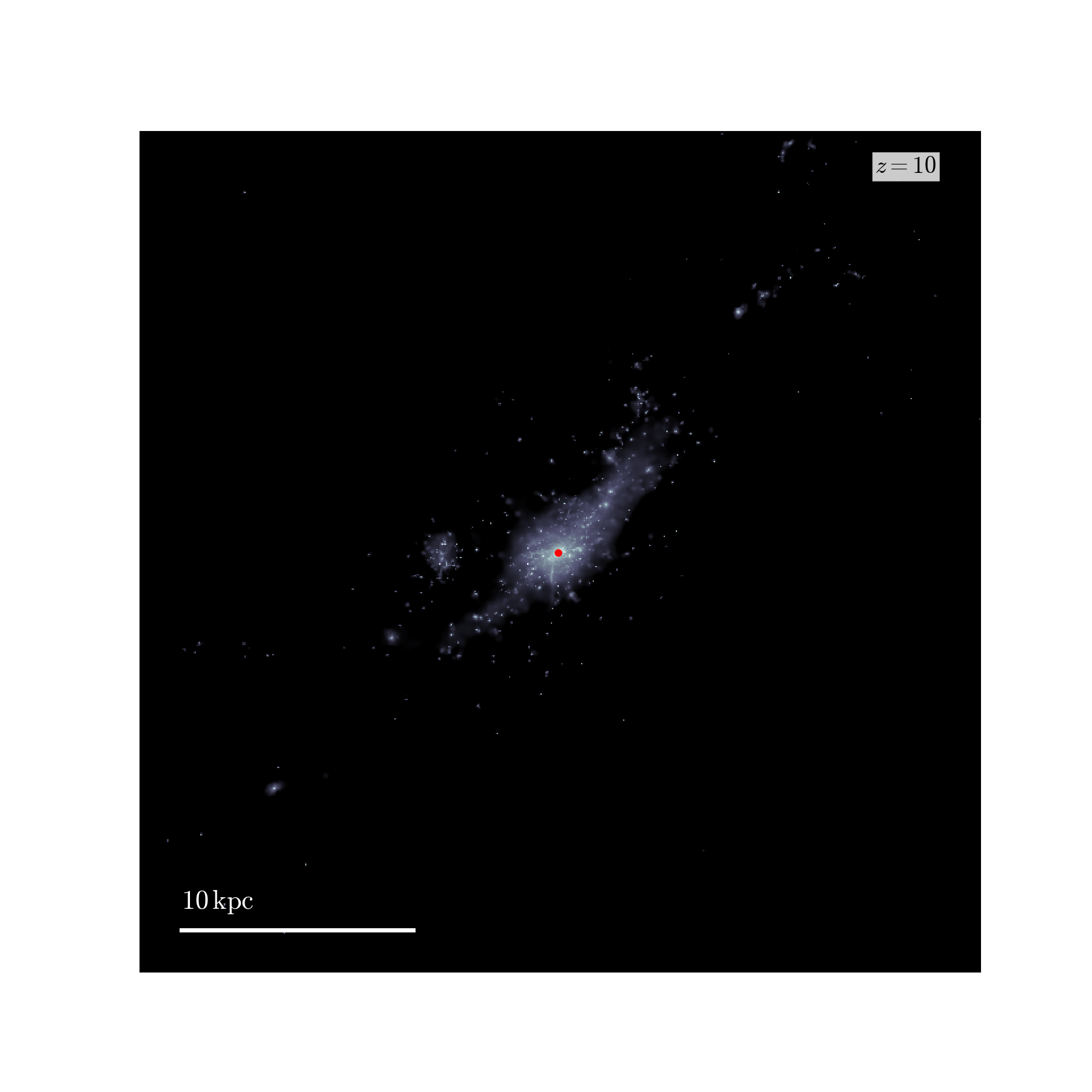}\\
\includegraphics[width=0.38\textwidth,trim={5.7cm 4.9cm 4.5cm 5.3cm},clip]{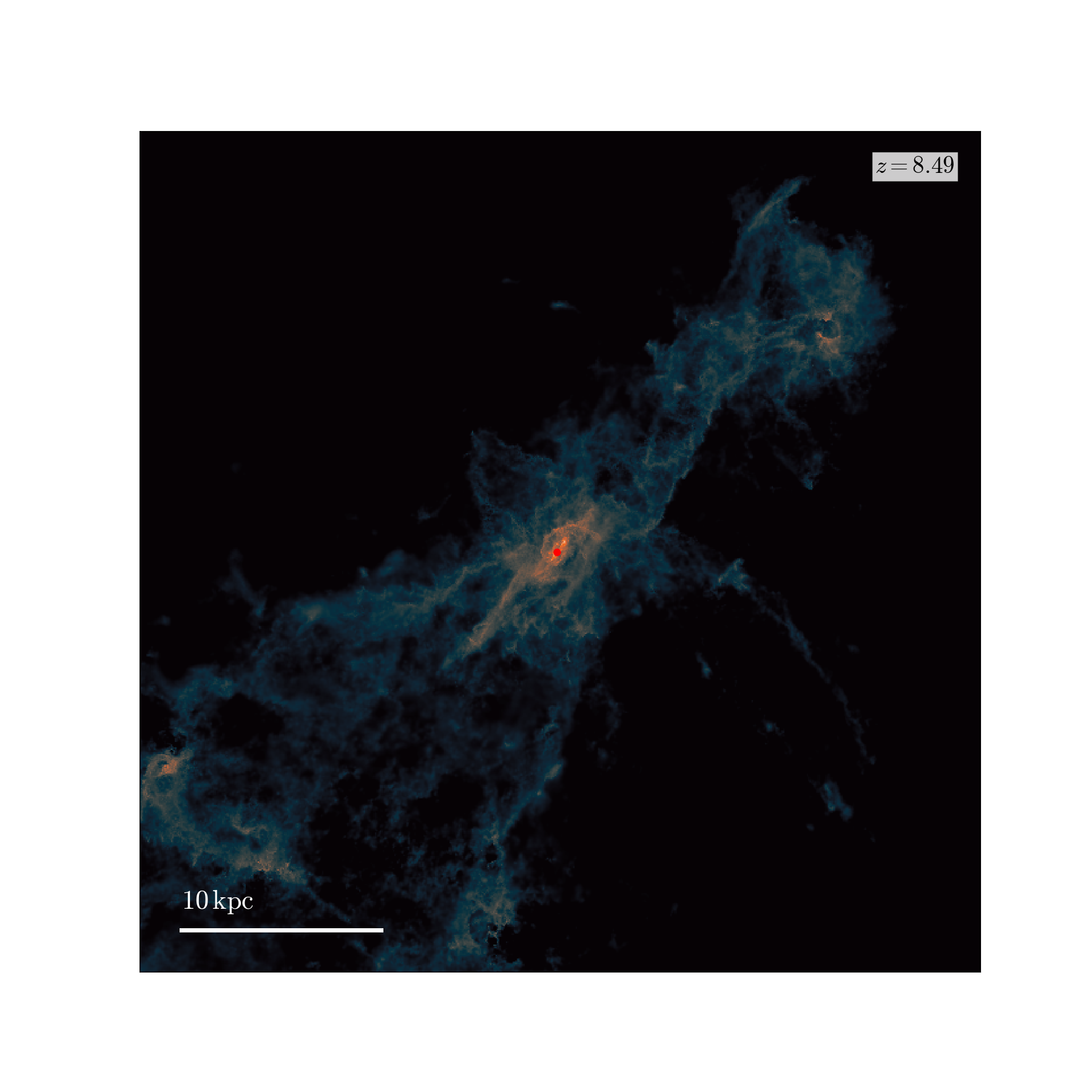}
\includegraphics[width=0.38\textwidth,trim={5.7cm 4.9cm 4.5cm 5.3cm},clip]{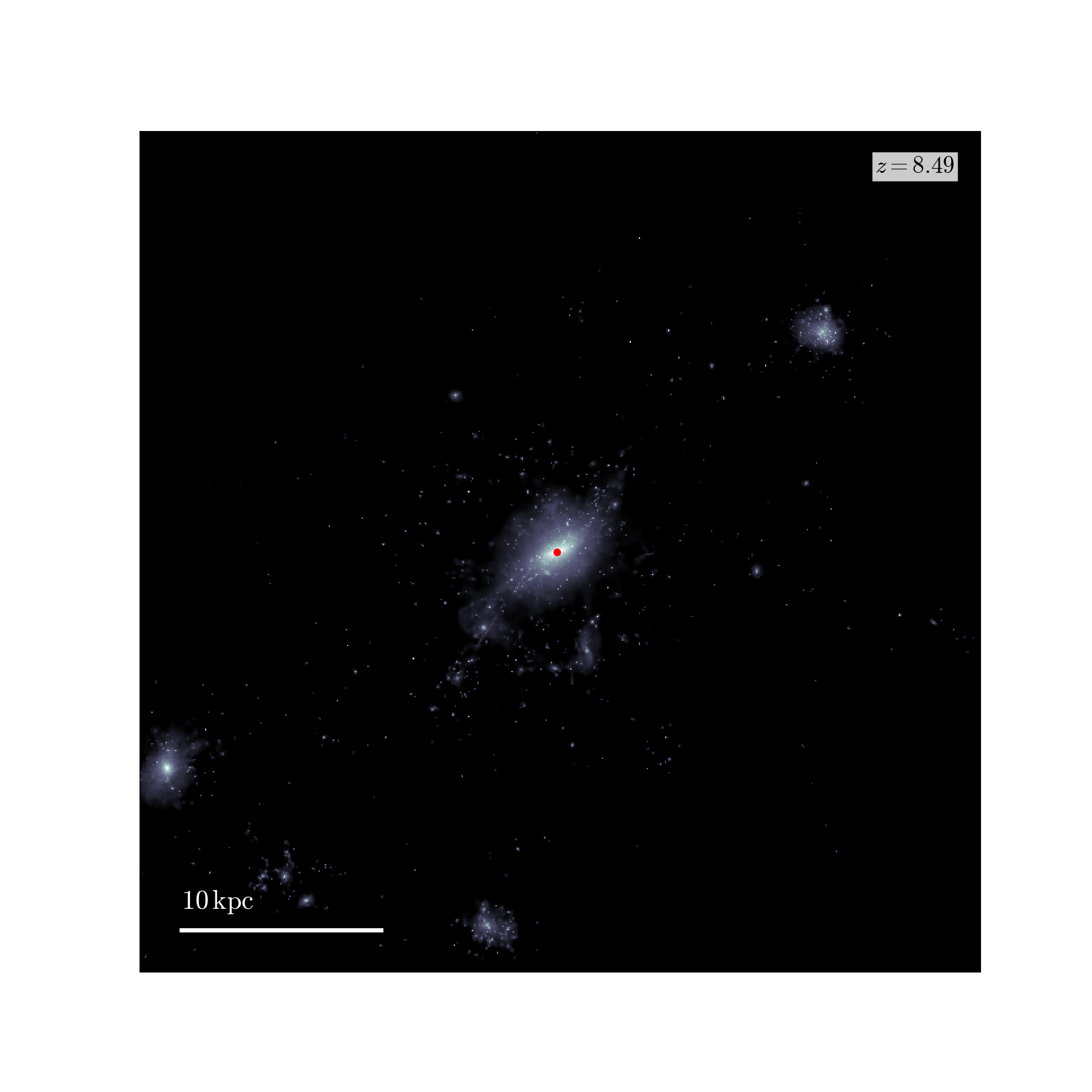}\\
\includegraphics[width=0.38\textwidth,trim={5.7cm 4.9cm 4.5cm 5.3cm},clip]{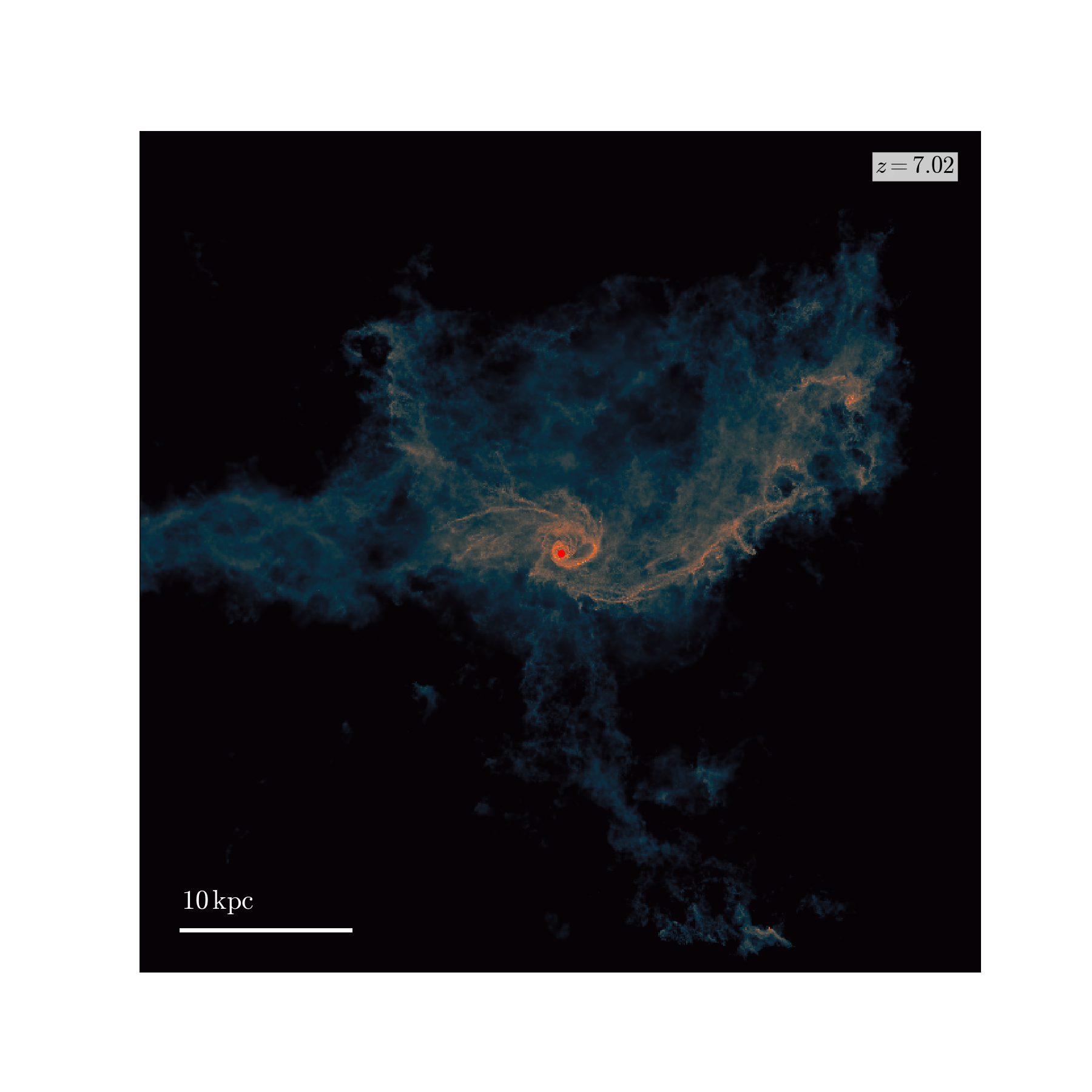}
\includegraphics[width=0.38\textwidth,trim={5.7cm 4.9cm 4.5cm 5.3cm},clip]{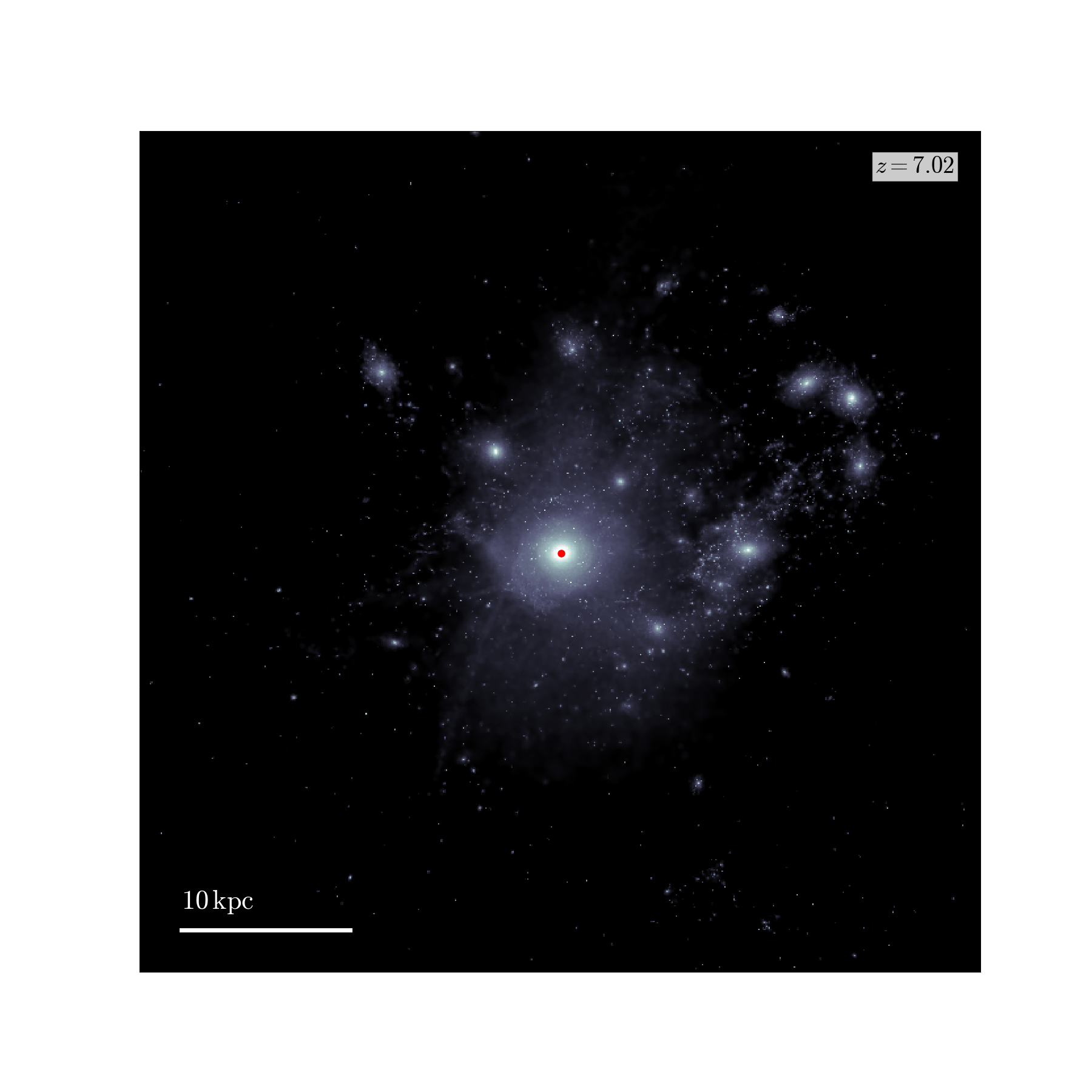}\\
\includegraphics[width=0.38\textwidth,trim={8.5cm 20.7cm 8.5cm 1.5cm},clip]{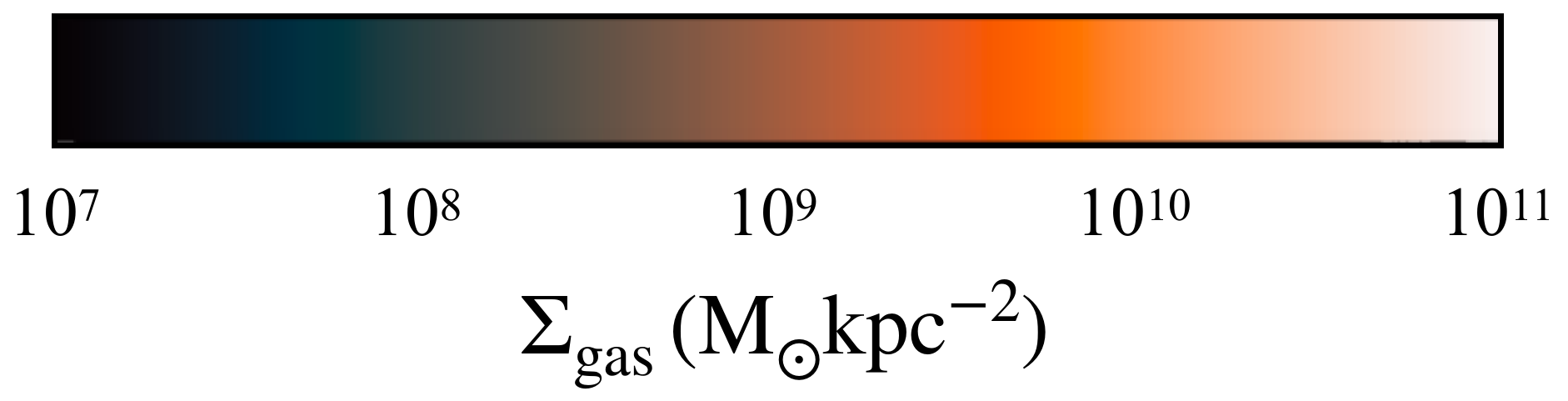}
\includegraphics[width=0.38\textwidth,trim={8.5cm 2.6cm 8.5cm 19.6cm},clip]{Pictures/maps/cbars.pdf}\\
\caption{Column density maps in a 400~comoving kpc box around the target galaxy at $z\approx 10,8.5,$ and $7$ from top to bottom. Gas is shown in the left-hand panels and stars in the right-hand ones. Gas flows continuously on to the galaxy from the cosmic filaments, providing fresh material for the build-up of the stellar and MBH mass. Despite the large number of mergers, most of them are minors and do not contribute significantly to the mass growth of the galaxy (see middle and bottom panels).}
\label{fig:dmaps}
\end{center}
\end{figure*}

As already discussed, after the first 400~Myr, the galaxy mainly grows via smooth accretion from the filaments around the halo, which continuously provide fresh material, as visible in Fig.~\ref{fig:dmaps}, where we show the gaseous (left-hand panels) and stellar (right-hand panels) distribution within 400~comoving~kpc from the central galaxy at $z\approx10,8.5,$ and $7$. The maps correspond to the column density integrated along the z-direction in the 400~comoving kpc box, with the colormap spanning column densities in the range of $10^7\,\msun kpc^{-2}<\Sigma_{\rm gas}<10^{11}\,\msun kpc^{-2}$ for gas and $5\times 10^6\,\msun kpc^{-2}<\Sigma_{\rm star}<5\times 10^{10}\,\msun kpc^{-2}$ for stars.
The left-hand panels show a large-scale gaseous disc extending over a few kpc,  whose orientation changes with time following the angular momentum evolution of the massive gas inflows. The stellar distribution also exhibits a disc-like structure, with a denser concentration in the central region, where the  MBH, shown as a red dot, is located.

\subsubsection{The central MBH}

\begin{figure}
\includegraphics[width=\columnwidth]{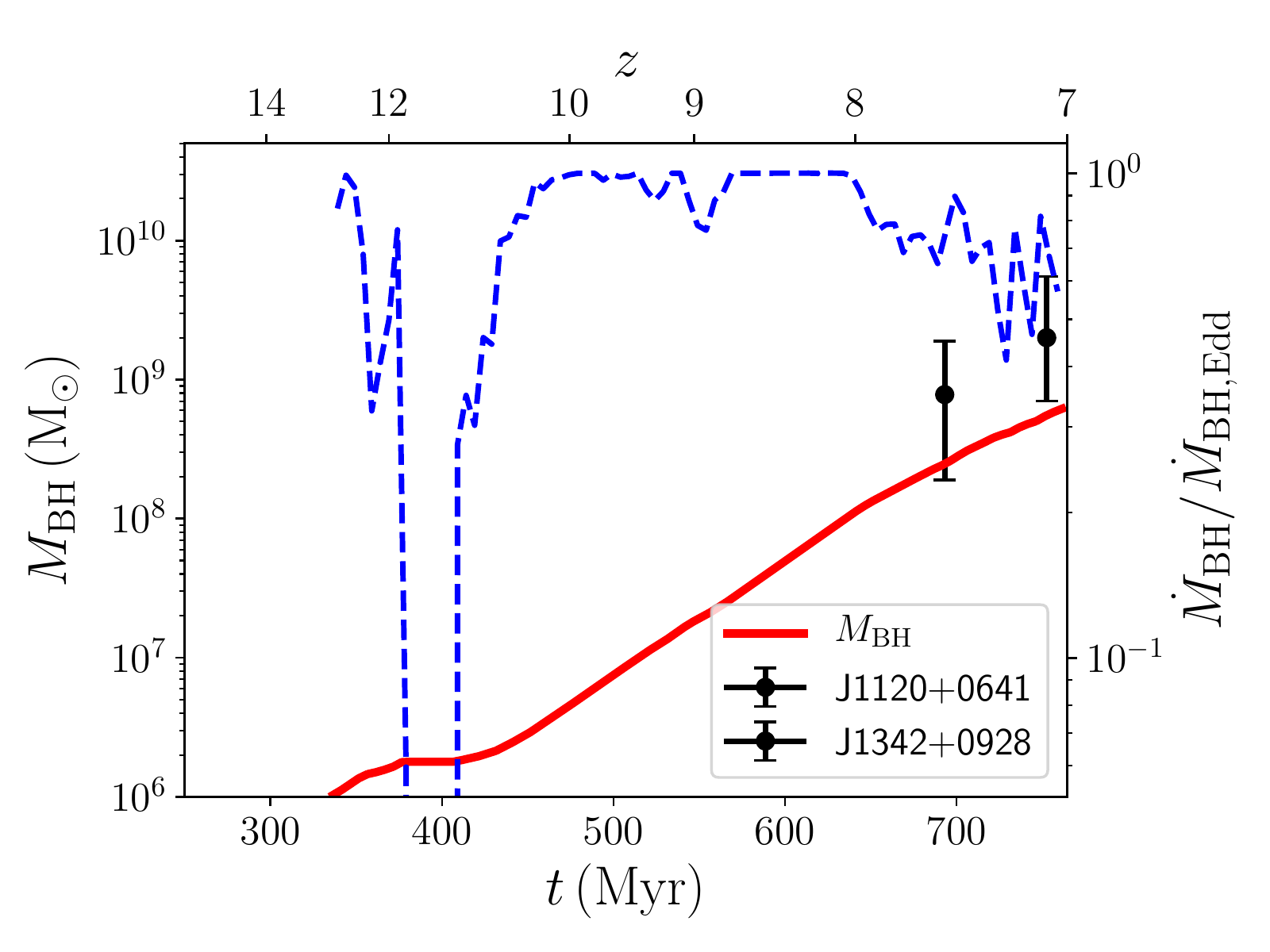}
\caption{Evolution of the MBH at the centre of the target galaxy as a function of time. The red solid line corresponds to the MBH mass, whereas the blue dashed line is the accretion rate in units of the Eddington accretion rate. When the MBH forms, the large amount of gas around it triggers an initial accretion event that, however, is very quickly halted by the SN feedback that sweeps away the gas from the centre and, partially, by the AGN heating. Then, the growth is suppressed until the galaxy becomes massive enough to retain a significant gas mass, and the accretion can restart, this time almost continuously at the Eddington limit. Only in the last 100~Myr, the combined effect of stellar feedback and AGN feedback are able to slow down the accretion to about 60 per cent Eddington.}
\label{fig:mbh_time}
\end{figure}

In Fig.~\ref{fig:mbh_time} we show the mass (red solid) and the average accretion rate (blue dashed) in units of the Eddington luminosity for the MBH, for $\eta_{\rm acc}=0.1$, as a function of time. Immediately after its formation, the gas density around the MBH is high enough to trigger a short accretion phase. This initial short accretion phase is a consequence of the seeding criterion employed here, which creates a new MBH at the highest gas density peak. However, after a few Myr, SN events (together with the AGN feedback) completely evacuate the region around the MBH, halting its growth. After this initial burst, the MBH experiences a long phase where accretion is suppressed by SN feedback \citep[the galaxy stellar mass is below $\sim 10^{9-10}\rm\,\msun$,][]{dubois14}, and only later it starts to grow at the Eddington limit until it reaches a few $10^8\,\msun$.  At this point, because of the larger energy injection powered by accretion, the gas around the MBH is heated up to more than $10^6$~K and the growth slows down, with an average accretion rate oscillating between 0.5 and 0.8 times the Eddington limit, and the MBH reaches $M_{\rm BH}=8.6\times 10^8\,\msun$ by $z\approx 7$. The behaviour observed for the MBH growth is consistent with several previous studies \citep{dubois14,habouzit17,prieto17,anglesalcazar17FIRE}, where the MBH was found to starve until the galaxy reached a critical mass between $10^9\,\msun$ and $10^{10}\,\msun$, or a halo mass/temperature about $10^{11}-10^{12}$ and $10^{5.6}$ K \citep{bower17,mcalpine18}. This result implies that our initial seed mass (see Section \ref{sec:seeding}) is not crucial for the MBH evolution, which is instead regulated by the galaxy's ability to funnel gas towards the centre. \footnote{To further confirm this statement we have rerun our simulation with a $10^5\,\msun$ initial BH mass, and evolved it down to $z=9$. The results are reported in Appendix \ref{app:seed}.} whose results Finally, the lack of MBH growth in low-mass galaxies enables to reconcile different observations of the high-z MBH and quasar populations, as initially suggested in \citet{volonteri11a}. 

In Fig.~\ref{fig:mbh_time}, we also report the MBH mass estimates for the two most distant quasars known to date, J1120+0641 at $z=7.085$ \citep{mortlock11} and J1342+0928 at $z=7.541$ \citep{banados18}. Our result is in reasonable agreement with observations of the most distant quasars to date. Compared to a previous similar study by \citet{smidt18}, our MBH forms later and, despite a larger seed mass, its growth is suppressed for the first 100~Myr, unlike in \citet{smidt18}, who use a less efficient thermal SN feedback. 

If we measure the accretion rate on the MBH directly computed by the code using the BHL formula before applying the cap at the Eddington limit, in many cases we get super-Eddington accretion rates, with Eddington ratios up to about 100. Since our sub-grid modelling for BH feedback does not take into account the super-critical regime where powerful jets are produced \citep[e.g.][]{sadowski16}, we opted for capping the accretion rate at the Eddington limit. Nevertheless, if such accretion rates are reached in massive galaxies at high redshift, even for a short time, the initial MBH growth could be accelerated, increasing the final mass and possibly alleviating the tight constraints on the seed BH masses.

\begin{figure}
\includegraphics[width=\columnwidth]{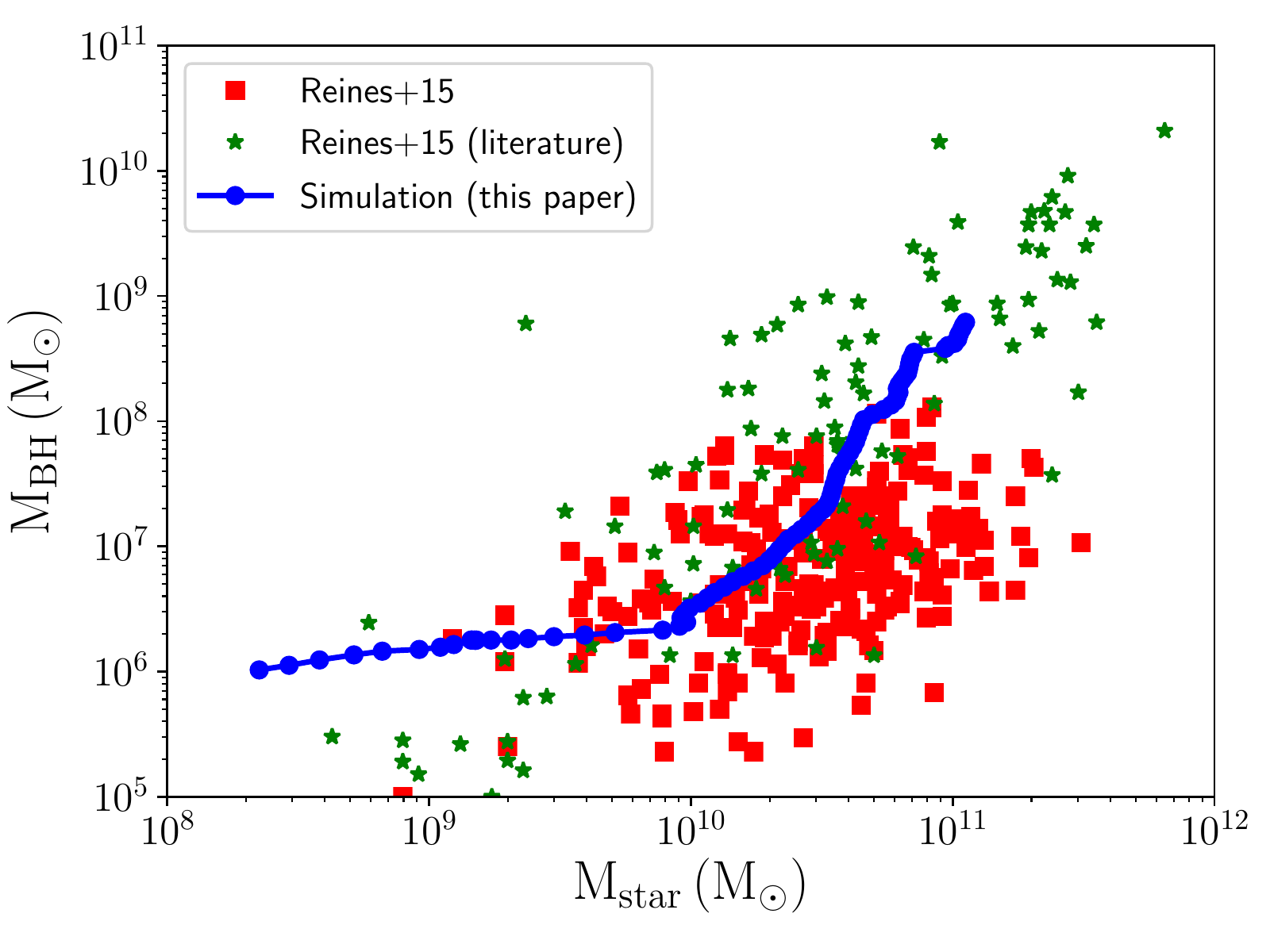}
\caption{Evolution of the MBH at the centre of the target galaxy with the stellar mass of its host. The blue solid line (with the blue dots) corresponds to our simulation, whereas the red squares are the broad line AGN sample from \citet{reines15}. The green stars are the additional data taken from the literature reported in \citet{reines15}. Despite the difference in redshift, the evolution of the MBH is in good agreement with the evolution expected from observations.}
\label{fig:mbh_mstar}
\end{figure}

In Fig.~\ref{fig:mbh_mstar}, we compare our simulation (as blue dots connected by a solid line) with local data from \citet{reines15}. The red squares are broad line AGN with single-epoch mass estimates, whereas the green stars correspond to MBHs with measurements from stellar or gas dynamics, masers and reverberation mapping. Interestingly, after the initial phase where SN feedback hampers MBH growth and its growth lags the galaxy's, the MBH in our simulation grows along with its host, always remaining within the scatter of the observations of low-redshift MBHs.  This suggests that MBHs in high-redshift galaxies are not different from their local counterparts. Analyses of observations at high redshift argued that MBHs grow faster than their galaxy host, resulting in MBHs overmassive with respect to those hosted in galaxies with the same masses at $z=0$. In Section~\ref{sec:dynmass}, we measure the dynamical masses from our simulation, in order to directly compare them with the observed data, and we discuss this apparent discrepancy. 

\begin{figure}
\includegraphics[width=\columnwidth,trim={0.1cm 1.2cm 0.5cm 0.2cm},clip]{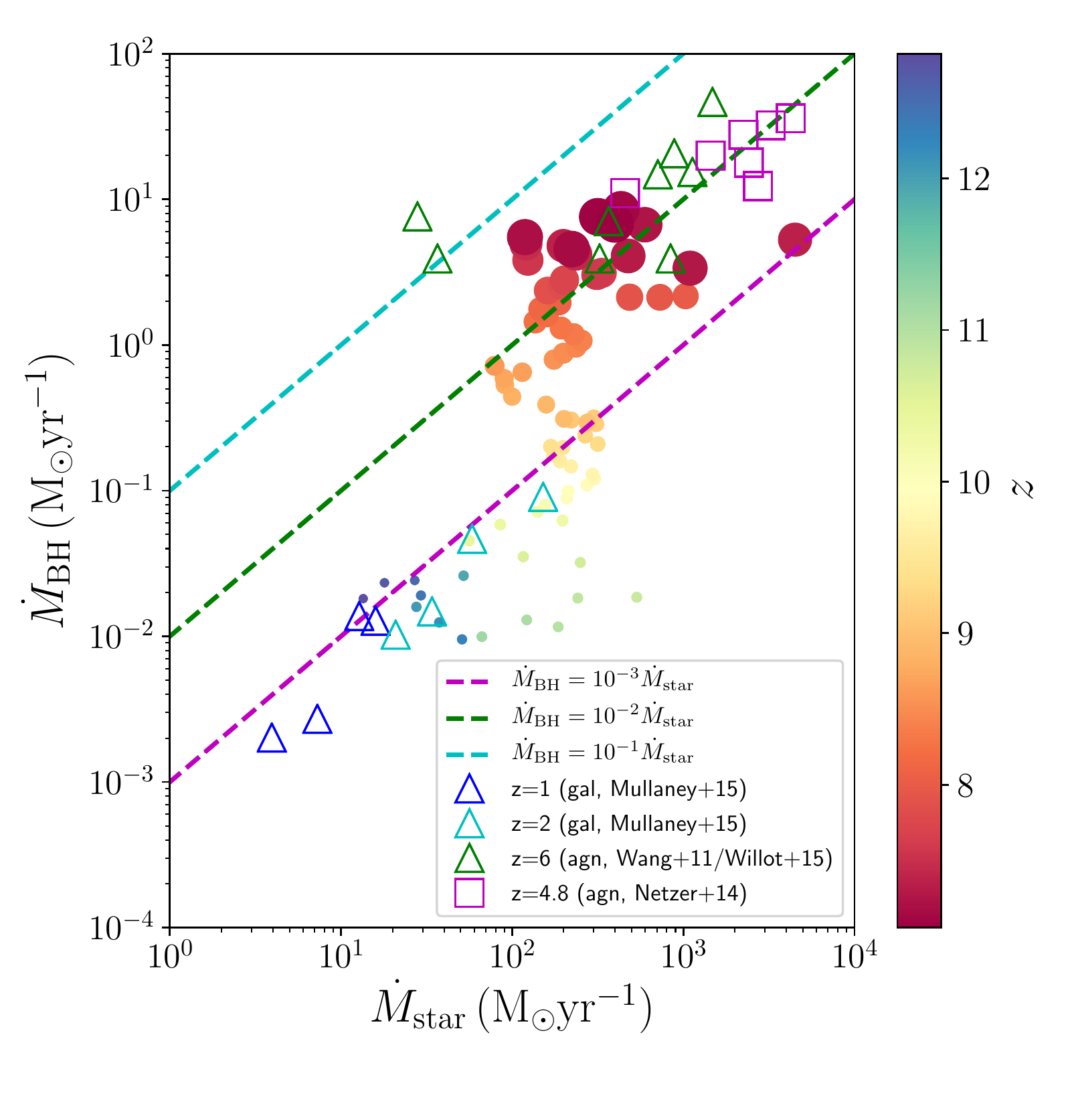}
\caption{MBH accretion rate versus SFR for the main galaxy. The empty symbols correspond to the observations by \citet{wang11}, \citet{mullaney12}, \citet{netzer14}, and \citet{willott15}, while the filled circles correspond to our simulation, with the colour representing the redshift and the size the MBH mass. At earlier times, the MBH does not grow significantly, and the simulation shows properties similar to those of galaxies at $z\sim 1-2$. At later times, instead, when the MBH starts to grow at the Eddington limit, the MBH accretion rate moves upward towards the data of high redshift quasars ($z=6$ and $z=4.8$).}
\label{fig:sfr_bhar}
\end{figure}

The relation between MBH and galaxy mass is built through how MBHs and galaxies gain their mass. The relation between MBH accretion rate and SFR is key to interpret this relation. In observations, MBH accretion rate and SFR appear to be uncorrelated on a source-by-source basis, but the ratio of the mean MBH accretion rate is about $10^{-3}$ the mean SFR \citep[][and references therein]{chen13}. For high redshift quasars, the observed ratios are consistently much higher, $\sim 10^{-2}$ \citep{wang11,netzer14,willott15}, or even larger, although we stress that these are not average values, and that they are obtained on sources selected for being quasars, i.e., where the MBH luminosity outshines the galaxy.  In Fig.~\ref{fig:sfr_bhar}, we compare these results with those in our simulation, shown as filled circles. The size of the circle corresponds to the MBH mass, whereas the colour is $z$. The dashed curves, from bottom to top, correspond to $\dot{M}_{\rm BH}/\dot{M}_{\rm star}=10^{-3},10^{-2},$~and~$10^{-1}$. At very high redshift, when the MBH is not growing much, the accretion rate is low, similar to, or lower than, $z\sim 1-2$ galaxies \citep{mullaney12}. At later times, after the MBH has entered the strong accretion regime, the data points move upwards, close to high redshift quasar measurements, although the observations are at somewhat lower redshift ($z=6$ and $z=4.8$) \citep{willott15,wang11,netzer14}. If we assume a baseline of about $10^{-3}$ for the average ratio of BH accretion rate to SFR, the interpretation is therefore that initially the MBH lags behind its galaxy ($z=10-11$), then it catches up and even surpasses it ($z<9$). Eventually, when BH feedback becomes important (at $t=700$~Myr, $z=7.5$), BH accretion slows down. 

\begin{figure}
\centering
\includegraphics[width=\columnwidth]{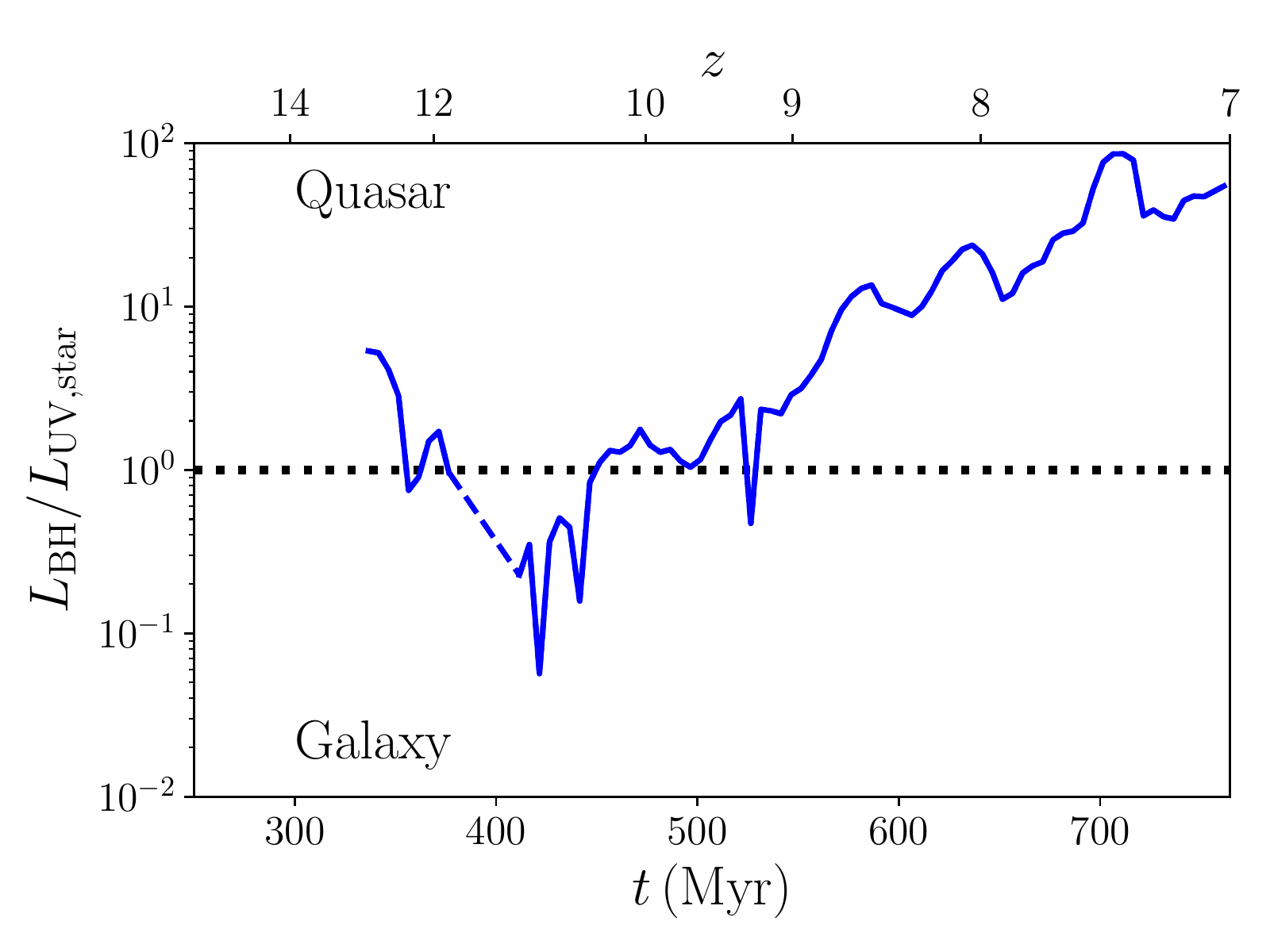}
\caption{BH accretion-powered luminosity relative to the rest-frame UV stellar luminosity of the galaxy as a function of time. The dashed line corresponds to an equal contribution to the luminosity, and  separates the galaxy regime (below the line) from the quasar phase (above the line). At very high redshift, before the Eddington-limited accretion phase of the MBH begins, the galaxy dominates the emission, whereas the situation completely changes below $z=10$, after the central MBH has started to quickly grow in mass. At $z\sim 7$, the luminosity is completely dominated by the central source (up to a hundred times), shining as a quasar. With JWST in the near future, we will finally be able to probe the galaxy hosts at high redshift, during the early phases of the MBH formation and initial growth.}
\label{fig:lum}
\end{figure}

Because of the large luminosity associated to the central MBH in the optical/UV band, that exceeds by far their stellar luminosity, quasar hosts at high redshift can currently only be detected via their molecular and dust component in the sub-millimetre band. In Fig.~\ref{fig:lum}, we show the evolution of the BH bolometric luminosity relative to the rest-frame UV luminosity of the galaxy as a blue line. The far-UV flux is computed from the up-to-date stellar population synthesis models by \citet{bruzual03} also employed for the stellar radiation sub-grid model. The UV luminosity has not been corrected for dust extinction, which is expected to be important, given the high metallicity of the gas \citep[see also][]{tenneti18b}.
The dashed region corresponds to the early phase when the MBH was not growing, hence only the galaxy would have been observable. For $z\lesssim 10$, the MBH has accreted enough mass to become more luminous than its host, i.e. a quasar. At $z\sim 7$, its luminosity has become almost a hundred times larger than that of the galaxy, completely dominating the emission. The advent of facilities that can observe the rest-frame optical/near-IR for high-redshift quasars with high angular resolution that permit good subtraction of the point spread function, like JWST, will soon allow us to measure the stellar properties of the host.

\subsection{Quasar hosts at $z\lesssim 7$}
In this section we focus on the properties of the MBH and its host at $z\sim 7$, and specifically how observational tracers and diagnostics compare to quantities directly measured in the simulation. Our aim is to aid interpretation of observations and identify possible biases caused by selection effects or limited spatial resolution. 

\begin{figure*}
\includegraphics[width=0.395\textwidth,trim={2cm 2.7cm 4.05cm 2.7cm},clip]{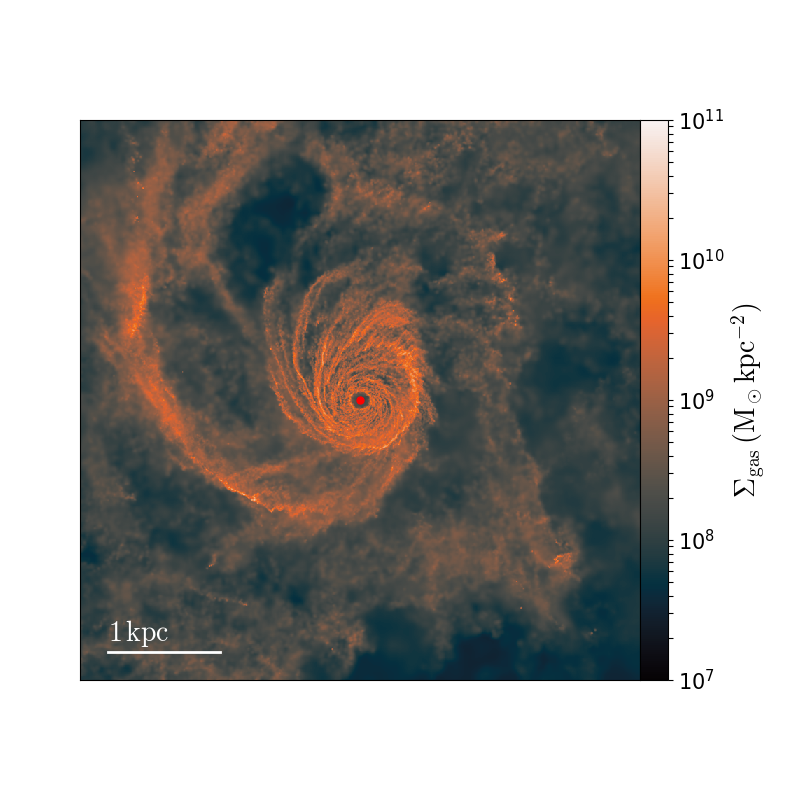}
\includegraphics[width=0.490\textwidth,trim={2cm 2.7cm 0.6cm 2.7cm},clip]{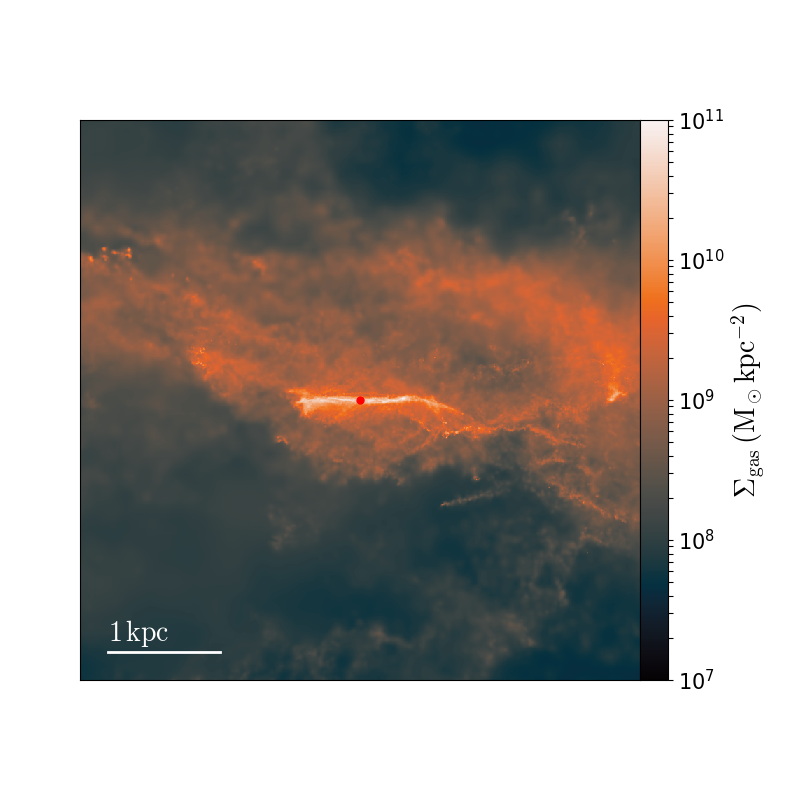}\\
\includegraphics[width=0.395\textwidth,trim={2cm 2.7cm 4.05cm 2.7cm},clip]{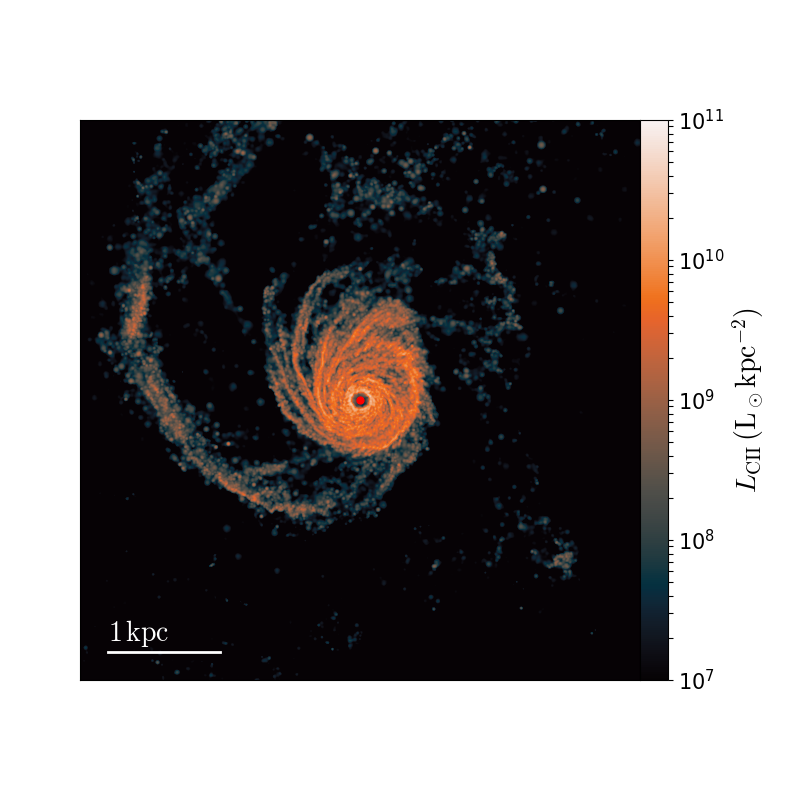}
\includegraphics[width=0.490\textwidth,trim={2cm 2.7cm 0.6cm 2.7cm},clip]{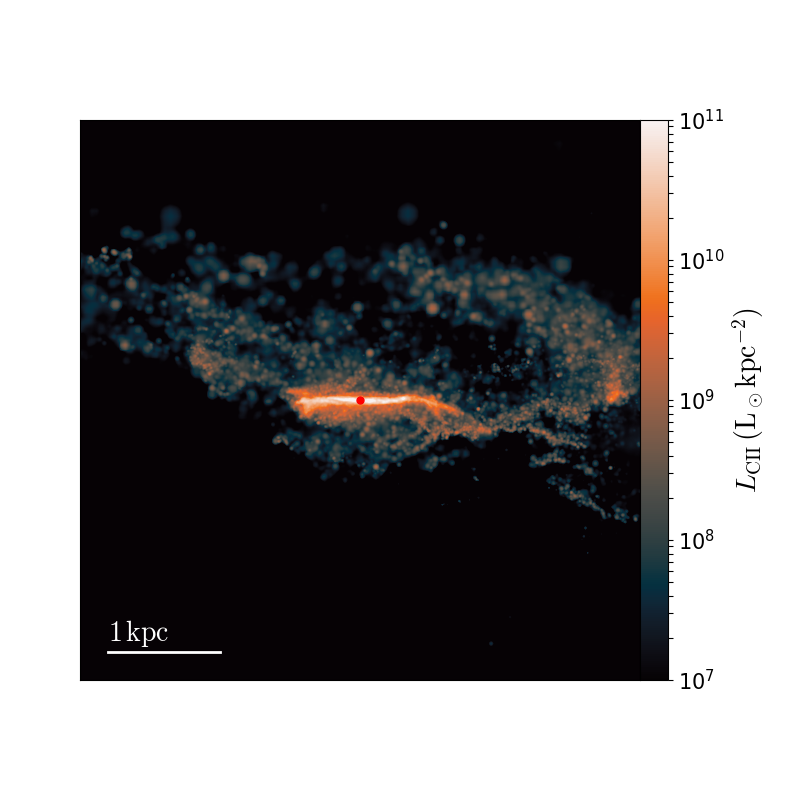}\\
\includegraphics[width=0.395\textwidth,trim={2cm 2.7cm 4.05cm 2.7cm},clip]{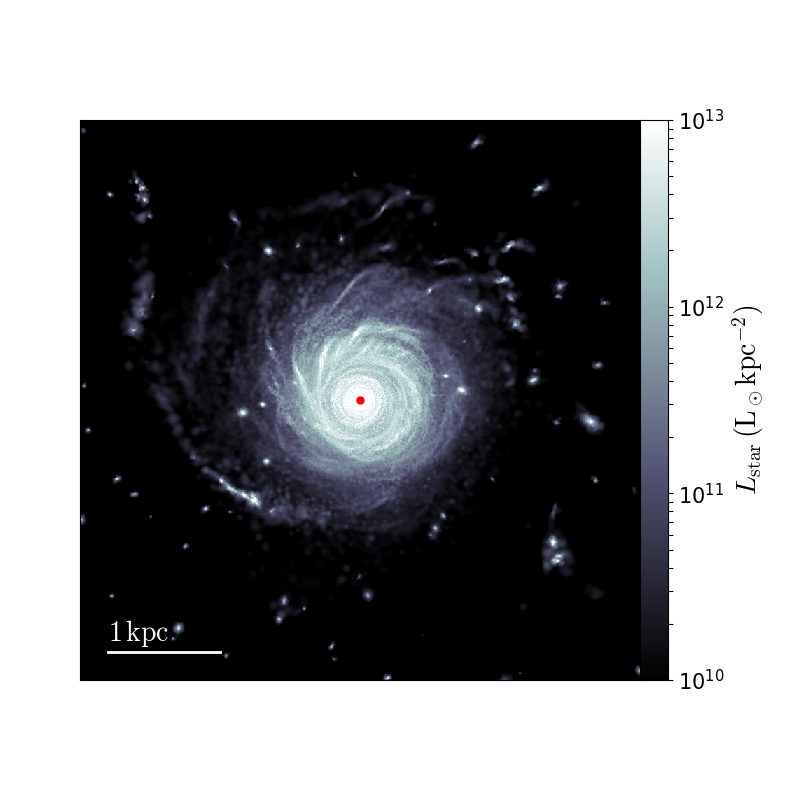}
\includegraphics[width=0.490\textwidth,trim={2cm 2.7cm 0.6cm 2.7cm},clip]{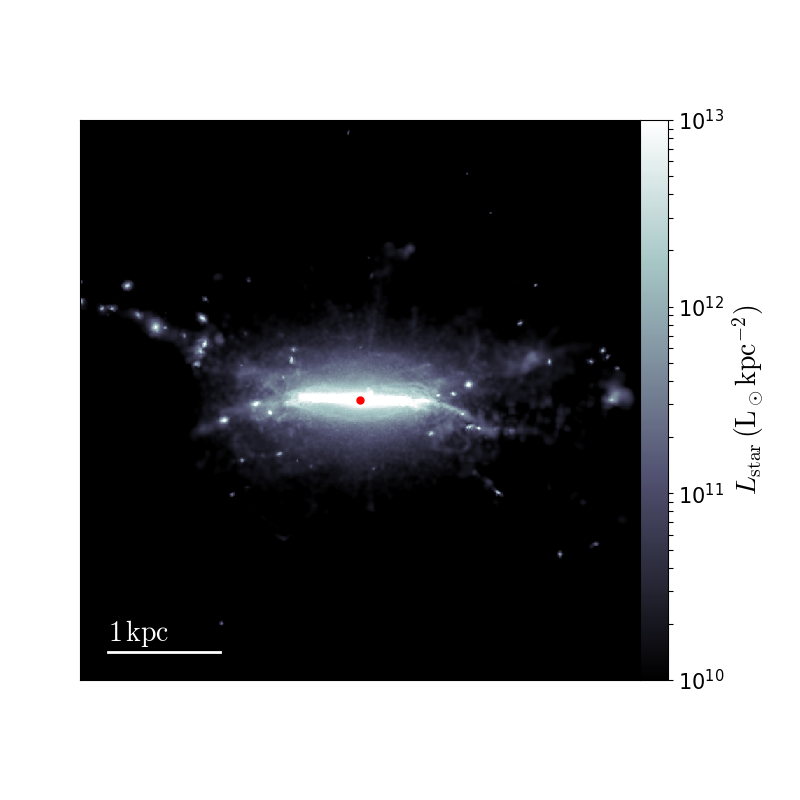}\\
\caption{Main galaxy at $z\approx 7$, face-on (left-hand panels) and edge-on (right-hand panels). We show the gas surface density in the top panels, the [CII] line flux in the central ones, and the stellar far-UV flux in the bottom ones. The galaxy exhibits a thin disc with well defined spiral arms, both in stars and in gas. However, a large gas reservoir is observed at larger distances, continuously flowing on to the galaxy in the form of dense filaments and clumps and providing fresh material for SF.}
\label{fig:distribution}
\end{figure*}

\subsubsection{Gas and stellar distribution}
Thanks to the very high resolution of the simulation, we can properly resolve the gas/star distribution in the galaxy. At $z\sim 7$, metal rich gas can quickly cool within the halo, and it settles in a well defined dense disc extending out to $\sim 1$~kpc. Continuous inflows and SN explosions within the system create a massive reservoir of mildly lower density gas extending out to several kpc with filaments and clumps (see top panels in Fig.~\ref{fig:distribution}, where we report the total gas surface density face-on and edge-on). The dense disc is relatively thin, and shows a well defined spiral structure, also observable in the stellar component (whose far-UV flux is reported in the bottom panels). A low density region can also be seen close to the central MBH (red dot), due to the effect of the BH feedback. However, because of the high gas densities around the MBH, BH feedback is not effective enough to affect the gas at larger distances. Around the central stellar disc, many stellar clusters and some small galaxies can be distinctly seen in the far-UV maps.  Although not visible in the young stellar component, a central bulge is already present in the galaxy. Following \citet{vogelsberger14}, we measure the B/T ratio from the circularity distribution of the stars, that results in $\sim 0.45$, suggesting that an already massive bulge has formed in the centre of the system. 

In the middle panels of Fig.~\ref{fig:distribution}, we show the expected luminosity from [CII] emission, assuming the [CII] emissivity defined in \citet{pallottini17} as 
\begin{equation}
\tilde{L}_{\rm [CII]} \left({\rm \frac{L_\odot}{M_\odot}}\right)=
\begin{cases}
 0.1\left(\frac{Z}{Z_\odot}\right) \left(\frac{n_{\rm gas}}{100\,\rm cm^{-3}}\right)  & n_{\rm gas}<10^3\rm\, cm^{-3}\\
 1.0\left(\frac{Z}{Z_\odot}\right) & \rm otherwise\\
 \end{cases}
\end{equation}
with $n_{\rm gas}=\rho_{\rm gas}/\rm m_p$.
As expected, [CII] traces very well the star-forming part of the galaxy, corresponding to the dense disc in the central kpc, and is present at larger distances only in the dense filaments or clumps flowing on to the galaxy or ejected from it. By integrating the [CII] flux in the disc, we obtain $L_{\rm [CII]} \sim \times 10^{9.5}\rm\, L_\odot$, in good agreement with the observed luminosities by D18. We also evaluated $L_{\rm [CII]}$ at different redshifts, and we found a reasonably good agreement with the $L_{\rm [CII]}$-SFR relation employed by D18, but with significant scatter, up to half a decade.

\subsubsection{Gas/stellar kinematics and dynamical masses}
\label{sec:dynmass}
\begin{figure*}
\large{\hspace{-1.3cm} $\Delta x = 50$~pc \hspace{2.6cm} $\Delta x = 200$~pc \hspace{2.6cm} $\Delta x = 2$~kpc}\\
\begin{center}
\includegraphics[width=.86\textwidth,trim={3cm 5.8cm 2.5cm 2cm},clip]{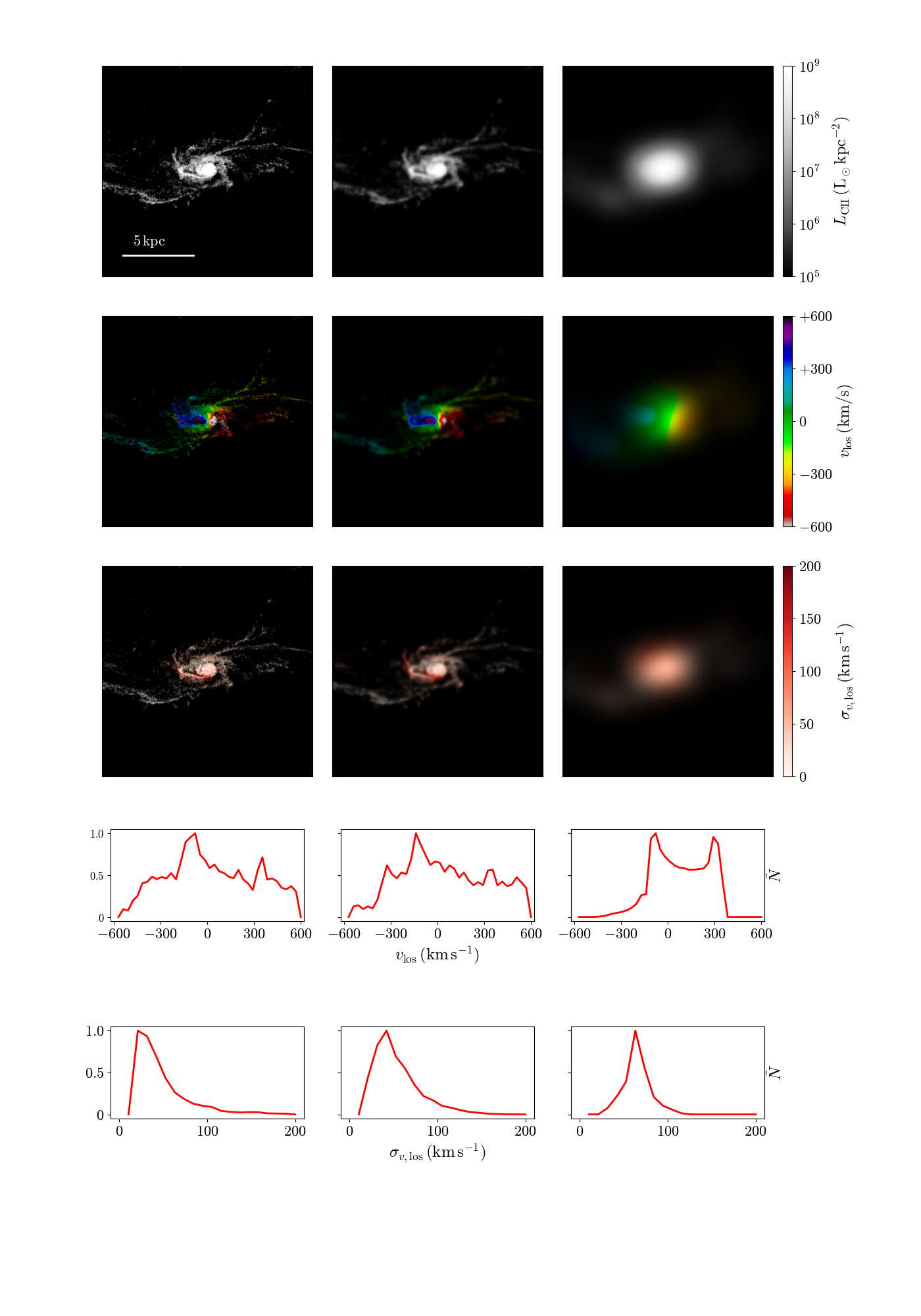}
\caption{Velocity maps from [CII] line emission in a box of 15~kpc (first three rows) and corresponding velocity profiles (last two rows) at $z=7$. From left to right, we report the maps/profiles at 50~pc, at 200~pc and 2~kpc resolutions. The first row shows the [CII] luminosity map, the second the line-of-sight velocity $v_{\rm los}$, and the third the velocity dispersion along the line-of-sight $\sigma_{v,\rm los}$. In the two bottom rows we report the profiles obtained from the velocity maps and the best-fit of a Gaussian profile. A double-peaked profile is observed for the line-of-sight velocity, due to the rotational motion, and it approaches a Gaussian distribution only at the lowest resolution. The velocity dispersion, instead, exhibits a high-velocity tail, that shifts the mean of the Gaussian fit as the resolution decreases.}
\end{center}
\label{fig:vmap}
\end{figure*}

At $z\sim 7$ and above, we are still unable to properly resolve the galaxy structure in observations, and furthermore for quasars the stellar distribution is hidden by the quasar emission in UV-based observations. This limitation results in the impossibility of accurately measuring the galaxy stellar masses. Dynamical masses are estimated from the gas properties in sub-mm observation,  from the emitting region size and the line width by assuming  virial equilibrium and either a rotationally supported disc or a dispersion dominated system. 

To consistently compare simulation results with the current observations, and to infer whether the dynamical mass estimates are good tracers of the entire system masses, we then need to mimic the observational approach in our numerical simulations. We summarise here the procedure we follow:
\begin{enumerate}
\item we estimate the galaxy angular momentum and we rotate the system by 55 degrees perpendicular to it;
\item we produce 2-dimensional average maps of the line-of-sight velocity $v_{\rm los}$ and the velocity dispersion $\sigma_{v,\rm los}$, weighted by the gas density (we limit to $\rho_{\rm gas}>1.67\times 10^{-24}\rm\, g\,cm^{-3}$ only to avoid considering low-density hot gas) (a) and the [CII] line intensity (b) at a resolution of 50~pc, the target resolution for observations to be able to resolve the Bondi region around the MBH;
\item we remove the low-density/luminosity outer regions by reducing the opacity of the map according to the gas column density (a) or the [CII] line intensity (b), in logarithmic steps over three orders of magnitude;
\item we degrade the spatial resolution of 50~pc by convolving the maps (and the opacity mask) with a Gaussian kernel with full width half maximum (FWHM) 200~pc and 2~kpc, respectively;
\item we compute the 1-dimension line profile from each map and estimate the width of the line; if the profile can be fitted by a Gaussian function, we estimate the FWHM from the best-fit, otherwise we assume the FWHM corresponds to the distance between the two peaks of the profile; for the velocity dispersion profile, we fit using a Gaussian function and take the mean value for $\sigma_{\rm fit}$;
\item we compute the emitting region size $R_x$ from the opacity mask, converting the 2-dimension image into a 1-dimension radial profile and taking the semi-major axis of the fitted Gaussian profile;
\item we estimate the dynamical mass from each channel (gas or [CII] luminosity) according to the virial theorem, as D18
\begin{equation}
M_{x,\rm dyn} = \begin{cases}
\frac{R_x}{\rm G}\left(\frac{0.75\rm FWHM}{\sin i}\right)^2 & \rm rotationally-supported \\
\frac{3R_x\sigma_{\rm FWHM}^2}{2\rm G} & \rm dispersion-dominated\\
\end{cases}
\end{equation}
where the $x$ subscript identifies the tracer employed (gas density or [CII] emission), $\rm G$ is the gravitational constant, $\sigma_{\rm FWHM} =\rm FWHM/2.35$ and $i$ is the inclination angle (55 degrees in the case of rotationally-supported systems). Since the gas in the galaxy is distributed mainly in a disc, the best estimate for the dynamical mass is that of the rotationally-supported case, but for sake of completeness, we also compute the expected dynamical mass assuming the FWHM as a measure of the dispersion of the system. 
\end{enumerate}

As an example, we show in Fig.~\ref{fig:vmap} the velocity maps and profiles obtained at $z\approx 7$ from the [CII] line intensity at resolutions of 50~pc (left column), 200~pc (middle column), and 2~kpc (right column), the latter two representing standard resolutions for ALMA observations of high-z quasars. In the top row we show the opacity mask, in the second row the line-of-sight velocity maps, and in the third row the line-of-sight velocity dispersion. In the bottom two rows we show the line-of-sight velocity and velocity dispersion profiles extracted from the [CII] maps, linearly sampled in bins 30~$\rm km\, s^{-1}$ wide. As we degrade the resolution, the velocity pattern of the spiral arms gets smoothed, resulting in a decrease of the peak velocity of a factor of a few relative to the original value. Similarly, at lower resolution the measured gas velocity dispersion is also suppressed. However, the fits to the profiles show that the FWHM of the Gaussian fit to the line-of-sight velocity is not significantly affected by resolution, although the approximation of a Gaussian distribution improves at lower resolution, while a double-peak is more clearly observed when the distribution is properly resolved. The mean of the velocity dispersion profile instead slightly rises with decreasing resolution, due to the increasing importance of the large-velocity tail resulting from the smoothing procedure.

\begin{table*}
\centering

\caption{Best-fit parameters from the velocity maps of [CII] line emission, resulting dynamical masses, and comparison with the intrinsic quantities. The first column identifies the tracer (and the resolution) used for the estimate and the second is the size of the emitting region. Columns 3-5 report the kinematical results;  the third column is the FWHM of the Gaussian fit to the line-of-sight velocity (when the approximation is valid) or the distance between the two peaks, the fourth column is the mean of the Gaussian profile fit to the velocity dispersion, and the fifth column is the stellar velocity dispersion, measured on the stellar particles. Columns 6-9 report the mass measurements; the sixth column is the dynamical mass assuming a rotationally-supported system, and the seventh assumes a dispersion-dominated system. In the last two columns, we report the total gas and stellar masses enclosed within the emitting radius $R_x$.}
\label{tab:gas_kin}
\begin{tabular}{lcccccccc}
Tracer/  & $R_x$& FWHM & $\sigma_{\rm fit}$ & $\sigma_{*}$ & $M_{\rm dyn, rot}$ & $M_{\rm dyn, sph}$ & $M_{\rm gas}(<R_x)$ & $M_{\rm star}(<R_x)$\\	
Resolution& (kpc) &(km~s$^{-1}$) & (km~s$^{-1}$)  & (km~s$^{-1}$) & $(10^{10}\msun)$ & $(10^{10}\msun)$ & $(10^{10}\msun)$ &$(10^{10}\msun)$\\
\hline
$\rho_{\rm gas, sim}$ & 1.10& 490 & 115& 350 & 5.1 & 1.7 & 3.3 & 9.1\\
$\rho_{\rm gas, 200~pc}$ & 1.10& 460 & 140 & 350 &4.5 & 1.5 & 3.3 & 9.1\\
$\rho_{\rm gas, 2~kpc}$ & 1.15 & 350 & 175 & 350 & 2.7 & 0.9 & 3.4 & 9.1\\
\hline
$L_{\rm [CII], sim}$ & 0.96& 430 & 35 &350 &$3.5$ &$1.1$ & 3.0 & 8.9 \\
$L_{\rm [CII], 200~pc}$ & 0.96 & 470& 55 &350 &$4.1$ & $1.3$ & 3.0 & 8.9\\
$L_{\rm [CII], 2~kpc}$ & 1.04 & 390 & 90 & 350 &$3.0$ & $1.0$ & 3.2 & 9.0\\
\hline
\end{tabular}
\end{table*}

In table \ref{tab:gas_kin} we report the relevant quantities extracted from the velocity maps of the target galaxy at $z\approx 7$ and compare them to the quantities measured directly from the simulation. 

Regarding gas kinematics, we find that rotational support is dominant. If we measure the actual rotation curve and velocity dispersion from the gas particle distribution, the velocity dispersion in the central kpc is about 35 per cent of the rotational velocity, and it is consistent with the values inferred from the velocity maps. We repeat the kinematical analysis detailed above on the stellar distribution: for the three resolutions, the FWHM is 430, 390, and 105~$\rm km~s^{-1}$, and $\sigma_{\rm fit,*}$ (measured as described in point (v) above, but for stars) is 350, 330, and 300~$\rm km~s^{-1}$, much larger than that of the gaseous disc, because of the presence of a central bulge and the outer stellar halo, that are dispersion-dominated. The ``true'' stellar velocity dispersion, measured on the stellar particles, $\sigma_*$, is 350~$\rm km~s^{-1}$, while the virial circular velocity of the halo $v_{\rm circ, vir} \approx 385\rm\, km~s^{-1}$. 
In summary, $\sigma_{\rm FWHM}$ from the gas maps underestimates the rotational velocity of the gas, the velocity dispersion of the stellar components as well as the circular velocity of the halo. The estimate of $\sigma_{\rm fit,*}$ from the stellar maps is instead consistent with the ``true" stellar velocity dispersion, and it is also consistent with the circular velocity of the halo. 

Regarding dynamical masses, our estimates of about $2~-~5\times~10^{10}\,\msun$ from the velocity maps are in good agreement with the observations by D18, but they underestimate the actual mass in the system (within the observed radius)  by about a factor of 2 to 5, for rotationally-supported systems, with an even larger discrepancy for dispersion-dominated systems. The total mass within the emitting radius is $\sim 1.2 \times 10^{11}\,\msun$, mostly in stars, with gas contributing to about 25 per cent.

\begin{figure}
\includegraphics[width=\columnwidth]{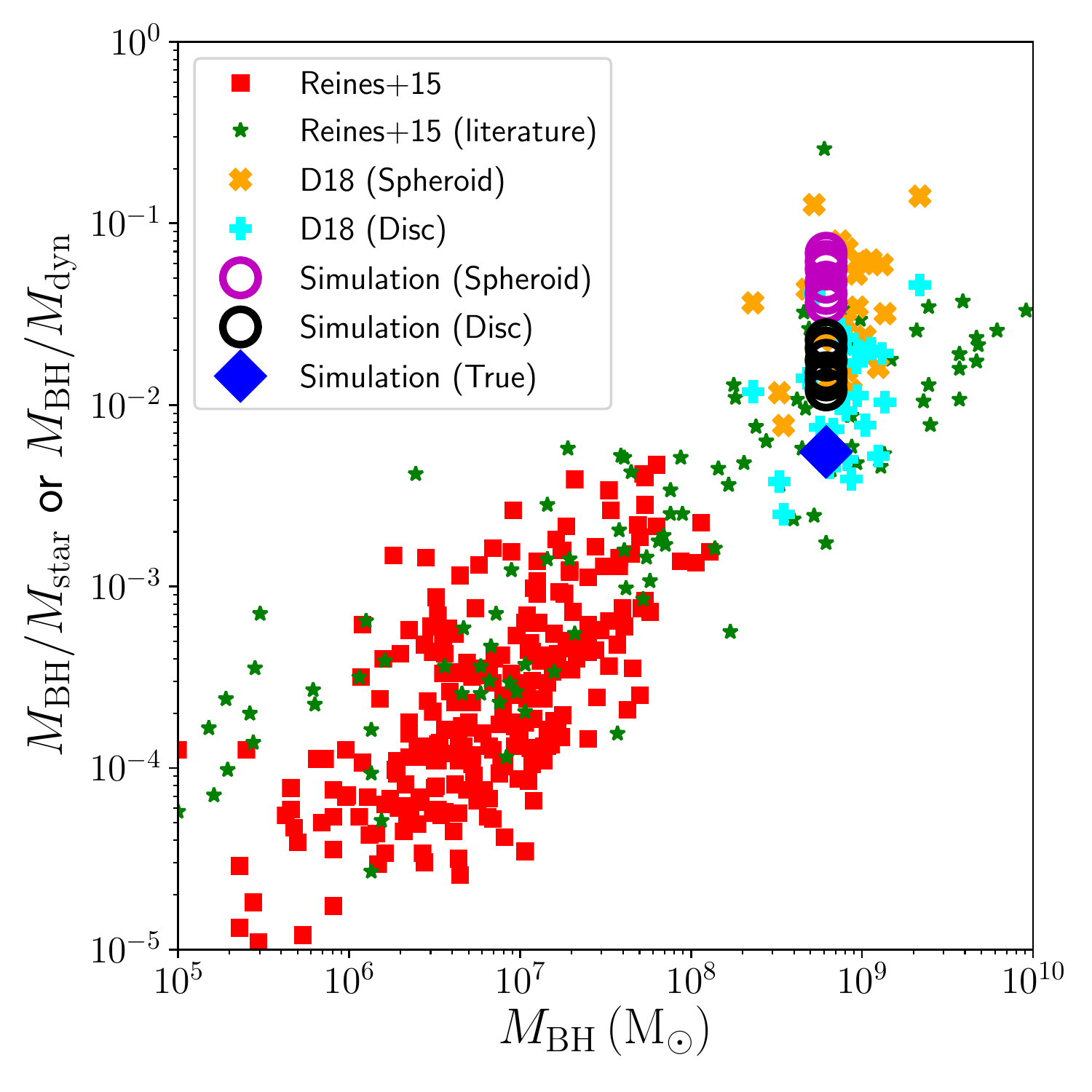}
\caption{MBH to stellar/dynamical mass ratio as a function of MBH mass. As in Fig.~\ref{fig:mbh_mstar}, red squares and green stars correspond to the data in \citet{reines15}, corresponding to local measurements, whereas the cyan plus signs and the orange crosses are the measurements by D18 of $z\sim 6$ quasars assuming rotationally-supported or dispersion-dominated systems, respectively. Our estimates from the simulation are shown as  black  (rotationally-supported) and magenta empty circles (dispersion-dominated), with the actual stellar mass as a blue diamond.}
\label{fig:MBH_Mdyn}
\end{figure}

\subsubsection{BH-galaxy correlations}

Benchmark correlations in local galaxies are based on bulge stellar masses and stellar velocity dispersions \citep[][and references therein]{kormendy13bh}. The observations of high redshift quasars, instead, provide dynamical masses and gas-based velocity dispersions, which, as detailed in the previous section, may differ from (bulge) stellar masses and stellar velocity dispersions. We therefore perform two comparisons:  with observations at $z>6$, using the high-redshift observables and with observations at $z=0$, using the low-redshift observables. This allows us on the one hand to verify that the simulation agrees with the observed properties of high-redshift quasars and validate our results, and on the other hand to compare correctly to the $z=0$ correlations. 

MBHs at these redshift are commonly considered overmassive compared to their local counterparts, based on the ratio between MBH and dynamical mass being much above the ``canonical'' value of $10^{-3}$. Although these mass ratios are well above the local values for low-mass MBHs, that settle around $\sim 10^{-4}-10^{-3}$, they are consistent with those  around $\sim 10^{-2}$ for high-mass MBHs (see Fig.~18 in \citealp{kormendy13bh} and Fig.~11 in \citealp{reines15}), as shown in Fig.~\ref{fig:MBH_Mdyn}, where we report the stellar (or dynamical) mass to MBH mass ratio as a function of MBH mass. This result is also consistent with \citet{lyu16}, who suggested that the BH-galaxy correlations could already be in place at $z\gsim 5$.

As in Fig.~\ref{fig:mbh_mstar}, we report the data by \citet{reines15} as red squares and green stars. Since at $z\sim 7$  bulge masses cannot be properly estimated in observations because of the limited angular resolution, even if there was no bright quasar, we compare to relations in local galaxies that use the total stellar mass, as in section 3.1.2. High-mass BHs, above $\sim 10^{8}\,\msun$, at $z=0$ are hosted in elliptical galaxies, and therefore the difference between total and bulge mass is negligible in that mass range. We also include the measures by D18 as cyan plus signs (rotationally-supported) and orange crosses (dispersion-dominated). Our dynamical mass measurements are shown as black (rotationally-supported) and magenta circles (dispersion-dominated); using the dynamical mass estimated through the gas tracers, the BH mass - dynamical mass ratio at $z\approx 7$ is $\sim 0.017$, in agreement with D18. The actual ratio, for the stellar mass in the galaxy, is shown with a blue diamond. 

The data points by D18, as well as our simulation results, are perfectly consistent with those of nearby massive ellipticals, without any sign of clear deviation. The observations would be slightly above the local data, but still consistent with them, if the galaxies were dispersion-dominated. Furthermore, for the simulation, the estimate from the dynamical mass is above the ``true'' value, based on the stellar mass. Since in the simulation, analysed in the same way observations are analysed, the dynamical mass underestimates the stellar mass in the system, and therefore even more the total mass, the inference is that MBHs at high redshift are in reality not different from their local counterparts, and they only represent the upper end of the $M_{\rm BH}-M_{\rm star}$ correlation, where massive elliptical galaxies lie today. This result is also in agreement with the semi-analytic model results by \citet{valiante11,valiante14}, who also suggested that dynamical masses could underestimate the actual galaxy mass.

\begin{figure}
\includegraphics[width=\columnwidth]{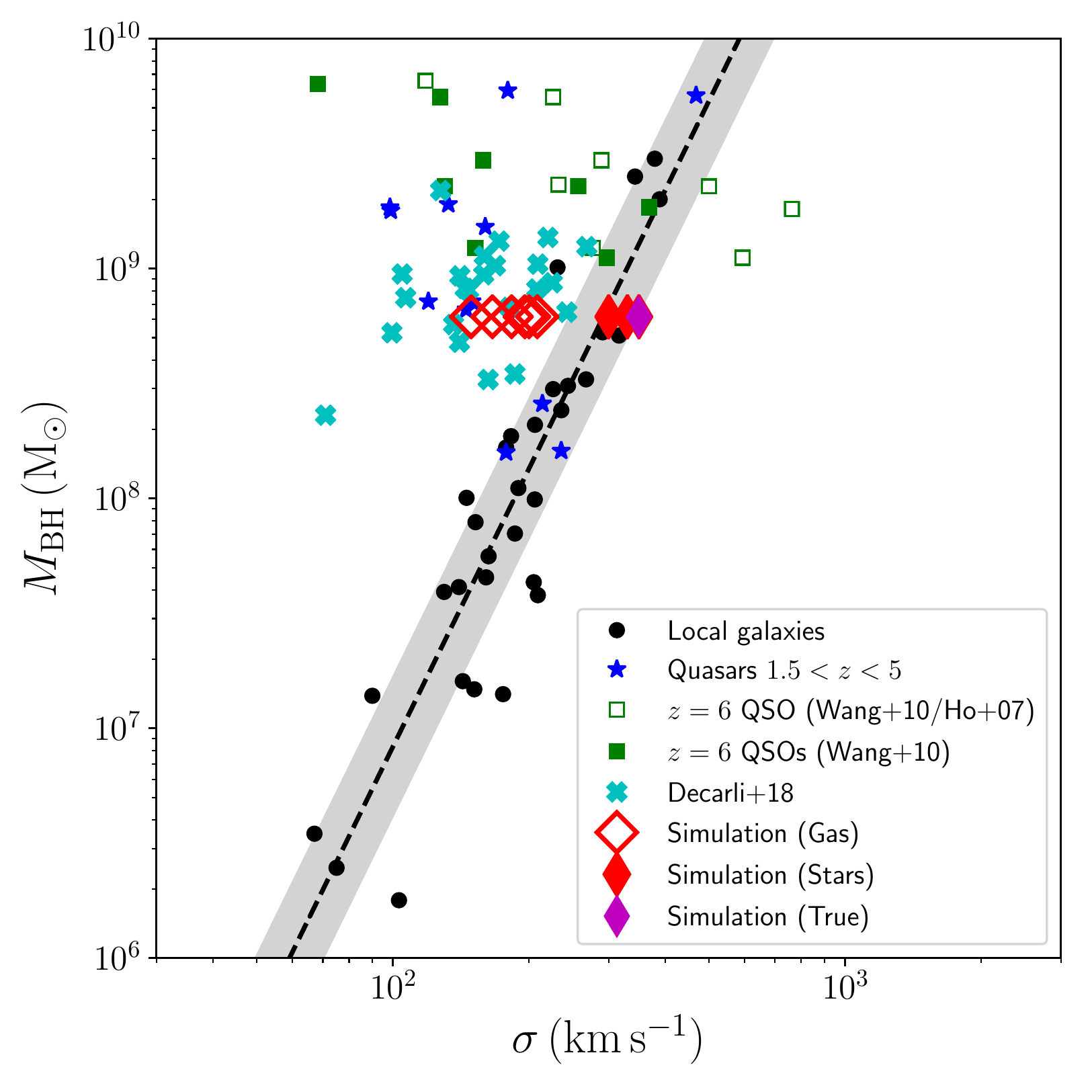}
\caption{Relation between MBH mass and velocity dispersion $\sigma$ in the galaxy. The black dots correspond to local galaxies, where $\sigma$ is measured from the velocity dispersion of the stellar bulge/spheroid \citep{tremaine02}, with the black dashed line representing the best-fit relation and the gray area the 0.3 decades nominal scatter. The blue stars are CO-detected quasars at $1.5<z<5$ \citep{shields06,coppin08a}, the green squares are quasars from \citet{wang10}, with the empty ones , and the cyan crosses are quasars from D18. Except for the empty green squares, based on the $\sigma$ estimate method in \citet{ho07}, in all other measurements  $\sigma \approx \rm FWHM/2.35$. The simulation data  from Table~\ref{tab:gas_kin} are shown with diamond-shaped symbols. The empty red diamonds assume that $\sigma=\rm FWHM/2.35$, as done for the high-redshift quasar samples, the filled red diamonds employ the velocity dispersions $\sigma_{\rm fit,*}$ of the stellar distribution from the velocity maps, while the filled magenta diamond shows the ``true'' stellar velocity dispersion $\sigma_*$, directly determined from the stellar particles in the galaxy. For the gas tracers, our results are in very good agreement with the bulk of the high-redshift quasar population, offset towards overmassive black holes with respect to the local relation. However,  when using the stellar velocity dispersion, i.e., the velocity dispersion used in the benchmark measurements for the local relation, the simulated MBH sits with the local measurements, and the MBH is not overmassive.}
\label{fig:mbh_sigma}
\end{figure}

In Fig.~\ref{fig:mbh_sigma}, we show  the $M_{\rm BH}$-$\sigma$ correlation. Observations cannot access the stellar velocity dispersion in  high-redshift quasars.  The velocity dispersion is estimated through the Gaussian fit to the line profile of the gas component, although it is a priori correct only for dispersion-dominated galaxies. CO-detected quasars at $1.5<z<5$ from \citet{shields06} and \citet{coppin08a} are shown as blue stars, high-redshift $z\sim 6$ quasars from \citet{wang10} as filled and empty green squares, with the empty squares based on the $\sigma$ estimate method by \citet{ho07}, and the D18 quasars as cyan crosses. For all these samples, $\sigma \approx \rm FWHM/2.35$ \citep{nelson00,shields06}. The results of the simulation, also based on $\sigma_{\rm FWHM}\equiv\rm FWHM/2.35$ as velocity dispersion (see Table~\ref{tab:gas_kin}), i.e., using the analogue of the quantity used for the high-redshift quasars, is shown with empty red diamonds. Independent of the method used to estimate $\sigma$ through gas tracers, the simulation agrees with observations, showing a shift with respect to the benchmark correlation in local galaxies, shown with black dashed line, $\log (M_{\rm BH}/\rm \msun) = 8.13 + 4.02 \log (\sigma/200\rm\, km\, s^{-1})$  \citep{tremaine02} and a gray area for the 0.3 decades nominal scatter. 

However, for local galaxies, shown as black dots, $\sigma$ is the stellar velocity dispersion, $\sigma_{*}$. The correct comparison to assess whether MBHs at high redshift are on the correlation is therefore with $\sigma_{\rm fit,*}$ derived from the stellar velocity dispersion maps (filled red diamonds) and $\sigma_*$ directly derived from the stellar particles (filled magenta diamond). As noted in Section~\ref{sec:dynmass}, the velocity dispersion obtained from the gas maps underestimates the stellar velocity dispersion. When we use $\sigma_{*}$ as $\sigma$, shown as filled diamonds in Fig.~\ref{fig:mbh_sigma}, the position occupied by the simulation's MBH is in very good agreement with local observations, which also adopt the stellar velocity dispersion as a tracer.

In summary, the MBH-galaxy relations appear to be consistent with $z=0$ values when the analogs of $z=0$ quantities are measured, and MBHs appear to be overmassive only because of the different tracers used.  This result is in agreement with simulations having lower resolution but larger statistics \citep{huang18}, which also find relations fully consistent with local relations. To this, one has to add the bias caused by selection of quasars with respect to quiescent galaxies \citep{lauer07}. Since the galaxy mass function is steep, at fixed quasar luminosity it is more probable to select a MBH more massive than the average in a galaxy of modest mass than a MBH with average mass for its host in a much rarer high-mass galaxy. For examples of the consequences of this bias, see \citet{volonteri11a} and \citet{volonteri16}.

\section{Discussion and conclusions}
\label{sec:conclusions}

We have investigated the evolution of a high-redshift quasar host with an extremely high resolution cosmological zoom-in simulation performed with the code \gizmo{} \citep{hopkins15}. In order to ensure that the target halo would be consistent with up-to-date constraints from both simulations \citep{dimatteo17,tenneti18,tenneti18b} and observations \citep{mazzucchelli17,uchiyama18}, we carefully selected the initial conditions to reproduce the desired halo properties and environmental conditions. In our simulation, we employed a state-of-the-art physically-motivated sub-grid modelling, including non-equilibrium chemistry for primordial elements, turbulence-based SF and stellar feedback (mechanical feedback from type II/Ia SNe and radiation from young stars),  as in \citet{lupi19}. Thanks to the extremely high resolution achieved (down to 5~pc for the gas), we were able to follow in great detail the evolution of the quasar host, and its central MBH, and accurately compare the results at $z \approx 7$ with recent observations of high-redshift quasars by \citet{decarli18}.

Two main caveats are present in the study. First, the lack of additional processes like HII regions, radiation pressure, stellar winds could result in a slightly less effective suppression of SF, as already shown by \citet{rosdahl17,lupi19}. Our choice of a fiducial boost factor to the radial momentum by SNe was motivated by this limitation, and also by the recent results by \citet{gentry17,gentry18}, and \citet{semenov18}, but it does not necessarily represent the ``true'' solution. Second, the MBH feedback model and the feedback efficiency employed could result in a slightly faster growth of the MBH and a subsequent less effective feedback on to the galaxy, that again could lead to a moderately higher stellar mass. Nonetheless, we do not expect these two limitations to significantly affect our conclusions: while a stronger SN/AGN feedback could reduce the stellar mass in the system, the resulting suppressed growth of the MBH due to a stronger AGN feedback could plausibly counterbalance this decrease, resulting in small differences compared to our results. Finally, our conclusions are based on one simulation of a particular halo, and additional simulations are needed before generalizing this result.

Our simulation showed that quasar hosts very quickly build up their stellar mass, exceeding $M_{\rm star}\sim 10^{10}\,\msun$ already at $z=8$, in agreement with previous results \citep{smidt18,barai18}. To reach these conditions, very high SFRs are required, in very good agreement with observations, and these SFRs also result in a great amount of metals produced (about twice the solar value), and a strong dust emission. Despite gas being very turbulent at these high redshift, such massive galaxies are able to settle in a disc-like structure with well defined spiral arms, and the system we have studied is more likely rotationally-supported rather than dispersion-dominated. To assess the presence of a bulge, we computed the disc to total ratio,  D/T, from the circularity of the stellar particles \citep{vogelsberger14,tenneti16,dimatteo17}. We placed the galaxy face-on, determined the specific angular momentum $j_z$ of the star particles along the z-direction and computed the circularity $c=j_z/j_{\rm circ}$, with $j_{\rm circ}$ the specific circular angular momentum. The disc component is then assumed to be composed of all the stellar particles with $c>0.7$ \citep{dimatteo17}.
This procedure produces D/T$\sim 0.55$, the galaxy, therefore presents a distinct bulge.  \cite{dimatteo17} suggest a predominance of  spheroids among the hosts of the most massive black holes, with two of the hosts having D/T$\sim 0.2$ and two having $\sim 0.55$; our MBH, hosted in a massive spiral galaxy with an important central bulge, has in fact a similar D/T to BH3 and BH4 in \citet{dimatteo17}, but our disc structure appears better resolved, likely because of the higher resolution in our simulation, leading to a better treatment of the disc vertical structure.

The large amount of gas available is also crucial for the growth of the central MBH, that accretes at the Eddington limit for most of the time, until its feedback becomes powerful enough to overcome the accretion, as already found by \citet{dimatteo17}. However, a crucial parameter for the growth of the MBH is the galaxy mass, as already pointed out by \citet[e.g.][]{dubois14b}, because of the impact of SN feedback. Indeed, in our simulation, MBH growth is strongly suppressed until the galaxy stellar mass reaches about $M_{\rm star}\sim 10^{10}\,\msun$, in good agreement with expectations. For instance, our result agrees with that of \citet{barai18}, that employ an effective kinetic SN feedback, but is very different from \citet{smidt18}, where the weaker thermal SN feedback probably allows the MBH to efficiently grow already at very early times (see, e.g., \citealt{rosdahl17} and \citealt{habouzit17} for discussions about the relative impact of different SN feedback models on the galaxy and the central MBHs).

Although not explicitly considered here, we also found that the central MBH could have experienced super-Eddington accretion phases during its life \citep[see, also,][where they found similar results with semi-analytic models]{pezzulli16}. If these phases really occur in the Universe, we could expect a more massive MBH, that would most likely be overmassive. However, this faster growth would also strongly depend on how the feedback from a super-Eddington accreting MBH would couple with the surrounding gas, as recently shown by \citet{regan18}.

We have examined the relative growth of the stellar component, via SFR, and of the MBH, via its accretion rate. If we consider a ``canonical'' value of $\sim 10^{-3}$ for keeping the relative growth at the pace needed for a symbiotic evolution maintaining the typical ratio between MBH and bulge mass, we find that at first the galaxy grows faster (ratio $<10^{-3}$), while later the MBH grows faster (ratio $>10^{-3}$). Towards the end of the simulation, once MBH feedback becomes important, the system is positioned close to the  $<10^{-3}$ value. This suggests a three-phase scenario. At first, when the galaxy is small, SN feedback regulates SF in the galaxy and suppresses MBH accretion. At intermediate galaxy mass SN feedback becomes inefficient at suppressing MBH growth, while AGN feedback cannot stop a runaway phase of MBH accretion. Finally, when the MBH has become sufficiently massive, it is its own feedback that limits growth. 

We have compared to observed MBH-galaxy correlations, local and at high redshift. We showed that gas-derived dynamical mass estimates typically underestimate the actual mass of the system, possibly explaining why these objects seem to be ``overmassive'' with respect to the host mass at high redshift, although the intrinsic relation is in good agreement with local measurements.  Similarly, when considering the relation with velocity dispersion, gas-based estimates underestimate the stellar velocity dispersion, leading to the MBHs being overmassive at a given velocity dispersion with respect to the local population.  In our simulation, when the ``true'' quantities are used, the discrepancy in the MBH-galaxy correlations completely disappears, simply leaving a population of ``normal'' MBHs in   massive (quickly evolving) galaxy. 

\bigskip
In summary, our main results are the following:
\begin{itemize}[topsep=0pt]
\item The host galaxy quickly builds up a stellar mass reaching $10^{11}\,\msun$  at $z=7$ (Fig.~\ref{fig:mass_z}). The galaxy itself is a very large galaxy at these redshifts. 
\item Initially, the BH growth lags behind the galaxy growth, in relative terms, while later the BH grows faster than the galaxy, also in relative terms (Fig.~\ref{fig:sfr_bhar}).
\item The BH-stellar mass evolution can be mapped on the sample of local MBHs, showing a similar behaviour as a function of galaxy mass (Fig.~\ref{fig:mbh_mstar})
\item Using the gas-based tracers for galaxy dynamical mass and velocity dispersion currently available for high-redshift quasars, MBHs appear overmassive with respect to the $z=0$ correlations  (Fig.~\ref{fig:MBH_Mdyn} and Fig.~\ref{fig:mbh_sigma}).
\item Using the stellar-based tracers for galaxy stellar mass and velocity dispersion used to derive the $z=0$ correlation the tension disappears, and no difference is found between the high-redshift and the $z=0$ MBHs (Fig.~\ref{fig:MBH_Mdyn} and Fig.~\ref{fig:mbh_sigma}).\\
\end{itemize}

Because current telescopes do not have high angular resolution at infrared wavelenghts (restframe optical), the actual stellar masses and velocity dispersion cannot be properly measured. JWST will improve our capabilities, but angular resolution will remain limited to a few hundred parsec.  Gas-based estimates of the velocity dispersion are a viable alternative, but they can easily lead to an underestimation of the actual velocity dispersion or dynamical mass:  our results can be used to optimize how to obtain better calibrated correction factors.

\section*{Acknowledgements}
We thank the anonymous referee for constructive comments that helped us to improve the quality of the paper. 
We acknowledge support from the European Research Council (Project No. 267117, `DARK', AL, JS; Project no. 614199, `BLACK', AL, MV). 
SB thanks for funding through Fondecyt Iniciacion 11170268, and PCI Redes Internacionales project number REDI170093. SB also thanks for funding via Conicyt PIA ACT172033 and BASAL Centro de Astrofısica y Tecnologıas Afines (CATA) PFB-06/2007. This work was granted access to the High Performance Computing resources of CINES under the allocations A0020406955, and A0040406955 by GENCI, and it has made use of the Horizon Cluster, hosted by Institut d'Astrophysique de Paris, for the analysis of the simulation results. The maps reported in this work have been created using \textsc{pynbody} \citep{pynbody}.
\bibliographystyle{mnras}
\bibliography{./Biblio}

\begin{thebibliography}{}
\makeatletter
\relax
\def\mn@urlcharsother{\let\do\@makeother \do\$\do\&\do\#\do\^\do\_\do\%\do\~}
\def\mn@doi{\begingroup\mn@urlcharsother \@ifnextchar [ {\mn@doi@}
  {\mn@doi@[]}}
\def\mn@doi@[#1]#2{\def\@tempa{#1}\ifx\@tempa\@empty \href
  {http://dx.doi.org/#2} {doi:#2}\else \href {http://dx.doi.org/#2} {#1}\fi
  \endgroup}
\def\mn@eprint#1#2{\mn@eprint@#1:#2::\@nil}
\def\mn@eprint@arXiv#1{\href {http://arxiv.org/abs/#1} {{\tt arXiv:#1}}}
\def\mn@eprint@dblp#1{\href {http://dblp.uni-trier.de/rec/bibtex/#1.xml}
  {dblp:#1}}
\def\mn@eprint@#1:#2:#3:#4\@nil{\def\@tempa {#1}\def\@tempb {#2}\def\@tempc
  {#3}\ifx \@tempc \@empty \let \@tempc \@tempb \let \@tempb \@tempa \fi \ifx
  \@tempb \@empty \def\@tempb {arXiv}\fi \@ifundefined
  {mn@eprint@\@tempb}{\@tempb:\@tempc}{\expandafter \expandafter \csname
  mn@eprint@\@tempb\endcsname \expandafter{\@tempc}}}

\bibitem[\protect\citeauthoryear{{Angl{\'e}s-Alc{\'a}zar},
  {Faucher-Gigu{\`e}re}, {Quataert}, {Hopkins}, {Feldmann}, {Torrey}, {Wetzel}
  \& {Kere{\v s}}}{{Angl{\'e}s-Alc{\'a}zar} et~al.}{2017}]{anglesalcazar17FIRE}
{Angl{\'e}s-Alc{\'a}zar} D.,  {Faucher-Gigu{\`e}re} C.-A.,  {Quataert} E.,
  {Hopkins} P.~F.,  {Feldmann} R.,  {Torrey} P.,  {Wetzel} A.,   {Kere{\v s}}
  D.,  2017, \mn@doi [\mnras] {10.1093/mnrasl/slx161}, \href
  {http://adsabs.harvard.edu/abs/2017MNRAS.472L.109A} {472, L109}

\bibitem[\protect\citeauthoryear{{Ba{\~n}ados} et~al.,}{{Ba{\~n}ados}
  et~al.}{2018}]{banados18}
{Ba{\~n}ados} E.,  et~al., 2018, \mn@doi [\nat] {10.1038/nature25180}, \href
  {http://adsabs.harvard.edu/abs/2018Natur.553..473B} {553, 473}

\bibitem[\protect\citeauthoryear{{Barai}, {Gallerani}, {Pallottini}, {Ferrara},
  {Marconi}, {Cicone}, {Maiolino}  \& {Carniani}}{{Barai}
  et~al.}{2018}]{barai18}
{Barai} P.,  {Gallerani} S.,  {Pallottini} A.,  {Ferrara} A.,  {Marconi} A.,
  {Cicone} C.,  {Maiolino} R.,   {Carniani} S.,  2018, \mn@doi [\mnras]
  {10.1093/mnras/stx2563}, \href
  {http://adsabs.harvard.edu/abs/2018MNRAS.473.4003B} {473, 4003}

\bibitem[\protect\citeauthoryear{{Beckmann}, {Slyz}  \& {Devriendt}}{{Beckmann}
  et~al.}{2018}]{beckmann18}
{Beckmann} R.~S.,  {Slyz} A.,   {Devriendt} J.,  2018, \mn@doi [\mnras]
  {10.1093/mnras/sty931}, \href
  {http://adsabs.harvard.edu/abs/2018MNRAS.478..995B} {478, 995}

\bibitem[\protect\citeauthoryear{{Behroozi}, {Wechsler}  \& {Wu}}{{Behroozi}
  et~al.}{2013}]{behroozi13}
{Behroozi} P.~S.,  {Wechsler} R.~H.,   {Wu} H.-Y.,  2013, \mn@doi [\apj]
  {10.1088/0004-637X/762/2/109}, \href
  {http://adsabs.harvard.edu/abs/2013ApJ...762..109B} {762, 109}

\bibitem[\protect\citeauthoryear{{Bellovary}, {Governato}, {Quinn}, {Wadsley},
  {Shen}  \& {Volonteri}}{{Bellovary} et~al.}{2010}]{bellovary10}
{Bellovary} J.~M.,  {Governato} F.,  {Quinn} T.~R.,  {Wadsley} J.,  {Shen} S.,
   {Volonteri} M.,  2010, \mn@doi [\apjl] {10.1088/2041-8205/721/2/L148}, \href
  {http://adsabs.harvard.edu/abs/2010ApJ...721L.148B} {721, L148}

\bibitem[\protect\citeauthoryear{{Bertschinger}}{{Bertschinger}}{1987}]{bertschinger87}
{Bertschinger} E.,  1987, \mn@doi [\apjl] {10.1086/185066}, \href
  {http://adsabs.harvard.edu/abs/1987ApJ...323L.103B} {323, L103}

\bibitem[\protect\citeauthoryear{{Biernacki}, {Teyssier}  \&
  {Bleuler}}{{Biernacki} et~al.}{2017}]{biernacki17}
{Biernacki} P.,  {Teyssier} R.,   {Bleuler} A.,  2017, \mn@doi [\mnras]
  {10.1093/mnras/stx845}, \href
  {http://adsabs.harvard.edu/abs/2017MNRAS.469..295B} {469, 295}

\bibitem[\protect\citeauthoryear{{Bondi}}{{Bondi}}{1952}]{bondi52}
{Bondi} H.,  1952, \mn@doi [\mnras] {10.1093/mnras/112.2.195}, \href
  {http://adsabs.harvard.edu/abs/1952MNRAS.112..195B} {112, 195}

\bibitem[\protect\citeauthoryear{{Bondi} \& {Hoyle}}{{Bondi} \&
  {Hoyle}}{1944}]{bondi44}
{Bondi} H.,  {Hoyle} F.,  1944, \mn@doi [\mnras] {10.1093/mnras/104.5.273},
  \href {http://adsabs.harvard.edu/abs/1944MNRAS.104..273B} {104, 273}

\bibitem[\protect\citeauthoryear{{Booth} \& {Schaye}}{{Booth} \&
  {Schaye}}{2009}]{booth09}
{Booth} C.~M.,  {Schaye} J.,  2009, \mn@doi [\mnras]
  {10.1111/j.1365-2966.2009.15043.x}, \href
  {http://adsabs.harvard.edu/abs/2009MNRAS.398...53B} {398, 53}

\bibitem[\protect\citeauthoryear{{Bovino}, {Grassi}, {Capelo}, {Schleicher}  \&
  {Banerjee}}{{Bovino} et~al.}{2016}]{bovino16}
{Bovino} S.,  {Grassi} T.,  {Capelo} P.~R.,  {Schleicher} D.~R.~G.,
  {Banerjee} R.,  2016, \mn@doi [\aap] {10.1051/0004-6361/201628158}, \href
  {http://adsabs.harvard.edu/abs/2016A%26A...590A..15B} {590, A15}

\bibitem[\protect\citeauthoryear{{Bower}, {Schaye}, {Frenk}, {Theuns},
  {Schaller}, {Crain}  \& {McAlpine}}{{Bower} et~al.}{2017}]{bower17}
{Bower} R.~G.,  {Schaye} J.,  {Frenk} C.~S.,  {Theuns} T.,  {Schaller} M.,
  {Crain} R.~A.,   {McAlpine} S.,  2017, \mn@doi [\mnras]
  {10.1093/mnras/stw2735}, \href
  {http://adsabs.harvard.edu/abs/2017MNRAS.465...32B} {465, 32}

\bibitem[\protect\citeauthoryear{{Bruzual} \& {Charlot}}{{Bruzual} \&
  {Charlot}}{2003}]{bruzual03}
{Bruzual} G.,  {Charlot} S.,  2003, \mn@doi [\mnras]
  {10.1046/j.1365-8711.2003.06897.x}, \href
  {http://adsabs.harvard.edu/abs/2003MNRAS.344.1000B} {344, 1000}

\bibitem[\protect\citeauthoryear{{Chabrier}}{{Chabrier}}{2003}]{chabrier03}
{Chabrier} G.,  2003, \mn@doi [\pasp] {10.1086/376392}, \href
  {http://adsabs.harvard.edu/abs/2003PASP..115..763C} {115, 763}

\bibitem[\protect\citeauthoryear{{Chen} et~al.,}{{Chen} et~al.}{2013}]{chen13}
{Chen} C.-T.~J.,  et~al., 2013, \mn@doi [\apj] {10.1088/0004-637X/773/1/3},
  \href {http://adsabs.harvard.edu/abs/2013ApJ...773....3C} {773, 3}

\bibitem[\protect\citeauthoryear{{Choi}, {Ostriker}, {Naab}  \&
  {Johansson}}{{Choi} et~al.}{2012}]{choi12}
{Choi} E.,  {Ostriker} J.~P.,  {Naab} T.,   {Johansson} P.~H.,  2012, \mn@doi
  [\apj] {10.1088/0004-637X/754/2/125}, \href
  {http://adsabs.harvard.edu/abs/2012ApJ...754..125C} {754, 125}

\bibitem[\protect\citeauthoryear{{Coppin} et~al.,}{{Coppin}
  et~al.}{2008}]{coppin08a}
{Coppin} K.~E.~K.,  et~al., 2008, \mn@doi [\mnras]
  {10.1111/j.1365-2966.2008.13553.x}, \href
  {http://adsabs.harvard.edu/abs/2008MNRAS.389...45C} {389, 45}

\bibitem[\protect\citeauthoryear{{Costa}, {Sijacki}, {Trenti}  \&
  {Haehnelt}}{{Costa} et~al.}{2014}]{costa14}
{Costa} T.,  {Sijacki} D.,  {Trenti} M.,   {Haehnelt} M.~G.,  2014, \mn@doi
  [\mnras] {10.1093/mnras/stu101}, \href
  {http://adsabs.harvard.edu/abs/2014MNRAS.439.2146C} {439, 2146}

\bibitem[\protect\citeauthoryear{{Decarli}, {Falomo}, {Treves}, {Labita},
  {Kotilainen}  \& {Scarpa}}{{Decarli} et~al.}{2010}]{decarli10}
{Decarli} R.,  {Falomo} R.,  {Treves} A.,  {Labita} M.,  {Kotilainen} J.~K.,
  {Scarpa} R.,  2010, \mn@doi [\mnras] {10.1111/j.1365-2966.2009.16049.x},
  \href {http://adsabs.harvard.edu/abs/2010MNRAS.402.2453D} {402, 2453}

\bibitem[\protect\citeauthoryear{{Decarli} et~al.,}{{Decarli}
  et~al.}{2018}]{decarli18}
{Decarli} R.,  et~al., 2018, \mn@doi [\apj] {10.3847/1538-4357/aaa5aa}, \href
  {http://adsabs.harvard.edu/abs/2018ApJ...854...97D} {854, 97}

\bibitem[\protect\citeauthoryear{{Di Matteo}, {Springel}  \& {Hernquist}}{{Di
  Matteo} et~al.}{2005}]{dimatteo05}
{Di Matteo} T.,  {Springel} V.,   {Hernquist} L.,  2005, \mn@doi [\nat]
  {10.1038/nature03335}, \href
  {http://adsabs.harvard.edu/abs/2005Natur.433..604D} {433, 604}

\bibitem[\protect\citeauthoryear{{Di Matteo}, {Croft}, {Feng}, {Waters}  \&
  {Wilkins}}{{Di Matteo} et~al.}{2017}]{dimatteo17}
{Di Matteo} T.,  {Croft} R.~A.~C.,  {Feng} Y.,  {Waters} D.,   {Wilkins} S.,
  2017, \mn@doi [\mnras] {10.1093/mnras/stx319}, \href
  {http://adsabs.harvard.edu/abs/2017MNRAS.467.4243D} {467, 4243}

\bibitem[\protect\citeauthoryear{{Dubois}, {Devriendt}, {Slyz}  \&
  {Teyssier}}{{Dubois} et~al.}{2012}]{dubois12}
{Dubois} Y.,  {Devriendt} J.,  {Slyz} A.,   {Teyssier} R.,  2012, \mn@doi
  [\mnras] {10.1111/j.1365-2966.2011.20236.x}, \href
  {http://adsabs.harvard.edu/abs/2012MNRAS.420.2662D} {420, 2662}

\bibitem[\protect\citeauthoryear{{Dubois}, {Volonteri}  \& {Silk}}{{Dubois}
  et~al.}{2014a}]{dubois14b}
{Dubois} Y.,  {Volonteri} M.,   {Silk} J.,  2014a, \mn@doi [\mnras]
  {10.1093/mnras/stu373}, \href
  {http://adsabs.harvard.edu/abs/2014MNRAS.440.1590D} {440, 1590}

\bibitem[\protect\citeauthoryear{{Dubois}, {Volonteri}, {Silk}, {Devriendt}  \&
  {Slyz}}{{Dubois} et~al.}{2014b}]{dubois14a}
{Dubois} Y.,  {Volonteri} M.,  {Silk} J.,  {Devriendt} J.,   {Slyz} A.,  2014b,
  \mn@doi [\mnras] {10.1093/mnras/stu425}, \href
  {http://adsabs.harvard.edu/abs/2014MNRAS.440.2333D} {440, 2333}

\bibitem[\protect\citeauthoryear{{Dubois} et~al.,}{{Dubois}
  et~al.}{2014c}]{dubois14}
{Dubois} Y.,  et~al., 2014c, \mn@doi [\mnras] {10.1093/mnras/stu1227}, \href
  {http://adsabs.harvard.edu/abs/2014MNRAS.444.1453D} {444, 1453}

\bibitem[\protect\citeauthoryear{{Dubois}, {Volonteri}, {Silk}, {Devriendt},
  {Slyz}  \& {Teyssier}}{{Dubois} et~al.}{2015}]{dubois15}
{Dubois} Y.,  {Volonteri} M.,  {Silk} J.,  {Devriendt} J.,  {Slyz} A.,
  {Teyssier} R.,  2015, \mn@doi [\mnras] {10.1093/mnras/stv1416}, \href
  {http://adsabs.harvard.edu/abs/2015MNRAS.452.1502D} {452, 1502}

\bibitem[\protect\citeauthoryear{{Fan} et~al.,}{{Fan} et~al.}{2006}]{fan06}
{Fan} X.,  et~al., 2006, \mn@doi [\aj] {10.1086/500296}, \href
  {http://adsabs.harvard.edu/abs/2006AJ....131.1203F} {131, 1203}

\bibitem[\protect\citeauthoryear{{Federrath} \& {Klessen}}{{Federrath} \&
  {Klessen}}{2012}]{federrath12}
{Federrath} C.,  {Klessen} R.~S.,  2012, \mn@doi [\apj]
  {10.1088/0004-637X/761/2/156}, \href
  {http://adsabs.harvard.edu/abs/2012ApJ...761..156F} {761, 156}

\bibitem[\protect\citeauthoryear{{Ferland} et~al.,}{{Ferland}
  et~al.}{2013}]{ferland13}
{Ferland} G.~J.,  et~al., 2013, \rmxaa, \href
  {http://adsabs.harvard.edu/abs/2013RMxAA..49..137F} {49, 137}

\bibitem[\protect\citeauthoryear{{Fiacconi}, {Mayer}, {Madau}, {Lupi}, {Dotti}
  \& {Haardt}}{{Fiacconi} et~al.}{2017}]{fiacconi17}
{Fiacconi} D.,  {Mayer} L.,  {Madau} P.,  {Lupi} A.,  {Dotti} M.,   {Haardt}
  F.,  2017, \mn@doi [\mnras] {10.1093/mnras/stx335}, \href
  {http://adsabs.harvard.edu/abs/2017MNRAS.467.4080F} {467, 4080}

\bibitem[\protect\citeauthoryear{{Gentry}, {Krumholz}, {Dekel}  \&
  {Madau}}{{Gentry} et~al.}{2017}]{gentry17}
{Gentry} E.~S.,  {Krumholz} M.~R.,  {Dekel} A.,   {Madau} P.,  2017, \mn@doi
  [\mnras] {10.1093/mnras/stw2746}, \href
  {http://adsabs.harvard.edu/abs/2017MNRAS.465.2471G} {465, 2471}

\bibitem[\protect\citeauthoryear{{Gentry}, {Krumholz}, {Madau}  \&
  {Lupi}}{{Gentry} et~al.}{2018}]{gentry18}
{Gentry} E.~S.,  {Krumholz} M.~R.,  {Madau} P.,   {Lupi} A.,  2018, preprint,
  \href {http://adsabs.harvard.edu/abs/2018arXiv180206860G} {} (\mn@eprint
  {arXiv} {1802.06860})

\bibitem[\protect\citeauthoryear{{Grassi}, {Bovino}, {Schleicher}, {Prieto},
  {Seifried}, {Simoncini}  \& {Gianturco}}{{Grassi} et~al.}{2014}]{grassi14}
{Grassi} T.,  {Bovino} S.,  {Schleicher} D.~R.~G.,  {Prieto} J.,  {Seifried}
  D.,  {Simoncini} E.,   {Gianturco} F.~A.,  2014, \mn@doi [\mnras]
  {10.1093/mnras/stu114}, \href
  {http://adsabs.harvard.edu/abs/2014MNRAS.439.2386G} {439, 2386}

\bibitem[\protect\citeauthoryear{{Haardt} \& {Madau}}{{Haardt} \&
  {Madau}}{2012}]{haardt12}
{Haardt} F.,  {Madau} P.,  2012, \mn@doi [\apj] {10.1088/0004-637X/746/2/125},
  \href {http://adsabs.harvard.edu/abs/2012ApJ...746..125H} {746, 125}

\bibitem[\protect\citeauthoryear{{Habouzit}, {Volonteri}  \&
  {Dubois}}{{Habouzit} et~al.}{2017}]{habouzit17}
{Habouzit} M.,  {Volonteri} M.,   {Dubois} Y.,  2017, \mn@doi [\mnras]
  {10.1093/mnras/stx666}, \href
  {http://adsabs.harvard.edu/abs/2017MNRAS.468.3935H} {468, 3935}

\bibitem[\protect\citeauthoryear{{Hahn} \& {Abel}}{{Hahn} \&
  {Abel}}{2013}]{hahn13}
{Hahn} O.,  {Abel} T.,  2013, {MUSIC: MUlti-Scale Initial Conditions},
  Astrophysics Source Code Library (\mn@eprint {ascl} {1311.011})

\bibitem[\protect\citeauthoryear{{Heiderman}, {Evans}, {Allen}, {Huard}  \&
  {Heyer}}{{Heiderman} et~al.}{2010}]{heiderman10}
{Heiderman} A.,  {Evans} II N.~J.,  {Allen} L.~E.,  {Huard} T.,   {Heyer} M.,
  2010, \mn@doi [\apj] {10.1088/0004-637X/723/2/1019}, \href
  {http://adsabs.harvard.edu/abs/2010ApJ...723.1019H} {723, 1019}

\bibitem[\protect\citeauthoryear{{Ho}}{{Ho}}{2007}]{ho07}
{Ho} L.~C.,  2007, \mn@doi [\apj] {10.1086/521917}, \href
  {http://adsabs.harvard.edu/abs/2007ApJ...669..821H} {669, 821}

\bibitem[\protect\citeauthoryear{{Hoffman} \& {Ribak}}{{Hoffman} \&
  {Ribak}}{1991}]{hoffman91}
{Hoffman} Y.,  {Ribak} E.,  1991, \mn@doi [\apjl] {10.1086/186160}, \href
  {http://adsabs.harvard.edu/abs/1991ApJ...380L...5H} {380, L5}

\bibitem[\protect\citeauthoryear{{Hopkins}}{{Hopkins}}{2015}]{hopkins15}
{Hopkins} P.~F.,  2015, \mn@doi [\mnras] {10.1093/mnras/stv195}, \href
  {http://adsabs.harvard.edu/abs/2015MNRAS.450...53H} {450, 53}

\bibitem[\protect\citeauthoryear{{Hopkins} et~al.,}{{Hopkins}
  et~al.}{2018}]{hopkins18}
{Hopkins} P.~F.,  et~al., 2018, \mn@doi [\mnras] {10.1093/mnras/sty1690}, \href
  {http://adsabs.harvard.edu/abs/2018MNRAS.480..800H} {480, 800}

\bibitem[\protect\citeauthoryear{{Hoyle} \& {Lyttleton}}{{Hoyle} \&
  {Lyttleton}}{1939}]{hoyle39}
{Hoyle} F.,  {Lyttleton} R.~A.,  1939, \mn@doi [Proceedings of the Cambridge
  Philosophical Society] {10.1017/S0305004100021150}, \href
  {http://adsabs.harvard.edu/abs/1939PCPS...35..405H} {35, 405}

\bibitem[\protect\citeauthoryear{{Huang}, {Di Matteo}, {Bhowmick}, {Feng}  \&
  {Ma}}{{Huang} et~al.}{2018}]{huang18}
{Huang} K.-W.,  {Di Matteo} T.,  {Bhowmick} A.~K.,  {Feng} Y.,   {Ma} C.-P.,
  2018, \mn@doi [\mnras] {10.1093/mnras/sty1329}, \href
  {http://adsabs.harvard.edu/abs/2018MNRAS.478.5063H} {478, 5063}

\bibitem[\protect\citeauthoryear{{Keller}, {Wadsley}, {Benincasa}  \&
  {Couchman}}{{Keller} et~al.}{2014}]{keller14}
{Keller} B.~W.,  {Wadsley} J.,  {Benincasa} S.~M.,   {Couchman} H.~M.~P.,
  2014, \mn@doi [\mnras] {10.1093/mnras/stu1058}, \href
  {http://adsabs.harvard.edu/abs/2014MNRAS.442.3013K} {442, 3013}

\bibitem[\protect\citeauthoryear{{Kim} \& {Ostriker}}{{Kim} \&
  {Ostriker}}{2015}]{kimostriker15}
{Kim} C.-G.,  {Ostriker} E.~C.,  2015, \mn@doi [\apj]
  {10.1088/0004-637X/802/2/99}, \href
  {http://adsabs.harvard.edu/abs/2015ApJ...802...99K} {802, 99}

\bibitem[\protect\citeauthoryear{{Kim} et~al.,}{{Kim} et~al.}{2014}]{kim14}
{Kim} J.-h.,  et~al., 2014, \mn@doi [\apjs] {10.1088/0067-0049/210/1/14}, \href
  {http://adsabs.harvard.edu/abs/2014ApJS..210...14K} {210, 14}

\bibitem[\protect\citeauthoryear{{Kim} et~al.,}{{Kim} et~al.}{2016}]{kim16}
{Kim} J.-h.,  et~al., 2016, \mn@doi [\apj] {10.3847/1538-4357/833/2/202}, \href
  {http://adsabs.harvard.edu/abs/2016ApJ...833..202K} {833, 202}

\bibitem[\protect\citeauthoryear{{Kim}, {Ostriker}  \& {Raileanu}}{{Kim}
  et~al.}{2017}]{kimostriker17}
{Kim} C.-G.,  {Ostriker} E.~C.,   {Raileanu} R.,  2017, \mn@doi [\apj]
  {10.3847/1538-4357/834/1/25}, \href
  {http://adsabs.harvard.edu/abs/2017ApJ...834...25K} {834, 25}

\bibitem[\protect\citeauthoryear{{Knollmann} \& {Knebe}}{{Knollmann} \&
  {Knebe}}{2009}]{knollmann09}
{Knollmann} S.~R.,  {Knebe} A.,  2009, \mn@doi [\apjs]
  {10.1088/0067-0049/182/2/608}, \href
  {http://adsabs.harvard.edu/abs/2009ApJS..182..608K} {182, 608}

\bibitem[\protect\citeauthoryear{{Kormendy} \& {Ho}}{{Kormendy} \&
  {Ho}}{2013}]{kormendy13bh}
{Kormendy} J.,  {Ho} L.~C.,  2013, \mn@doi [\araa]
  {10.1146/annurev-astro-082708-101811}, \href
  {http://adsabs.harvard.edu/abs/2013ARA%26A..51..511K} {51, 511}

\bibitem[\protect\citeauthoryear{{Lauer}, {Tremaine}, {Richstone}  \&
  {Faber}}{{Lauer} et~al.}{2007}]{lauer07}
{Lauer} T.~R.,  {Tremaine} S.,  {Richstone} D.,   {Faber} S.~M.,  2007, \mn@doi
  [\apj] {10.1086/522083}, \href
  {http://adsabs.harvard.edu/abs/2007ApJ...670..249L} {670, 249}

\bibitem[\protect\citeauthoryear{{Lupi}}{{Lupi}}{2019}]{lupi19}
{Lupi} A.,  2019, \mn@doi [\mnras] {10.1093/mnras/stz100}, \href
  {http://adsabs.harvard.edu/abs/2019MNRAS.484.1687L} {484, 1687}

\bibitem[\protect\citeauthoryear{Lupi, Bovino, Capelo, Volonteri  \& Silk}{Lupi
  et~al.}{2018}]{lupi18a}
Lupi A.,  Bovino S.,  Capelo P.~R.,  Volonteri M.,   Silk J.,  2018, \mn@doi
  [\mnras] {10.1093/mnras/stx2874}, 474, 2884

\bibitem[\protect\citeauthoryear{{Lyu}, {Rieke}  \& {Alberts}}{{Lyu}
  et~al.}{2016}]{lyu16}
{Lyu} J.,  {Rieke} G.~H.,   {Alberts} S.,  2016, \mn@doi [\apj]
  {10.3847/0004-637X/816/2/85}, \href
  {http://adsabs.harvard.edu/abs/2016ApJ...816...85L} {816, 85}

\bibitem[\protect\citeauthoryear{{Maoz}, {Mannucci}  \& {Brandt}}{{Maoz}
  et~al.}{2012}]{maoz12}
{Maoz} D.,  {Mannucci} F.,   {Brandt} T.~D.,  2012, \mn@doi [\mnras]
  {10.1111/j.1365-2966.2012.21871.x}, \href
  {http://adsabs.harvard.edu/abs/2012MNRAS.426.3282M} {426, 3282}

\bibitem[\protect\citeauthoryear{{Martizzi}, {Faucher-Gigu{\`e}re}  \&
  {Quataert}}{{Martizzi} et~al.}{2015}]{martizzi15}
{Martizzi} D.,  {Faucher-Gigu{\`e}re} C.-A.,   {Quataert} E.,  2015, \mn@doi
  [\mnras] {10.1093/mnras/stv562}, \href
  {http://adsabs.harvard.edu/abs/2015MNRAS.450..504M} {450, 504}

\bibitem[\protect\citeauthoryear{{Mazzucchelli}, {Ba{\~n}ados}, {Decarli},
  {Farina}, {Venemans}, {Walter}  \& {Overzier}}{{Mazzucchelli}
  et~al.}{2017}]{mazzucchelli17}
{Mazzucchelli} C.,  {Ba{\~n}ados} E.,  {Decarli} R.,  {Farina} E.~P.,
  {Venemans} B.~P.,  {Walter} F.,   {Overzier} R.,  2017, \mn@doi [\apj]
  {10.3847/1538-4357/834/1/83}, \href
  {http://adsabs.harvard.edu/abs/2017ApJ...834...83M} {834, 83}

\bibitem[\protect\citeauthoryear{{McAlpine}, {Bower}, {Rosario}, {Crain},
  {Schaye}  \& {Theuns}}{{McAlpine} et~al.}{2018}]{mcalpine18}
{McAlpine} S.,  {Bower} R.~G.,  {Rosario} D.~J.,  {Crain} R.~A.,  {Schaye} J.,
   {Theuns} T.,  2018, \mn@doi [\mnras] {10.1093/mnras/sty2489}, \href
  {http://adsabs.harvard.edu/abs/2018MNRAS.481.3118M} {481, 3118}

\bibitem[\protect\citeauthoryear{{Merloni} et~al.,}{{Merloni}
  et~al.}{2010}]{merloni10}
{Merloni} A.,  et~al., 2010, \mn@doi [\apj] {10.1088/0004-637X/708/1/137},
  \href {http://adsabs.harvard.edu/abs/2010ApJ...708..137M} {708, 137}

\bibitem[\protect\citeauthoryear{{Mortlock} et~al.,}{{Mortlock}
  et~al.}{2011}]{mortlock11}
{Mortlock} D.~J.,  et~al., 2011, \mn@doi [\nat] {10.1038/nature10159}, \href
  {http://adsabs.harvard.edu/abs/2011Natur.474..616M} {474, 616}

\bibitem[\protect\citeauthoryear{{Moster}, {Naab}  \& {White}}{{Moster}
  et~al.}{2018}]{moster18}
{Moster} B.~P.,  {Naab} T.,   {White} S.~D.~M.,  2018, \mn@doi [\mnras]
  {10.1093/mnras/sty655}, \href
  {http://adsabs.harvard.edu/abs/2018MNRAS.477.1822M} {477, 1822}

\bibitem[\protect\citeauthoryear{{Mullaney} et~al.,}{{Mullaney}
  et~al.}{2012}]{mullaney12}
{Mullaney} J.~R.,  et~al., 2012, \mn@doi [\apjl] {10.1088/2041-8205/753/2/L30},
  \href {http://adsabs.harvard.edu/abs/2012ApJ...753L..30M} {753, L30}

\bibitem[\protect\citeauthoryear{{Negri} \& {Volonteri}}{{Negri} \&
  {Volonteri}}{2017}]{negri17}
{Negri} A.,  {Volonteri} M.,  2017, \mn@doi [\mnras] {10.1093/mnras/stx362},
  \href {http://adsabs.harvard.edu/abs/2017MNRAS.467.3475N} {467, 3475}

\bibitem[\protect\citeauthoryear{{Nelson}}{{Nelson}}{2000}]{nelson00}
{Nelson} C.~H.,  2000, \mn@doi [\apjl] {10.1086/317314}, \href
  {http://adsabs.harvard.edu/abs/2000ApJ...544L..91N} {544, L91}

\bibitem[\protect\citeauthoryear{{Netzer}, {Mor}, {Trakhtenbrot}, {Shemmer}  \&
  {Lira}}{{Netzer} et~al.}{2014}]{netzer14}
{Netzer} H.,  {Mor} R.,  {Trakhtenbrot} B.,  {Shemmer} O.,   {Lira} P.,  2014,
  \mn@doi [\apj] {10.1088/0004-637X/791/1/34}, \href
  {http://adsabs.harvard.edu/abs/2014ApJ...791...34N} {791, 34}

\bibitem[\protect\citeauthoryear{{Ostriker}, {Choi}, {Ciotti}, {Novak}  \&
  {Proga}}{{Ostriker} et~al.}{2010}]{ostriker10}
{Ostriker} J.~P.,  {Choi} E.,  {Ciotti} L.,  {Novak} G.~S.,   {Proga} D.,
  2010, \mn@doi [\apj] {10.1088/0004-637X/722/1/642}, \href
  {http://adsabs.harvard.edu/abs/2010ApJ...722..642O} {722, 642}

\bibitem[\protect\citeauthoryear{{Padoan} \& {Nordlund}}{{Padoan} \&
  {Nordlund}}{2011}]{padoan11}
{Padoan} P.,  {Nordlund} {\AA}.,  2011, \mn@doi [\apj]
  {10.1088/0004-637X/730/1/40}, \href
  {http://adsabs.harvard.edu/abs/2011ApJ...730...40P} {730, 40}

\bibitem[\protect\citeauthoryear{{Pallottini}, {Ferrara}, {Gallerani},
  {Vallini}, {Maiolino}  \& {Salvadori}}{{Pallottini}
  et~al.}{2017}]{pallottini17}
{Pallottini} A.,  {Ferrara} A.,  {Gallerani} S.,  {Vallini} L.,  {Maiolino} R.,
    {Salvadori} S.,  2017, \mn@doi [\mnras] {10.1093/mnras/stw2847}, \href
  {http://adsabs.harvard.edu/abs/2017MNRAS.465.2540P} {465, 2540}

\bibitem[\protect\citeauthoryear{{Pezzulli}, {Valiante}  \&
  {Schneider}}{{Pezzulli} et~al.}{2016}]{pezzulli16}
{Pezzulli} E.,  {Valiante} R.,   {Schneider} R.,  2016, \mn@doi [\mnras]
  {10.1093/mnras/stw505}, \href
  {http://adsabs.harvard.edu/abs/2016MNRAS.458.3047P} {458, 3047}

\bibitem[\protect\citeauthoryear{{Pfister}, {Volonteri}, {Dubois}, {Dotti}  \&
  {Colpi}}{{Pfister} et~al.}{2019}]{pfister19}
{Pfister} H.,  {Volonteri} M.,  {Dubois} Y.,  {Dotti} M.,   {Colpi} M.,  2019,
  arXiv e-prints, \href {http://adsabs.harvard.edu/abs/2019arXiv190201297P} {}

\bibitem[\protect\citeauthoryear{{Planck Collaboration} et~al.,}{{Planck
  Collaboration} et~al.}{2016}]{planck16}
{Planck Collaboration} et~al., 2016, \mn@doi [A&A]
  {10.1051/0004-6361/201525830}, 594, A13

\bibitem[\protect\citeauthoryear{{Pontzen}, {Ro{\v s}kar}, {Stinson}, {Woods},
  {Reed}, {Coles}  \& {Quinn}}{{Pontzen} et~al.}{2013}]{pynbody}
{Pontzen} A.,  {Ro{\v s}kar} R.,  {Stinson} G.~S.,  {Woods} R.,  {Reed} D.~M.,
  {Coles} J.,   {Quinn} T.~R.,  2013, {pynbody: Astrophysics Simulation
  Analysis for Python}

\bibitem[\protect\citeauthoryear{{Prieto}, {Escala}, {Volonteri}  \&
  {Dubois}}{{Prieto} et~al.}{2017}]{prieto17}
{Prieto} J.,  {Escala} A.,  {Volonteri} M.,   {Dubois} Y.,  2017, \mn@doi
  [\apj] {10.3847/1538-4357/aa5be5}, \href
  {http://adsabs.harvard.edu/abs/2017ApJ...836..216P} {836, 216}

\bibitem[\protect\citeauthoryear{{Regan}, {Downes}, {Volonteri}, {Beckmann},
  {Lupi}, {Trebitsch}  \& {Dubois}}{{Regan} et~al.}{2018}]{regan18}
{Regan} J.~A.,  {Downes} T.~P.,  {Volonteri} M.,  {Beckmann} R.,  {Lupi} A.,
  {Trebitsch} M.,   {Dubois} Y.,  2018, preprint, \href
  {http://adsabs.harvard.edu/abs/2018arXiv181104953R} {} (\mn@eprint {arXiv}
  {1811.04953})

\bibitem[\protect\citeauthoryear{{Reines} \& {Volonteri}}{{Reines} \&
  {Volonteri}}{2015}]{reines15}
{Reines} A.~E.,  {Volonteri} M.,  2015, \mn@doi [\apj]
  {10.1088/0004-637X/813/2/82}, \href
  {http://adsabs.harvard.edu/abs/2015ApJ...813...82R} {813, 82}

\bibitem[\protect\citeauthoryear{{Richardson}, {Scannapieco}, {Devriendt},
  {Slyz}, {Thacker}, {Dubois}, {Wurster}  \& {Silk}}{{Richardson}
  et~al.}{2016}]{richardson16}
{Richardson} M.~L.~A.,  {Scannapieco} E.,  {Devriendt} J.,  {Slyz} A.,
  {Thacker} R.~J.,  {Dubois} Y.,  {Wurster} J.,   {Silk} J.,  2016, \mn@doi
  [\apj] {10.3847/0004-637X/825/2/83}, \href
  {http://adsabs.harvard.edu/abs/2016ApJ...825...83R} {825, 83}

\bibitem[\protect\citeauthoryear{{Rosdahl}, {Schaye}, {Dubois}, {Kimm}  \&
  {Teyssier}}{{Rosdahl} et~al.}{2017}]{rosdahl17}
{Rosdahl} J.,  {Schaye} J.,  {Dubois} Y.,  {Kimm} T.,   {Teyssier} R.,  2017,
  \mn@doi [\mnras] {10.1093/mnras/stw3034}, \href
  {http://adsabs.harvard.edu/abs/2017MNRAS.466...11R} {466, 11}

\bibitem[\protect\citeauthoryear{{Salmon} et~al.,}{{Salmon}
  et~al.}{2015}]{salmon15}
{Salmon} B.,  et~al., 2015, \mn@doi [\apj] {10.1088/0004-637X/799/2/183}, \href
  {http://adsabs.harvard.edu/abs/2015ApJ...799..183S} {799, 183}

\bibitem[\protect\citeauthoryear{{S{\c a}dowski}, {Lasota}, {Abramowicz}  \&
  {Narayan}}{{S{\c a}dowski} et~al.}{2016}]{sadowski16}
{S{\c a}dowski} A.,  {Lasota} J.-P.,  {Abramowicz} M.~A.,   {Narayan} R.,
  2016, \mn@doi [\mnras] {10.1093/mnras/stv2854}, \href
  {http://adsabs.harvard.edu/abs/2016MNRAS.456.3915S} {456, 3915}

\bibitem[\protect\citeauthoryear{{Schaye} et~al.,}{{Schaye}
  et~al.}{2015}]{schaye15}
{Schaye} J.,  et~al., 2015, \mn@doi [\mnras] {10.1093/mnras/stu2058}, \href
  {http://adsabs.harvard.edu/abs/2015MNRAS.446..521S} {446, 521}

\bibitem[\protect\citeauthoryear{{Semenov}, {Kravtsov}  \& {Gnedin}}{{Semenov}
  et~al.}{2018}]{semenov18}
{Semenov} V.~A.,  {Kravtsov} A.~V.,   {Gnedin} N.~Y.,  2018, \mn@doi [\apj]
  {10.3847/1538-4357/aac6eb}, \href
  {http://adsabs.harvard.edu/abs/2018ApJ...861....4S} {861, 4}

\bibitem[\protect\citeauthoryear{{Shen}, {Wadsley}  \& {Stinson}}{{Shen}
  et~al.}{2010}]{shen10}
{Shen} S.,  {Wadsley} J.,   {Stinson} G.,  2010, \mn@doi [\mnras]
  {10.1111/j.1365-2966.2010.17047.x}, \href
  {http://adsabs.harvard.edu/abs/2010MNRAS.407.1581S} {407, 1581}

\bibitem[\protect\citeauthoryear{{Shen}, {Madau}, {Guedes}, {Mayer},
  {Prochaska}  \& {Wadsley}}{{Shen} et~al.}{2013}]{shen13}
{Shen} S.,  {Madau} P.,  {Guedes} J.,  {Mayer} L.,  {Prochaska} J.~X.,
  {Wadsley} J.,  2013, \mn@doi [\apj] {10.1088/0004-637X/765/2/89}, \href
  {http://adsabs.harvard.edu/abs/2013ApJ...765...89S} {765, 89}

\bibitem[\protect\citeauthoryear{{Shields}, {Menezes}, {Massart}  \& {Vanden
  Bout}}{{Shields} et~al.}{2006}]{shields06}
{Shields} G.~A.,  {Menezes} K.~L.,  {Massart} C.~A.,   {Vanden Bout} P.,  2006,
  \mn@doi [\apj] {10.1086/500542}, \href
  {http://adsabs.harvard.edu/abs/2006ApJ...641..683S} {641, 683}

\bibitem[\protect\citeauthoryear{{Sijacki}, {Springel}, {Di Matteo}  \&
  {Hernquist}}{{Sijacki} et~al.}{2007}]{sijacki07}
{Sijacki} D.,  {Springel} V.,  {Di Matteo} T.,   {Hernquist} L.,  2007, \mn@doi
  [\mnras] {10.1111/j.1365-2966.2007.12153.x}, \href
  {http://adsabs.harvard.edu/abs/2007MNRAS.380..877S} {380, 877}

\bibitem[\protect\citeauthoryear{{Sijacki}, {Springel}  \&
  {Haehnelt}}{{Sijacki} et~al.}{2009}]{sijacki09}
{Sijacki} D.,  {Springel} V.,   {Haehnelt} M.~G.,  2009, \mn@doi [\mnras]
  {10.1111/j.1365-2966.2009.15452.x}, \href
  {http://adsabs.harvard.edu/abs/2009MNRAS.400..100S} {400, 100}

\bibitem[\protect\citeauthoryear{{Smidt}, {Whalen}, {Johnson}, {Surace}  \&
  {Li}}{{Smidt} et~al.}{2018}]{smidt18}
{Smidt} J.,  {Whalen} D.~J.,  {Johnson} J.~L.,  {Surace} M.,   {Li} H.,  2018,
  \mn@doi [\apj] {10.3847/1538-4357/aad7b8}, \href
  {http://adsabs.harvard.edu/abs/2018ApJ...865..126S} {865, 126}

\bibitem[\protect\citeauthoryear{{Soltan}}{{Soltan}}{1982}]{soltan82}
{Soltan} A.,  1982, \mnras, \href
  {http://adsabs.harvard.edu/abs/1982MNRAS.200..115S} {200, 115}

\bibitem[\protect\citeauthoryear{{Song} et~al.,}{{Song} et~al.}{2016}]{song16}
{Song} M.,  et~al., 2016, \mn@doi [\apj] {10.3847/0004-637X/825/1/5}, \href
  {http://adsabs.harvard.edu/abs/2016ApJ...825....5S} {825, 5}

\bibitem[\protect\citeauthoryear{{Springel}}{{Springel}}{2005}]{springel05}
{Springel} V.,  2005, \mn@doi [\mnras] {10.1111/j.1365-2966.2005.09655.x},
  \href {http://adsabs.harvard.edu/abs/2005MNRAS.364.1105S} {364, 1105}

\bibitem[\protect\citeauthoryear{{Springel}}{{Springel}}{2010}]{springel10}
{Springel} V.,  2010, \mn@doi [\mnras] {10.1111/j.1365-2966.2009.15715.x},
  \href {http://adsabs.harvard.edu/abs/2010MNRAS.401..791S} {401, 791}

\bibitem[\protect\citeauthoryear{{Springel}, {Di Matteo}  \&
  {Hernquist}}{{Springel} et~al.}{2005}]{springel05a}
{Springel} V.,  {Di Matteo} T.,   {Hernquist} L.,  2005, \mn@doi [\mnras]
  {10.1111/j.1365-2966.2005.09238.x}, \href
  {http://adsabs.harvard.edu/abs/2005MNRAS.361..776S} {361, 776}

\bibitem[\protect\citeauthoryear{{Tacconi} et~al.,}{{Tacconi}
  et~al.}{2010}]{tacconi10}
{Tacconi} L.~J.,  et~al., 2010, \mn@doi [\nat] {10.1038/nature08773}, \href
  {http://adsabs.harvard.edu/abs/2010Natur.463..781T} {463, 781}

\bibitem[\protect\citeauthoryear{{Tenneti}, {Mandelbaum}  \& {Di
  Matteo}}{{Tenneti} et~al.}{2016}]{tenneti16}
{Tenneti} A.,  {Mandelbaum} R.,   {Di Matteo} T.,  2016, \mn@doi [\mnras]
  {10.1093/mnras/stw1823}, \href
  {http://adsabs.harvard.edu/abs/2016MNRAS.462.2668T} {462, 2668}

\bibitem[\protect\citeauthoryear{{Tenneti}, {Wilkins}, {Matteo}, {Croft}  \&
  {Feng}}{{Tenneti} et~al.}{2018a}]{tenneti18b}
{Tenneti} A.,  {Wilkins} S.~M.,  {Matteo} T.~D.,  {Croft} R.~A.~C.,   {Feng}
  Y.,  2018a, \mn@doi [\mnras] {10.1093/mnras/sty3161}, \href
  {http://adsabs.harvard.edu/abs/2018MNRAS.tmp.3003T} {}

\bibitem[\protect\citeauthoryear{{Tenneti}, {Di Matteo}, {Croft}, {Garcia}  \&
  {Feng}}{{Tenneti} et~al.}{2018b}]{tenneti18}
{Tenneti} A.,  {Di Matteo} T.,  {Croft} R.,  {Garcia} T.,   {Feng} Y.,  2018b,
  \mn@doi [\mnras] {10.1093/mnras/stx2788}, \href
  {http://adsabs.harvard.edu/abs/2018MNRAS.474..597T} {474, 597}

\bibitem[\protect\citeauthoryear{{Trebitsch}, {Volonteri}, {Dubois}  \&
  {Madau}}{{Trebitsch} et~al.}{2018}]{trebitsch18}
{Trebitsch} M.,  {Volonteri} M.,  {Dubois} Y.,   {Madau} P.,  2018, \mn@doi
  [\mnras] {10.1093/mnras/sty1406}, \href
  {http://adsabs.harvard.edu/abs/2018MNRAS.478.5607T} {478, 5607}

\bibitem[\protect\citeauthoryear{{Tremaine} et~al.,}{{Tremaine}
  et~al.}{2002}]{tremaine02}
{Tremaine} S.,  et~al., 2002, \mn@doi [\apj] {10.1086/341002}, \href
  {http://adsabs.harvard.edu/abs/2002ApJ...574..740T} {574, 740}

\bibitem[\protect\citeauthoryear{{Tremmel}, {Governato}, {Volonteri}  \&
  {Quinn}}{{Tremmel} et~al.}{2015}]{tremmel15}
{Tremmel} M.,  {Governato} F.,  {Volonteri} M.,   {Quinn} T.~R.,  2015, \mn@doi
  [\mnras] {10.1093/mnras/stv1060}, \href
  {http://adsabs.harvard.edu/abs/2015MNRAS.451.1868T} {451, 1868}

\bibitem[\protect\citeauthoryear{{Tremmel}, {Karcher}, {Governato},
  {Volonteri}, {Quinn}, {Pontzen}, {Anderson}  \& {Bellovary}}{{Tremmel}
  et~al.}{2017}]{tremmel17}
{Tremmel} M.,  {Karcher} M.,  {Governato} F.,  {Volonteri} M.,  {Quinn} T.~R.,
  {Pontzen} A.,  {Anderson} L.,   {Bellovary} J.,  2017, \mn@doi [\mnras]
  {10.1093/mnras/stx1160}, \href
  {http://adsabs.harvard.edu/abs/2017MNRAS.470.1121T} {470, 1121}

\bibitem[\protect\citeauthoryear{{Uchiyama} et~al.,}{{Uchiyama}
  et~al.}{2018}]{uchiyama18}
{Uchiyama} H.,  et~al., 2018, \mn@doi [\pasj] {10.1093/pasj/psx112}, \href
  {http://adsabs.harvard.edu/abs/2018PASJ...70S..32U} {70, S32}

\bibitem[\protect\citeauthoryear{{Valiante}, {Schneider}, {Salvadori}  \&
  {Bianchi}}{{Valiante} et~al.}{2011}]{valiante11}
{Valiante} R.,  {Schneider} R.,  {Salvadori} S.,   {Bianchi} S.,  2011, \mn@doi
  [\mnras] {10.1111/j.1365-2966.2011.19168.x}, \href
  {http://adsabs.harvard.edu/abs/2011MNRAS.416.1916V} {416, 1916}

\bibitem[\protect\citeauthoryear{{Valiante}, {Schneider}, {Salvadori}  \&
  {Gallerani}}{{Valiante} et~al.}{2014}]{valiante14}
{Valiante} R.,  {Schneider} R.,  {Salvadori} S.,   {Gallerani} S.,  2014,
  \mn@doi [\mnras] {10.1093/mnras/stu1613}, \href
  {http://adsabs.harvard.edu/abs/2014MNRAS.444.2442V} {444, 2442}

\bibitem[\protect\citeauthoryear{{Venemans} et~al.,}{{Venemans}
  et~al.}{2017}]{venemans17}
{Venemans} B.~P.,  et~al., 2017, \mn@doi [\apjl] {10.3847/2041-8213/aa943a},
  \href {http://adsabs.harvard.edu/abs/2017ApJ...851L...8V} {851, L8}

\bibitem[\protect\citeauthoryear{{Vestergaard}, {Fan}, {Tremonti}, {Osmer}  \&
  {Richards}}{{Vestergaard} et~al.}{2008}]{vestergaard08}
{Vestergaard} M.,  {Fan} X.,  {Tremonti} C.~A.,  {Osmer} P.~S.,   {Richards}
  G.~T.,  2008, \mn@doi [\apjl] {10.1086/528981}, \href
  {http://adsabs.harvard.edu/abs/2008ApJ...674L...1V} {674, L1}

\bibitem[\protect\citeauthoryear{{Vogelsberger} et~al.,}{{Vogelsberger}
  et~al.}{2014}]{vogelsberger14}
{Vogelsberger} M.,  et~al., 2014, \mn@doi [\mnras] {10.1093/mnras/stu1536},
  \href {http://adsabs.harvard.edu/abs/2014MNRAS.444.1518V} {444, 1518}

\bibitem[\protect\citeauthoryear{{Volonteri} \& {Reines}}{{Volonteri} \&
  {Reines}}{2016}]{volonteri16}
{Volonteri} M.,  {Reines} A.~E.,  2016, \mn@doi [\apjl]
  {10.3847/2041-8205/820/1/L6}, \href
  {http://adsabs.harvard.edu/abs/2016ApJ...820L...6V} {820, L6}

\bibitem[\protect\citeauthoryear{{Volonteri} \& {Stark}}{{Volonteri} \&
  {Stark}}{2011}]{volonteri11a}
{Volonteri} M.,  {Stark} D.~P.,  2011, \mn@doi [\mnras]
  {10.1111/j.1365-2966.2011.19391.x}, \href
  {http://adsabs.harvard.edu/abs/2011MNRAS.417.2085V} {417, 2085}

\bibitem[\protect\citeauthoryear{{Walter}, {Carilli}, {Bertoldi}, {Menten},
  {Cox}, {Lo}, {Fan}  \& {Strauss}}{{Walter} et~al.}{2004}]{walter04}
{Walter} F.,  {Carilli} C.,  {Bertoldi} F.,  {Menten} K.,  {Cox} P.,  {Lo}
  K.~Y.,  {Fan} X.,   {Strauss} M.~A.,  2004, \mn@doi [\apjl] {10.1086/426017},
  \href {http://adsabs.harvard.edu/abs/2004ApJ...615L..17W} {615, L17}

\bibitem[\protect\citeauthoryear{{Wang} et~al.,}{{Wang} et~al.}{2010}]{wang10}
{Wang} R.,  et~al., 2010, \mn@doi [\apj] {10.1088/0004-637X/714/1/699}, \href
  {http://adsabs.harvard.edu/abs/2010ApJ...714..699W} {714, 699}

\bibitem[\protect\citeauthoryear{{Wang} et~al.,}{{Wang} et~al.}{2011}]{wang11}
{Wang} R.,  et~al., 2011, \mn@doi [\aj] {10.1088/0004-6256/142/4/101}, \href
  {http://adsabs.harvard.edu/abs/2011AJ....142..101W} {142, 101}

\bibitem[\protect\citeauthoryear{{Willott}, {Bergeron}  \& {Omont}}{{Willott}
  et~al.}{2015}]{willott15}
{Willott} C.~J.,  {Bergeron} J.,   {Omont} A.,  2015, \mn@doi [\apj]
  {10.1088/0004-637X/801/2/123}, \href
  {http://adsabs.harvard.edu/abs/2015ApJ...801..123W} {801, 123}

\makeatother
\end{thebibliography}

\appendix
\section{Validation of the black hole accretion and feedback model}
\label{app:validation} 
The limited resolution in large-scale simulation does not allow a proper modelling of the accretion process on to MBHs. In many cases, even the influence radius of the MBH is not properly resolved. Therefore, to model gas accretion and the MBH growth, different sub-grid prescriptions have been proposed in the literature. These prescriptions exhibit different sensitivity to the mass/spatial resolution achieved and to the thermodynamic modelling of the gas.

When MBHs accrete mass, a fraction of the mass-energy is released in the form of radiation, with the accretion-powered luminosity $L_{\rm BH} = \eta_{\rm rad}\dot{M}_{\rm BH} c^2$, where $\eta_{\rm rad}$ is the radiative efficiency. The exact value of $\eta_{\rm rad}$ depends on the radius of the innermost stable circular orbit, which is tied to the MBH spin $a$, and ranges from $\sim 0.057$ (for a Schwarzchild MBH, i.e. $a=0$) up to $\sim 0.42$ for a maximally rotating Kerr MBH ($a=1.0$), but the commonly assumed value, that we also employ here, is 0.1 \citep{soltan82}. However, only a fraction of the produced radiation actually couples with the surrounding gas, resulting in a feedback effect from the MBH. In numerical simulations, this coupling efficiency $\eta_{\rm fbk}$ is sensitive to the resolution, the numerical technique employed and also depends on the way feedback is modelled on unresolved scales. The common approach is to employ ad-hoc values calibrated to reproduce the local galaxy-MBH correlations. The coupling efficiencies thus calibrated  range from a few $10^{-3}$ up to 0.15 \citep[e.g.][]{dimatteo05,booth09,ostriker10}. Obviously, different choices result in different accretion histories and different feedback effects on the galaxy host.

The accretion prescription employed in this study has been chosen among several others after a detailed investigation in idealised conditions. In particular, we tested three different prescriptions on an idealised Milky-Way like galaxy with a MBH with $M_{\rm BH}=4.6\times 10^6\,\msun$, evolved for 1~Gyr using the same sub-grid model employed in the cosmological run. The initial conditions are those of the AGORA isolated disc comparison \citep{kim16}.

In particular, we tested the following models:
\begin{enumerate}
\item Common BHL accretion, based on the kernel-weighted quantities around the MBH, where
\begin{equation}
\dot{M}_{\rm BHL}= \frac{4\pi\grav^2 M_{\bullet}^2\langle\rho_{\rm gas}\rangle}{(\langle v_{\rm rel}\rangle^2+\langle c_{\rm s}\rangle^2)^{3/2}},
\end{equation}
where $M_\bullet$ is the BH mass, $\langle\rho_{\rm gas}\rangle$ is the average gas density, $\langle v_{\rm rel}\rangle$ the average gas-BH relative velocity, and $\langle c_{\rm s}\rangle$ the average sound speed.
\item Weighted BHL accretion, i.e. our fiducial model, where the accretion rate is the kernel average of the local accretion rate from each gas neighbour \citep{choi12}, where
\begin{equation}
\dot{M}_{\rm accr}= \left\langle\frac{4\pi\grav^2 M_{\bullet}^2\rho_{\rm gas}}{(v_{\rm rel}^2+c_{\rm s}^2)^{3/2}}\right\rangle.
\end{equation}
\item Flux accretion, where the accretion rate comes from the continuity equation as the mass flux within the MBH accretion radius, i.e.
\begin{equation}
\dot{M}_{\rm accr} = -\int_{\Delta V} \nabla\cdot(\rho_{\rm gas}\mathbf{v}_{\rm rel}).
\end{equation}
\end{enumerate}
For the BH feedback, we explore different values for $\eta_{\rm fbk}$, respectively 0.5, 2.0, and 5.0 per cent, to assess how sensitive the different models were to this choice.

\begin{figure}
\includegraphics[width=\columnwidth]{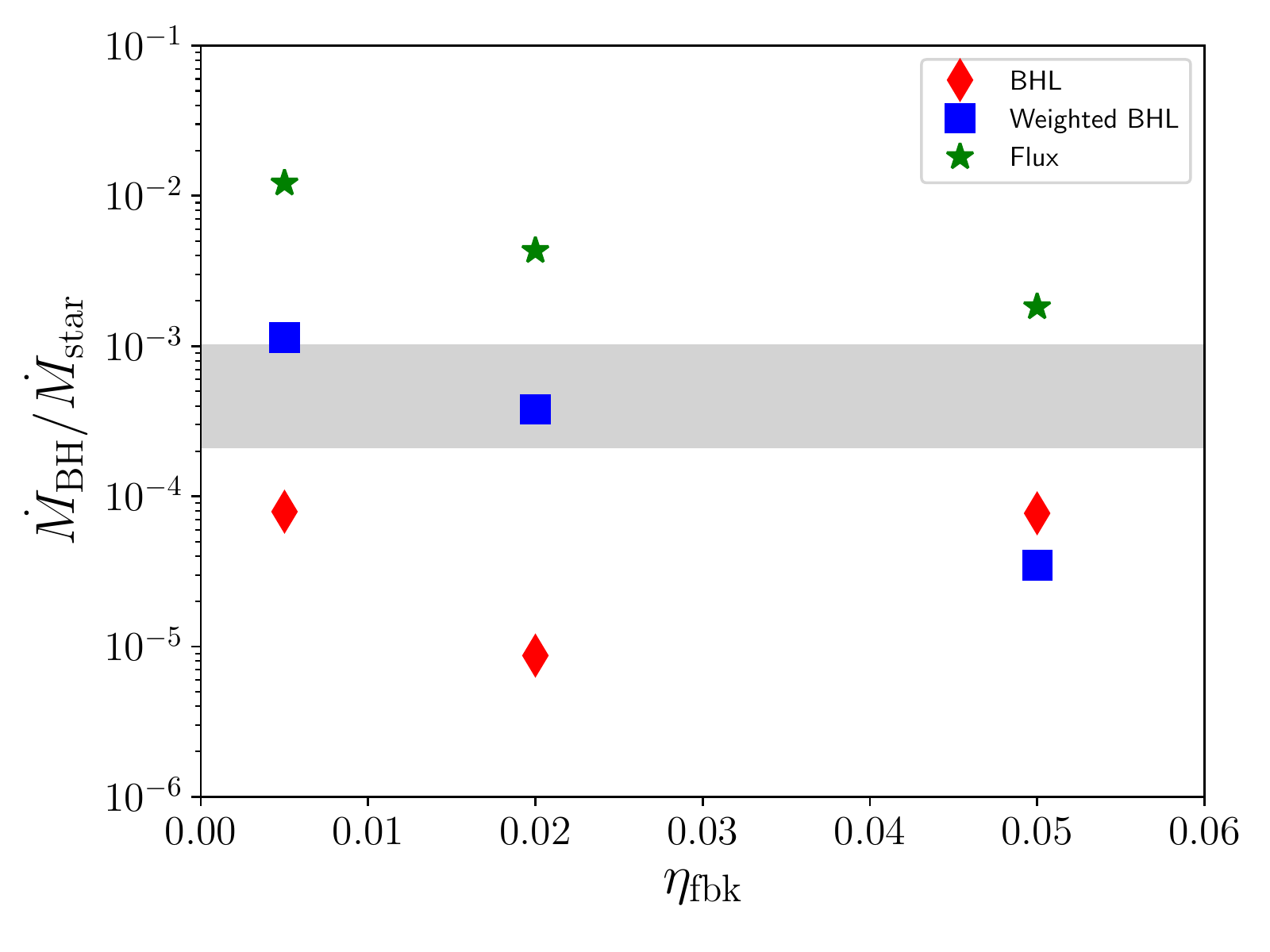}
\caption{Ratio of BH accretion rate and SFR for a suite of isolated Milky-Way like galaxies with a central MBH as a function of the MBH feedback efficiency. Model (i) is shown as red diamonds, model (ii) as blue squares, and model (iii) as green stars. The grey shaded area corresponds to the observational constraints from \citet{mullaney12} (upper edge) and \citet{chen13} (lower edge). Model (ii) is the least sensitive to both resolution and $\eta_{\rm fbk}$, and falls within the observational range for $0.005\lesssim \eta_{\rm fbk}\lesssim 0.02$. Flux accretion, that does not take into account the gas thermodynamical state, always overestimates the accretion rate, whereas the `standard' BHL tends to underestimate it, because of the stronger sensitivity to higher sound speeds and lower densities.}
\label{fig:test_accr}
\end{figure}

All simulations produce galaxies that remain on the correlation between MBH and galaxy mass. We have therefore explored additional ways to test the model, by looking at the relation between MBH accretion rate and SFR. The results are shown in Fig.~\ref{fig:test_accr},
where we report the ratio between MBH accretion rate and SFR for the different models, averaged over the last 500~Myr of each run (checking that different choices for the period do not affect our conclusions), and compare it with the observational constraints (grey shaded area) determined from \citet{mullaney12} (upper edge) and \citet{chen13} (lower edge). The red diamonds correspond to model (i), blue squares to model (ii), and green stars to model (iii). These results show that our fiducial model, applied on a local galaxy, produces a better agreement with local correlations of the $\dot{M}_{\rm BH}-\dot{M}_{\rm star}$ ratio. 


Furthermore, as demonstrated by \citet{negri17}, the feedback efficiency is dependent on resolution, with higher resolutions requiring a lower efficiency. This is due to the fact that, for a fixed number of elements affected by BH feedback and energy injected, heating is more effective at higher resolution, because the mass to be heated and swept up is smaller.

This can be also confirmed via simple analytical arguments. Let's assume that we have two simulations with different resolutions (a 'low' resolution case and a 'high' resolution case). We consider that both simulations have a MBH with mass $M_{\rm BH}$ surrounded by gas with the same properties, so that they measure the same $\dot{M}_{\rm BH}$. Assuming the BHL prescription, this ensures that the sound speed $c_s$ and the gas density $\rho$ are fixed for both cases.

In the two simulations, the accretion time-step can be defined by the Courant condition, i.e. $\Delta t = H/c_S$, with $H = \left(\frac{3N_{\rm ngb}m_{\rm gas}}{4\pi\rho}\right)^{1/3}$ the kernel size, with $N_{\rm ngb}$ the number of encompassed neighbours and $m_{\rm gas}$ the gas particle mass. The only parameters that change in the simulations are $N_{\rm ngb}$ and $m_{\rm gas}$.

In a single time-step, the accretion powered energy injected can be written as
\begin{equation}
\Delta E = L_{\rm BH} \Delta t = \eta_{\rm fbk}\eta_{\rm rad} \dot{M}_{\rm BH} c^2 \left(\frac{3N_{\rm ngb}m_{\rm gas}}{4\pi\rho c_s^3}\right)^{1/3}.
\end{equation}

Now, assuming for simplicity that every neighbour in the kernel receives the same amount of energy (let's keep in mind that the injection is usually done via a kernel-weighting procedure, which gives more weight to the closest particles), the average temperature increase is
\begin{equation}
\Delta T = \frac{\Delta E}{N_{\rm ngb}m_{\rm gas}} \frac{(\gamma-1) m_{\rm p}}{k_{\rm B}} = \zeta \frac{\eta_{\rm c}}{(N_{\rm ngb}m_{\rm gas})^{2/3}},
\end{equation}
where $\zeta=\left(4/3\pi\rho\right)^{-1/3}[\eta_{\rm rad}\dot{M}_{\rm BH}c^2(\gamma-1)m_{\rm p}]/[c_s k_{\rm B}]$ is constant for both resolutions.

If we now compute the temperature variation ratio between low- and high-resolution simulations, we get
\begin{equation}
r_{\Delta T}=\frac{\Delta T_{\rm low}}{\Delta T_{\rm high}} = \frac{\eta_{\rm fbk,low}}{\eta_{\rm fbk,high}}\left(\frac{N_{\rm ngb,high}m_{\rm gas,high}}{N_{\rm ngb,low}m_{\rm gas,low}}\right)^{2/3},
\end{equation}
where `low' and `high' identify the two resolutions.

If we require the same temperature increase observed in our isolated-galaxy, where $m_{\rm gas,low}=8.6\times 10^4\,\msun$ and $0.005\lesssim \eta_{\rm fbk} \lesssim 0.02$, in the cosmological run ($m_{\rm gas, high}\sim 1.5\times 10^4\,\msun$), then we should set $0.001\lesssim \eta_{\rm fbk,high}\lesssim 0.006$, consistent with our choice. This interval is also consistent with that proposed by \citet{choi12} for kinetic feedback models.

In general, if we compare our choice ($m_{\rm gas}\sim 1.5\times 10^4\rm M_\odot$, $N_{\rm ngb}\sim 32$, and $\eta_{\rm fbk}=0.005$) to other cosmological simulations \citep[e.g.][]{schaye15,dimatteo17}, where $m_{\rm gas} \sim 2-3\times 10^6\rm\, M_\odot$, $N_{\rm ngb}\sim 64$ and $\eta_{\rm fbk}=0.05$ were employed, we get
\begin{equation}
r_{\Delta T} = 10\left(\frac{32}{64}\frac{1.5\times 10^4}{2\times 10^6}\right)^{2/3} = 0.24.
\end{equation}

Since $r_{\Delta T}<1$, for fixed $M_{\rm BH}$, $c_s$, and $\rho$, our feedback typically heats up the gas more than that in low-resolution simulations. Obviously, the combination of the idealised simulations and the simple analytical arguments do not assure that the value we employed is the right one to reproduce the local correlations in a full cosmological simulation, but unfortunately a full box simulation evolved down to $z=0$ with the resolution we achieve is currently unfeasible.

\section{The effect of the BH coupling efficiency in the cosmological run}
\label{app:efficiency}

\begin{figure}
\includegraphics[width=\columnwidth]{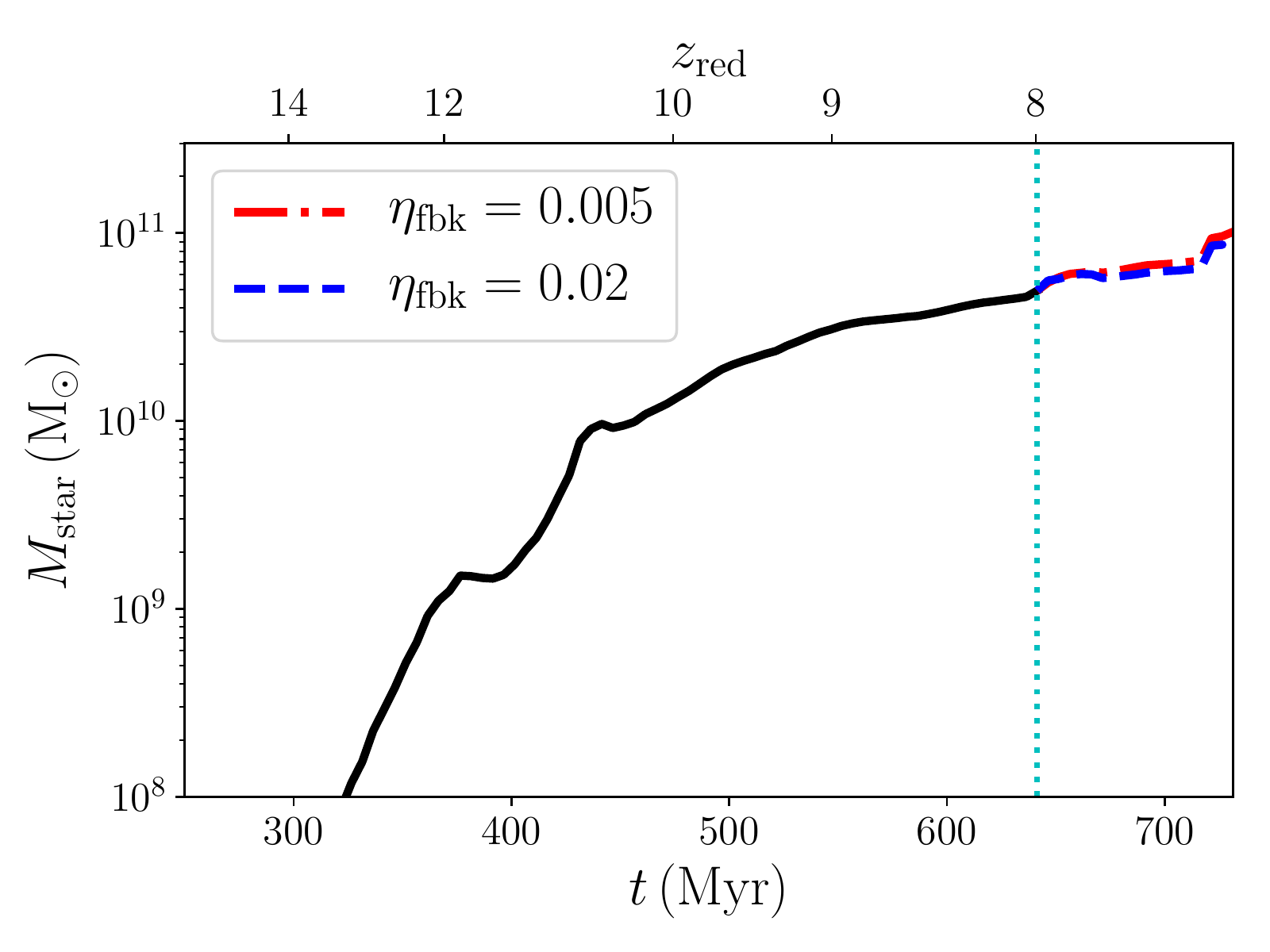}
\includegraphics[width=\columnwidth]{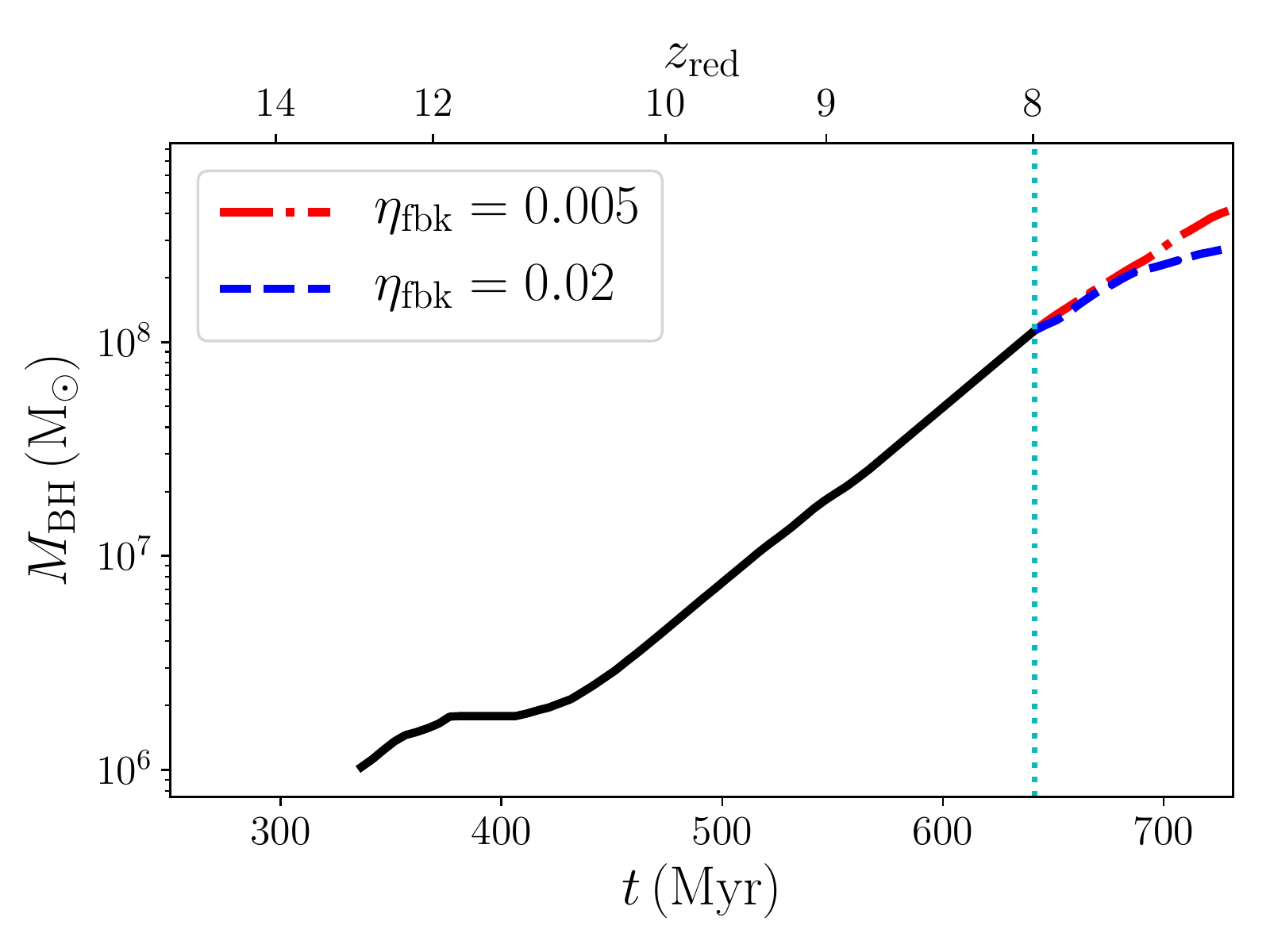}
\caption{Evolution of $M_{\rm star}$ (top panel) and $M_{\rm BH}$ (bottom panel) for different MBH feedback efficiencies$\eta_{\rm fbk}$. The black solid line corresponds to the fiducial evolution down to $z=8$, with the cyan dotted vertical line indicating the initial redshift of the comparison; the red dot-dashed and the blue dashed lines correspond to $\eta_{\rm fbk}=0.005$ and $\eta_{\rm fbk}=0.02$, from $z=8$ to $z=7.24$, corresponding to about 100~Myr. The stellar mass is not significantly affected by the different efficiency, with a small suppression by about 10 per cent. The MBH growth instead changes significantly when a higher $\eta_{\rm fbk}$ is employed, but only after the first $\sim 45-50$~Myr (about an {\it e}-folding time), when the MBH has become massive enough for its feedback to dominate against the radiative cooling of the gas. This suggests that at earlier times the two choices should not play any role in the evolution. Anyway, the difference in the mass of the MBH after the 100~Myr run is still within a factor of two.}
\label{fig:star_comp}
\end{figure}

In our simulation, we employ a low efficiency value $\eta_{\rm fbk}=5\times 10^{-3}$, finding a good  agreement with observations of high-redshift quasars. To assess how the star formation history of the galaxy and accretion history of the MBH would change if a more powerful feedback was used, we restarted our simulation at $z=8$ and evolved it for about 100~Myr with an efficiency four times higher, i.e. $\eta_{\rm fbk}=0.02$, as in \citet{tremmel15}. This change in the injected energy results in a moderately larger low-density region around the MBH only, but its impact on the galaxy host as a whole is small, as shown in the top panel of Fig.~\ref{fig:star_comp}, where we show the evolution of the galaxy host stellar mass in the two cases (the red dot-dashed line is the fiducial run and the blue dashed one the higher-efficiency case). The black solid line corresponds to the redshift range with only the fiducial case, with the cyan dotted vertical line identifying the initial redshift of the comparison. After 100 Myr, the difference in the stellar mass is about 10 per cent. In the bottom panel, we show instead the MBH mass in the two cases (with the same color/line-style scheme). Because of the stronger feedback, gas becomes hotter, and the growth is suppressed. However, during the first 45-50~Myr (about an $e$-folding time), when the MBH is less massive and the Eddington limit lower, only a small difference in accretion is observed, with the mass still increasing by about a factor of 2 and a difference of about 20 per cent. This is due to the balance between the energy injected by the MBH and the very high densities around it, that power strong radiative cooling, moderately inhibiting the MBH feedback effect. At later times, instead, the effect of the stronger feedback builds-up, keeping the gas hotter and quickly dropping the accretion rate to about 0.1 times the Eddington limit. Nevertheless, the difference in the MBH mass after 100~Myr is about a factor of two.
According to these results, we can expect that, because of the very high densities in high-redshift quasar hosts, a higher $\eta_{\rm fbk}$ would only play a role at late times, between $z=6$ and $z=8$, when the MBH has already exceeded $M_{\rm BH}\sim 10^8\,\msun$. The main growth phase, instead, would still be at the Eddington limit.

\section{The role of the initial BH mass}
\label{app:seed}
\begin{figure}
\includegraphics[width=\columnwidth]{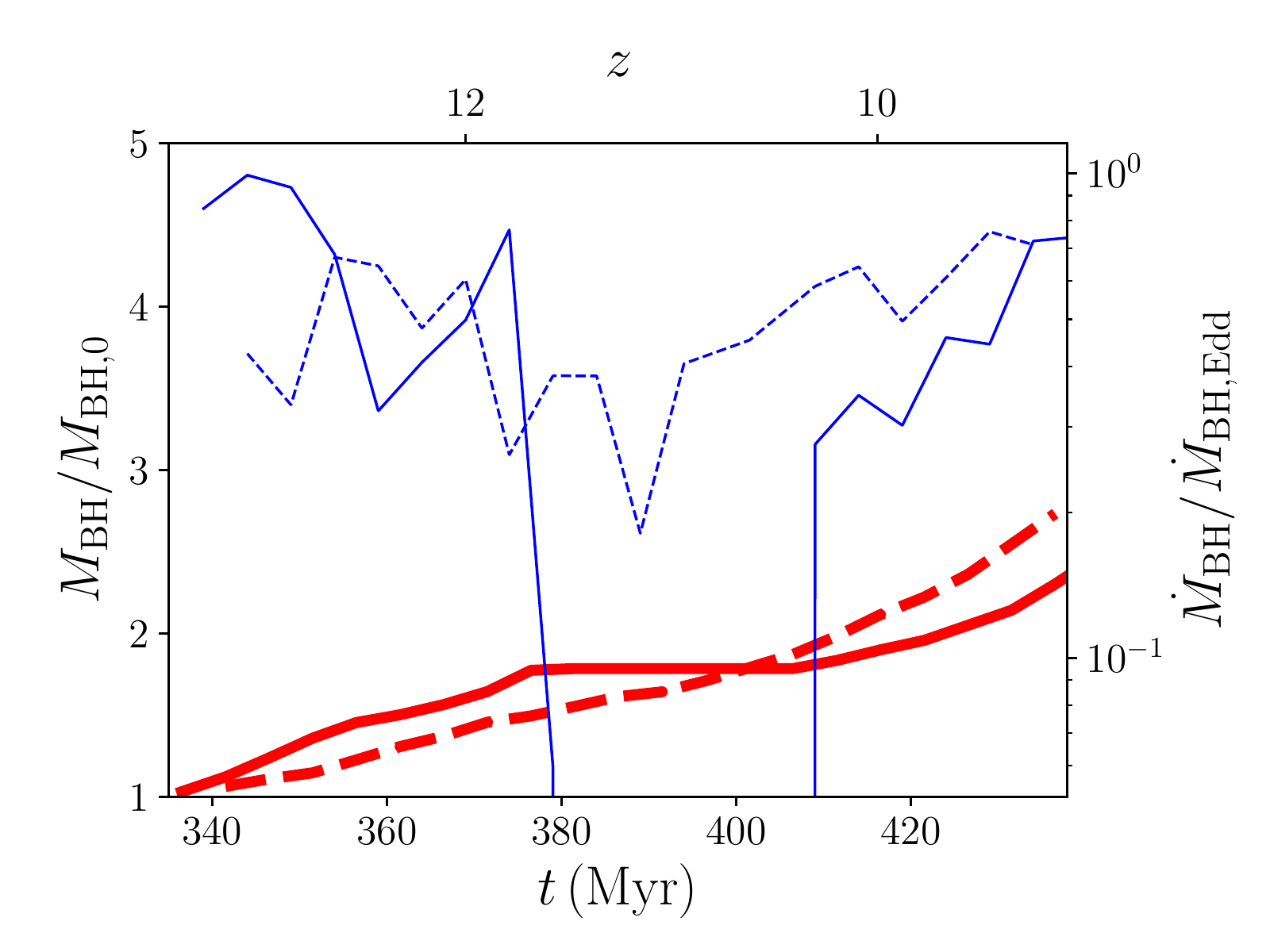}
\caption{BH masses (thick lines) and accretion rates (thin lines) for the fiducial run (solid lines) and the low-mass seed case ($M_{\rm BH,0}=10^5\,\msun$, dashed lines) from $z=14$ down to $z\sim10.6$. In the first few tens of Myr the growth is similar between the two BHs, i.e. it is regulated by the galaxy's ability to funnel gas towards the centre and not by the BH mass. At later times, when the fiducial BH stops accreting, the low-mass BH grows steadily, and reaches the Eddington rate earlier. This slightly reduces the mass difference between the two BHs. Although the computational cost of the run did not allow us to run this additional simulation down to $z=7$, we expect that the low-mass BH, remaining less massive than the fiducial case during the Eddington-limited phase, would reach the AGN feedback-dominated regime later (below $z=8$), likely catching up with the more massive one by $z=7$.}
\label{fig:bh_comp}
\end{figure}

In order to test the impact of the initial BH mass ($M_{\rm BH,0}$) on the initial BH growth, we have performed an additional run with a seed mass of $10^5\,\msun$ down to $z\sim 10.5$ and compared it with our fiducial case. In Fig.~\ref{fig:bh_comp}, we report the BH masses and accretion rates for the two runs. Immediately after formation, the lower mass seed accretes at a similar pace relative to the fiducial case. However, the situation changes when the BH in the fiducial case stops growing after a few tens of Myr (because of the combined effect of the SN feedback and the AGN heating), and the low-mass seed grows almost undisturbed. This confirms our conclusion that the initial accretion phases are mainly determined by the galaxy's ability to funnel gas towards the BH, and not by the initial BH mass. At later times, when the galaxy mass has exceeded a critical mass, the growth would proceed at the Eddington rate for both BHs, until AGN feedback could take over. In the lower-mass case, the AGN-dominated regime would be reached below $z=8$, the redshift where it occurs in the fiducial run, because of the lower BH mass, and this would likely result in similar masses observed by $z=7$.

\label{lastpage}
\end{document}